\documentclass[journal]{IEEEtran}
\usepackage{graphicx}
\usepackage{amsmath}
\usepackage{array}
\newcommand{\PreserveBackslash}[1]{\let\temp=\\#1\let\\=\temp}
\newcolumntype{C}[1]{>{\PreserveBackslash\centering}p{#1}}
\newcolumntype{R}[1]{>{\PreserveBackslash\raggedleft}p{#1}}
\newcolumntype{L}[1]{>{\PreserveBackslash\raggedright}p{#1}}
\usepackage[usenames]{color}
\usepackage{colortbl,booktabs}
\usepackage{tabularx}
\usepackage{multicol}
\usepackage{booktabs}
\usepackage[Symbol]{upgreek}
\usepackage{subfigure}
\usepackage{bm}
\usepackage{cite}
\usepackage{balance}
\usepackage[boxed]{algorithm2e}
\usepackage{url}
\usepackage{graphicx,amssymb,lineno}
\usepackage{amsmath,amsfonts,amssymb}
\usepackage{epsfig}
\usepackage{epstopdf}
\usepackage{colortbl}
\usepackage{color}
\usepackage{makecell}
\usepackage{threeparttable}
\usepackage{dcolumn}
\usepackage{multirow}
\usepackage{booktabs}
\usepackage{amsfonts}
\usepackage{stfloats}
\newcommand{\tabincell}[2]{\begin{tabular}{@{}#1@{}}#2\end{tabular}}  
\newcolumntype{d}[1]{D{.}{.}{#1}}

\usepackage{nomencl}
\makenomenclature

\begin{document}

\bibliographystyle{IEEEtran}

\title{Edge Intelligence: Architectures, Challenges, and Applications}

\author{Dianlei Xu,
        Tong Li,
        ~Yong Li,~\IEEEmembership{Senior Member, IEEE},
        ~Xiang Su,~\IEEEmembership{Member, IEEE},
        ~Sasu Tarkoma,~\IEEEmembership{Senior Member, IEEE},
        ~Tao Jiang,~\IEEEmembership{Fellow, IEEE},
      ~Jon~Crowcroft,~\IEEEmembership{Fellow, IEEE},
       and~Pan~Hui,~\IEEEmembership{Fellow, IEEE}

\thanks{D. Xu, T. Li, X. Su, and P. Hui are with the Department of Computer Science, University of Helsinki, 00014 Helsinki, Finland.\protect (e-mail: dianlei.xu@helsinki.fi, t.li@connect.ust.hk, xiang.su@helsinki.fi, sasu.tarkoma@helsinki.fi, panhui@cse.ust.hk.)} %
\thanks{D. Xu and Y. Li are with Beijing National Research Center for Information Science and Technology (BNRist), Department of Electronic Engineering, Tsinghua University, Beijing 100084, China.\protect (e-mail: liyong07@tsinghua.edu.cn.)} %
\thanks{T. Li and P. Hui are also with the Department of Computer Science and Engineering, Hong Kong University of Science and Technology, Hong Kong.}
\thanks{T. Jiang is with the School of Electronics Information and Communications, Huazhong University of Science and Technology, Wuhan 430074, China. (e-mail: taojiang@ieee.org)}
\thanks{J. Crowcroft is with the Computer Laboratory, University of Cambridge, William Gates Building, 15 JJ Thomson Avenue, Cambridge CB3 0FD, UK. (e-mail: Jon.Crowcroft@cl.cam.ac.uk.)}
}

\maketitle

\begin{abstract}
Edge intelligence refers to a set of connected systems and devices for data collection, caching, processing, and analysis proximity to where data is captured based on artificial intelligence. Edge intelligence aims at enhancing data processing and protect the privacy and security of the data and users.
Although recently emerged, spanning the period from 2011 to now, this field of research has shown explosive growth over the past five years.
In this paper, we present a thorough and comprehensive survey on the literature surrounding edge intelligence. We first identify four fundamental components of edge intelligence, i.e. edge caching, edge training, edge inference, and edge offloading based on theoretical and practical results pertaining to proposed and deployed systems. We then aim for a systematic classification of the state of the solutions by examining research results and observations for each of the four components and present a taxonomy that includes practical problems, adopted techniques, and application goals. For each category, we elaborate, compare and analyse the literature from the perspectives of adopted techniques, objectives, performance, advantages and drawbacks, etc.
This article provides a comprehensive survey to edge intelligence and its application areas. In addition, we summarise the development of the emerging research fields and the current state-of-the-art and discuss the important open issues and possible theoretical and technical directions.
\end{abstract}


\printnomenclature

\IEEEpeerreviewmaketitle

\section{Introduction}\label{S1}

\IEEEPARstart{W}{ith} the breakthrough of Artificial Intelligence (AI), we are witnessing a booming increase in AI-based applications and services. AI technology, e.g., machine learning (ML) and deep learning (DL), achieves state-of-the-art performance in various fields, ranging from facial recognition \cite{parkhi2015deep,sun2015deepid3}, natural language processing \cite{bengio2003neural, collobert2008unified}, computer vision \cite{kendall2017uncertainties,alhaija2017augmented}, traffic prediction \cite{huang2014deep,lv2014traffic}, and anomaly detection \cite{fang2016abnormal,potes2016ensemble}. Benefiting from the services provided by these intelligent applications and services, our lifestyles have been dramatically changed.

However, existing intelligent applications are computation-intensive, which present strict requirements on resources, e.g., CPU, GPU, memory, and network, which makes it impossible to be available anytime and anywhere for end users. Although current end devices are increasingly powerful, it is still insufficient to support some deep learning models. For example, most voice assistants, e.g., Apple Siri and Google Microsoft's Cortana, are based on cloud computing and they would not function if the network is unavailable. Moreover, existing intelligent applications generally adopt centralised data management, which requires users to upload their data to central cloud based data-centre.
However, there is giant volume of data which has been generated and collected by billions of mobile users and Internet of Thing (IoT) devices distributed at the network edge. According to Cisco's forecast, there will be 850 ZB of data generated by mobile users and IoT devices by 2021 \cite{cisco2016}. Uploading such volume of data to the cloud consumes significant bandwidth resources, which would also result in unacceptable latency for users. On the other hand, users increasingly concern their privacy. The European Union has promulgated General Data Protection Regulation (GDPR) to protect private information of users \cite{voigt2017eu}. If mobile users upload their personal data to the cloud for a specific intelligent application, they would take the risk of privacy leakage, i.e., the personal data might be extracted by malicious hackers or companies for illegal purposes.

Edge computing \cite{7488250,tran2017collaborative,garcia2015edge,hu2015mobile,bonomi2012fog} emerges as an extension of cloud computing to push cloud services to the proximity of end users. Edge computing offers virtual computing platforms which provide computing, storage, and networking resources, which are usually located at the edge of networks. The devices that provide services for end devices are referred to as edge servers, which could be IoT gateways, routers, and micro data centres in mobile network base stations, on vehicles, and amongst other places. End devices, such as mobile phones, IoT devices, and embedded devices that requests services from edge servers are called edge devices. The main advantages of the edge computing paradigm could be summarised into three aspects. ($\romannumeral 1$) Ultra-low latency: computation usually takes place in the proximity of the source data, which saves substantial amounts of time on data transmission. Edge servers provides nearly real-time responses to end devices. ($\romannumeral 2$) Saving energy for end devices: since end devices could offload computing tasks to edge servers, the energy consumption on end devices would significantly shrink. Consequently, the battery life of end devices would be extended. ($\romannumeral 3$) Scalability: cloud computing is still available if there are no enough resource on edge devices or edge servers. In such a case, the cloud server would help to perform tasks. In addition, end devices with idle resources could communicate amongst themselves to collaboratively finish a task. The capability of the edge computing paradigm is flexible to accommodate different application scenarios.


Edge computing addresses the critical challenges of AI based applications and the combination of edge computing and AI provides a promising solution.
This new paradigm of intelligence is called edge intelligence \cite{wang2019edge,li2018edge}, also named mobile intelligence \cite{wang2019first}. Edge intelligence refers to a set of connected systems and devices for data collection, caching, processing, and analysis proximity to where data is collected, with the purpose of enhancing the quality and speed of data processing and to protect the privacy and security of data. Compared with traditional cloud-based intelligence that requires end devices to upload generated or collected data to the remote cloud, edge intelligence processes and analyses data locally, which effectively protects users' privacy, reduces response time, and saves on bandwidth resources \cite{khelifi2018bringing,lane2018deep}. Moreover, users could also customise intelligent applications by training ML/DL models with self-generated data \cite{chen2018federated,chen2019fedhealth}. It is predicted that edge intelligence will be a vital component in 6G network \cite{peltonen20206g}. It is also worth noting that AI could also be a powerful assistance for edge computing. This paradigm is called intelligent edge \cite{han2019convergence,8970161}, which is different from edge intelligence. The emphasis of edge intelligence is to realize intelligent applications in edge environment with the assistance of edge computing and protect users' privacy, while intelligent edge focuses on solving problems of edge computing with AI solutions, e.g., resource allocation optimization. Intelligent edge is out of our scope in this survey.


\begin{figure*}[tp!]
    \centering
    \subfigure[Centralized intelligence]{\label{intro1}
    \includegraphics[width=0.46\linewidth]{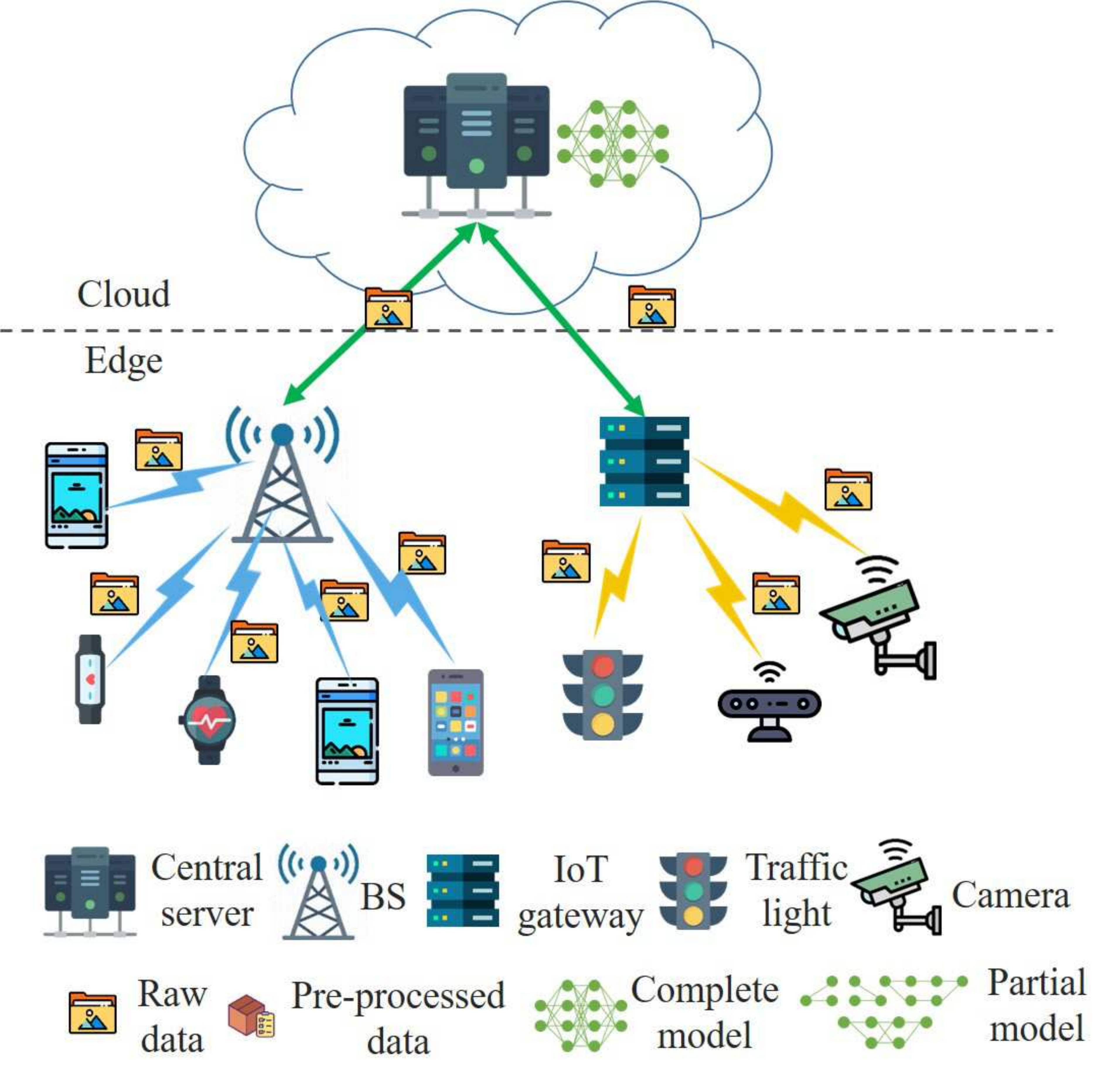}}
    \subfigure[Edge intelligence]{\label{intro2}
    \includegraphics[width=0.46\linewidth]{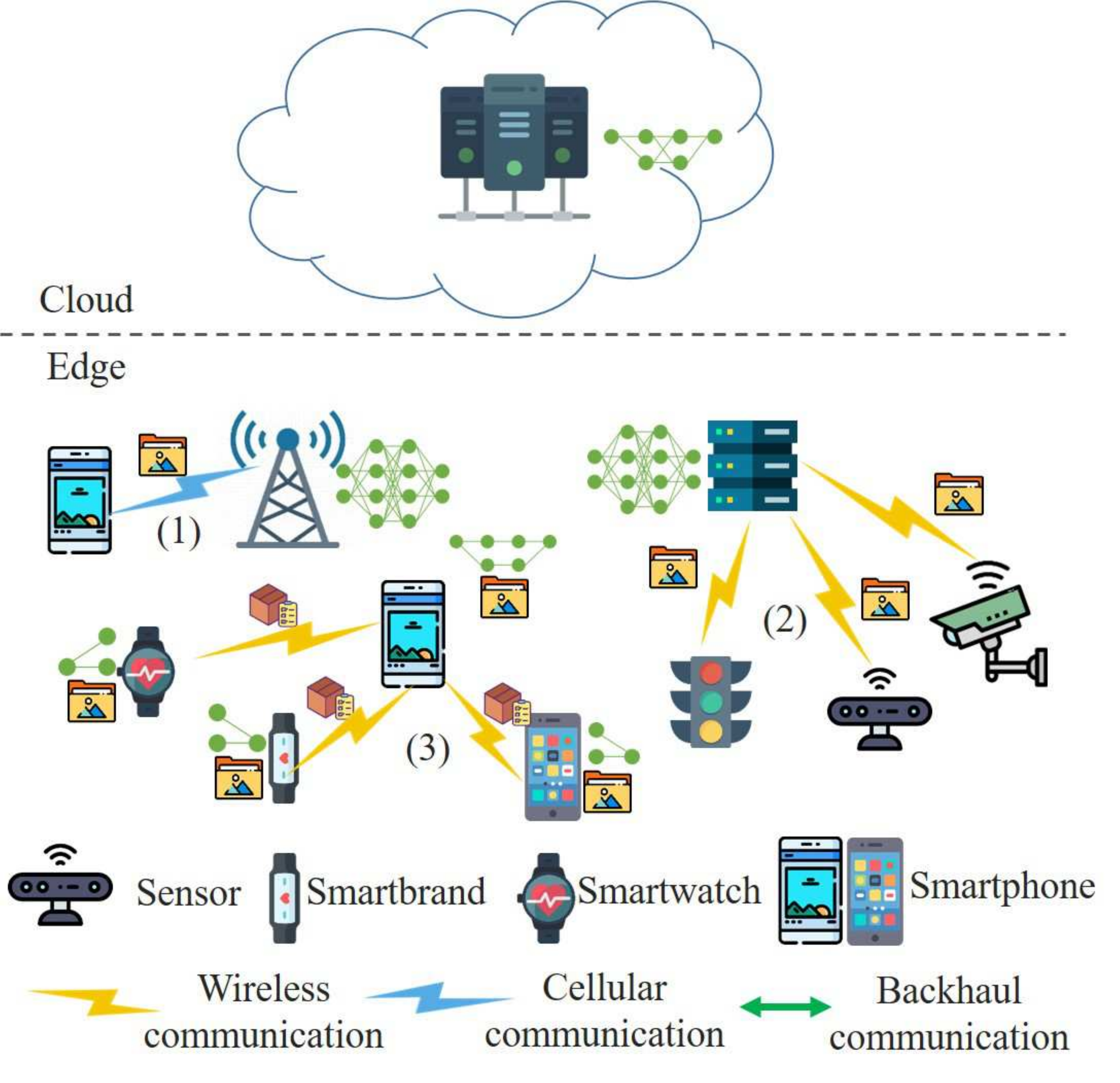}}
    \caption{The comparison of traditional intelligence and edge intelligence from the perspective of implementation. In traditional intelligence, all data must be uploaded to a central cloud server, whilst in edge intelligence, intelligent application tasks are done at the edge with locally-generated data in a distributed manner.}
    \label{intro}
    \end{figure*}

There exists lots of works which have proved the feasibility of edge intelligence by applying an edge intelligence paradigm to practical application areas. Yi \emph{et al.} implement a face recognition application across a smartphone and edge server \cite{yi2015fog}. Results show that the latency is reduced from 900ms to 169ms, compared with cloud based paradigm. Ha \emph{et al.} use a cloudlet to help a wearable cognitive assistance execute recognition tasks, which saves energy consumption by 30\%-40\% \cite{ha2014towards}. Some researchers pay attention to the performance of AI in the context of edge computing. Lane \emph{et al.} successfully implement a constrained DL model on smartphones for activity recognition \cite{lane2017squeezing}. The demo achieves a better performance than shallow models, which demonstrates that ordinary smart devices are qualified for simple DL models. Similar verification is also done on wearable devices \cite{radu2018multimodal} and embedded devices \cite{lane2015early}. The most famous edge intelligence application is Google G-board, which uses federated learning \cite{mcmahan2017federated} to collaboratively train the typing prediction model on smartphones. Each user uses their own typing records to train G-board. Hence, the trained G-board could be used immediately, powering experiences personalised by the way users use this application.

This paper aims at providing a comprehensive survey to the development and the state-of-the-art of edge intelligence. As far as we know, there exist few recent efforts \cite{yang2019federated,wang2018deep,mohammadi2018deep,zhang2019deep,han2019convergence,zhou2019edge} in this direction, but they have very different focuses from our survey. Table \ref{relatedwork} summarizes the comparison among these works. Specifically, Yang \emph{et al.} provide a survey on federated learning, in which they mainly focus on the architecture and applications of federated learning \cite{yang2019federated}. The authors divide literature of federated learning into three classifications: horizontal federated learning, vertical federated learning, and federated transfer learning. Federated learning is also involved as a collaborative training structure in our survey. We present how federated learning is applied in edge environment with the consideration of communication and privacy/security issues. The focus of \cite{wang2018deep} is how to realize the training and inference of DL models on a single mobile device. They briefly introduce some challenges and existing solutions from the perspective of training and inference. By contrast, we provide a more comprehensive and deeper review on solutions from the perspective of model design, model compression, and model acceleration. We also survey how to realize model training and inference with collaboration of edge devices and edge servers, even the assistance from the cloud server, in addition to solo training and inference at edge. Mohammadi \emph{et al.} review works on IoT big data analytic with DL approaches \cite{mohammadi2018deep}. Edge intelligence is not necessary in this work. The emphasis of survey \cite{zhang2019deep} is how to use DL techniques to deal with the problems in wireless networks, e.g., spectrum resource allocation, which has no overlap with our work.

To our best knowledge, ref. \cite{han2019convergence} and \cite{zhou2019edge} are two most relevant articles to our survey. The focus of \cite{han2019convergence} is the inter-availability between edge computing and DL. Hence the scope of ref. \cite{han2019convergence} includes two parts: DL for edge computing, and edge computing for DL. The former part focuses on some optimisation problems at edge with DL approaches, whilst the latter part focus on applying DL in the context of edge computing (i.e., techniques to perform DL at edge). The authors analyse these two parts from a macro view. By contrast, we pay more attention to the implementation of AI based applications and services (including ML and DL) with the assistance of edge resources from the micro view. More specifically, we provide more comprehensive and detailed classification and comparison on existing works of this research area from multi-dimensions. Not only the implementation of AI based applications and services (including both training and inference), but also the management of edge data and the required computing power are involved in our work. Moreover, statistically, there are only 40 coincident surveyed papers between \cite{han2019convergence} and our work. Similarly, the survey \cite{zhou2019edge} analyses the implementation of edge intelligence on different layers from a macro view. They propose a six-level rating to describe edge intelligence. This is also involved in our work. Different from \cite{zhou2019edge}, we analyse its implementation from a micro view, e.g., offloading strategies and caching strategies.
\begin{table*}[ht]
\footnotesize
\centering
\caption{Comparison of relative surveys.}
\label{relatedwork}
\vspace{-0.1in}
\begin{tabular}{c|c|c|c|c}
\hline
\bf{Ref.} & \bf{Year} & \bf{Domain} & \bf{Scope} &  \bf{Analysing perspective
}\\
\hline
\cite{yang2019federated} & 2019 & Federated learning  & \tabincell{c}{Horizontal federated learning, vertical federated learning,\\ and federated transfer learning} & Macro-perspective \\
\hline
\cite{wang2018deep} & 2018 & DL-based mobile applications  & \tabincell{c}{Training and inference on single mobile device} & Micro-perspective\\
\hline
\cite{mohammadi2018deep} & 2018 & IoT big data &  \tabincell{c}{DL in IoT applications, and DL on IoT devices} &Micro-perspective\\
\hline
\cite{zhang2019deep} & 2019 & Intelligent wireless network &  \tabincell{c}{Algorithms that enables DL in wireless networks\\ Applications ranging from traffic analytic to security} & Micro-perspective \\
\hline
\cite{han2019convergence} & 2019 & \tabincell{c}{Edge intelligence \\ Intelligent edge} & \tabincell{c}{Training and inference systems\\  DL for optimizing edge, and DL application on edge} & Macro-perspective \\
\hline
\cite{zhou2019edge} & 2019 & Edge intelligence  & \tabincell{c}{Cloud-edge-device coordination architecture\\  Optimisation technologies in training and inference}  & Macro-perspective\\
\hline
Our work & 2020 & Edge intelligence  & \tabincell{c}{Edge caching, edge training, edge inference, and edge offloading}  & Micro-perspective \\
\hline
\end{tabular}
\end{table*}

Our survey focuses on how to realise edge intelligence in a systematic way. There exist three key components in AI, i.e. data, model/algorithm\footnote{model and algorithm are interchangeable in this article}, and computation. A complete process of implementing AI applications involves data collection and management, model training, and model inference. Computation plays an essential role throughout the whole process.
Hence, we limit the scope of our survey on four aspects, including how to cache data to fuel intelligent applications (i.e., edge caching), how to train intelligent applications at the edge (i.e., edge training), how to infer intelligent applications at the edge (edge inference), and how to provide sufficient computing power for intelligent applications at the edge (edge offloading). Our contributions are summarized as following:
\begin{itemize}
    \item We survey recent research achievements on edge intelligence and identify four key components: edge caching, edge training, edge inference, and edge offloading. For each component, we outline a systematical and comprehensive classification from a multi-dimensional view, e.g., practical challenges, solutions, optimisation goals, etc.
    \item We present thorough discussion and analysis on relevant papers in the field of edge intelligence from multiple views, e.g., applicable scenarios, methodology, performance, etc. and summarise their advantages and shortcomings.
    \item We discuss and summarise open issues and challenges in the implementation of edge intelligence, and outline five important future research directions and development trends, i.e., data scarcity, data consistency, adaptability of model/algorithms, privacy and security, and incentive mechanisms.
\end{itemize}

The remainder of this article is organized as follow. Section II overviews the research on edge intelligence, with considerations of the essential elements of edge intelligence, as well as the development situation of this research field. We present detailed introduction, discussion, and analysis on the development and recent advances of edge caching, edge training, edge inference, and edge offloading in Section III to Section VI, respectively. Finally, we discuss the open issue and possible solutions for future research in Section VII, and conclude the paper in Section VIII.

\section{Overview}\label{S2}

As an emerging research area, edge intelligence has received broad interests in the past few years. With benefits from edge computing and artificial intelligence techniques, combination of their contributions enables easy-to-use intelligent applications for users in daily lives and less dependent on the centralised cloud.

For convenience, we present the comparison between traditional centralised intelligence with edge intelligence from the perspective of implementation in Fig.~\ref{intro}. Traditional centralised intelligence is shown in Fig.~\ref{intro1}, where all edge devices first upload data to the central server for intelligent tasks, e.g., model training or inference. The central server/data-centre is usually, but not necessarily, located in remote cloud. After the processing on the central server, results, e.g., recognition or prediction results, are transmitted back to edge devices. Fig.~\ref{intro2} demonstrates the implementation of edge intelligence, where a task, e.g., recognition and prediction is either done by edge servers and peer devices, or with the edge-cloud cooperation paradigm. A very small amount, or none of the data is uploaded to the cloud. For example, in area $(1)$ and $(2)$, cloudlet, i.e. BS and IoT gateway could run complete intelligent models/algorithms to provide services for edge devices. In area $(3)$, a model is divided into several parts with different functions, which are performed by several edge devices. These edge devices work together to finish the task.

\begin{figure*}[tp!]
    \begin{center}
    \includegraphics[width=\linewidth]{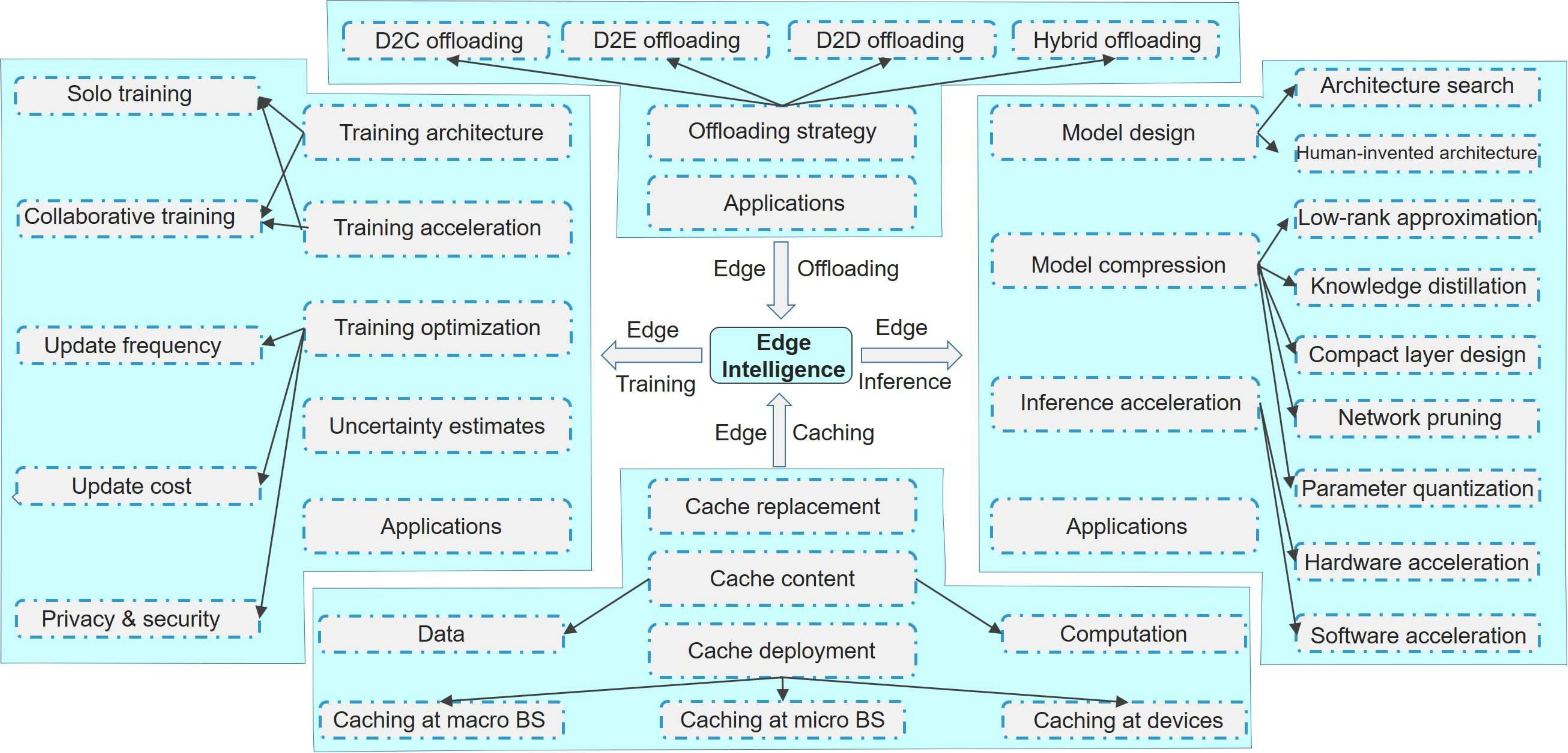}
    \end{center}
    \caption{The classification of edge intelligence literature.}
    \label{overview}
    \end{figure*}

It is known that three most important elements for an intelligent application are: data, model, and computation. Suppose that an intelligent application is a `human', model would be the `body', and computation is the `heart' which powers the `body'. Data is then the `book'. The `human' improves their abilities by learning knowledge extracted from the `book'. After learning, the `human' starts to work with the learned knowledge. Correspondingly, the complete deployment of most intelligent applications (unsupervised learning based application is not included) includes three components: data collection and management (preparing the `book'), training (learning), and inference (working). Computation is a hidden component that is essential for the other three components. Combined with an edge environment, these three obvious  components turn into edge cache (data collection and storage at edge), edge training (training at edge), and edge inference (inference at edge), respectively. Note that edge devices and edge servers are usually not powerful. Computation at edge usually is done via offloading. Hence, the hidden component turn into edge offloading (computation at edge). Our classification is organised around these four components, each of which features multidimensional analysis and discussion. The global outline of our proposed classification is shown in Fig.~\ref{overview}. For each component, we identify key problems in practical implementation and further break down these problems into multiple specific issues to outline a tiered classification. Next, we present an overview of these modules shown as Fig.~\ref{overview}.


\subsection{Edge Caching}

\begin{figure}[tp!]
    \begin{center}
    \includegraphics[width=0.99\columnwidth]{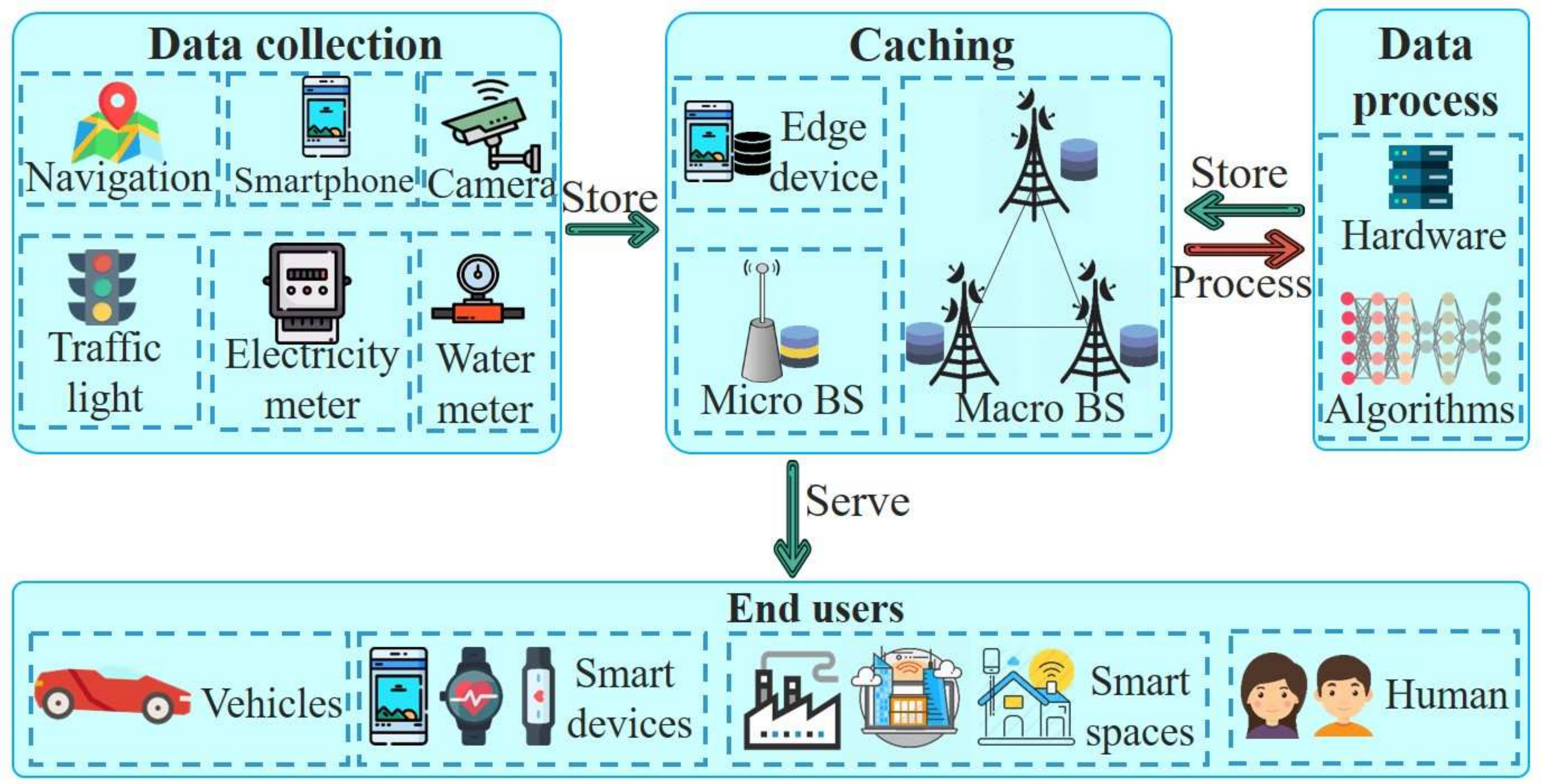}
    \end{center}
    \caption{The illustration of edge caching. Data generated by mobile users and collected from surrounding environments is collected and stored on edge devices, micro BSs, and macro BSs. Such data is processed and analysed by intelligent algorithms to provide services for end users.}
    \label{caching}
    \end{figure}

In edge intelligence, edge caching refers to a distributed data system proximity to end users, which collects and stores the data generated by edge devices and surrounding environments, and the data received from the Internet to support intelligent applications for users at the edge. Fig.~\ref{caching} presents the essential idea of edge caching. Data is distributed at the edge. For example, mobile users' information generated by themselves is stored in their smartphones. Edge devices such as monitoring devices and sensors record the environmental information. Such data is stored at reasonable places and used for processing and analysis by intelligent algorithms to provide services for end users. For example, the video captured by cameras could be cached on vehicles for aided driving \cite{xu2017accelerating}. BS caches the data that users recently accessed from the Internet to characterise users' interests for better a recommendation service \cite{liu2018learning}. To implement edge caching, we answer three questions: ($\romannumeral 1$) what to cache, ($\romannumeral 2$) where to cache, and ($\romannumeral 3$) how to cache. The structure of this section is organised as the bottom module in Fig.~\ref{overview}.

For the first problem, what to cache, we know that caching is based on the redundancy of requests. In edge caching, the collected data is inputted into intelligent applications and results are sent back to where data is cached. Hence, there are two kinds of redundancy: data redundancy and computation redundancy. Data redundancy, also named communication redundancy, means that the inputs of an intelligent application may be the same or partially the same. For example, in continuous mobile vision analysis, there are large amounts of similar pixels between consecutive frames. Some resource-constrained edge devices need to upload collected videos to edge servers or the cloud for further processing. With cache, edge devices only needs to upload different pixels or frames. For the repeated part, edge devices could reuse the results to avoid unnecessary computation. Ref.  \cite{crane2008robust,adar2008large,traverso2013temporal,dernbach2016cache} have investigated the pattern of data redundancy. Caching based on such redundancy could effectively reduce computation and accelerate the inference.
Computation redundancy means that the requested computing tasks of intelligent applications may be the same. For example, an edge server provides image recognition services for edge devices. The recognition tasks from the same context may be the same, e.g., the same tasks of flower recognition from different users of the same area. Edge servers could directly send the recognition results achieved previously back to users. Such kind of caching could significantly decrease computation and execution time. Some practical applications based on computation redundancy are developed in \cite{guo2018foggycache,GoogleStreetView,huitl2012tumindoor}.

For the second problem, where to cache, existing works mainly focus on three places to deploy caches: macro BSs, micro BSs, and edge devices. The work in \cite{liu2016caching,li2016performance} have discussed the advantages of caching at macro BSs from the perspective of covering range and hit probability. Some researchers also focus on the cached content at macro BSs. According to statistics, two kinds of content are considered: popular files \cite{blaszczyszyn2015optimal,andrews2011tractable,roadknight2000file,ahlehagh2014video,chatzieleftheriou2017caching} and intelligent models \cite{konevcny2016federated,li2019abnormal,li2019federated,abad2019hierarchical,lim2019federated}. In edge intelligence, macro BSs usually work as edge servers, which provide intelligent services with cached data. In addition, some works \cite{wang2014cache,peng2015backhaul,tran2017collaborative,xu2017understanding} consider how to improve the performance of caching with the cooperation among macro BSs. Compared with macro BSs, micro BSs provide smaller coverage but higher quality of experience \cite{xiao2016dynamic,xiao2016analytical,lai2019clustering,xiao2016resource,hamidouche2016mean,zhao2016communications,liu2016cache}. Existing efforts on this area mainly focus on two problems: how to deliver the cached content, and what to cache. For the aspect of delivery, research mainly focuses on two directions: delivery from single BS \cite{pantisano2014cache}, and delivery from multiple BSs based on the cooperation amongst them \cite{shanmugam2013femtocaching,liu2017energy,ao2015distributed,ao2018fast,chen2017cooperative}. Considering the small coverage of micro BSs and the mobility of mobile users, research on handover and users' mobility for better delivery service \cite{krishnan2017effect,guan2014mobicacher,Poularakis2017Code,Ozfatura2018Mobility} is also carried out. In addition, the optimal content to cache, i.e., data redundancy based content \cite{chang2018learn,kader2015leveraging,pantisano2015match,cheng2019localized,bacstuug2015transfer,el2018edge,quevedo2014case,sharma2017live} and computation redundancy based content \cite{drolia2017cachier,lv2007multi,guo2018foggycache,drolia2017precog,venugopal2018shadow,taylor2018adaptive,zhao2018privacy,ogden2018modi} is thoroughly investigated. Edge devices are usually of limited resources and high mobility, compared with macro BSs and micro BSs. Therefore, only few efforts pay attention to the problem of caching on a single edge device. For example, \cite{xu2017accelerating,cavigelli2019cbinfer,huynh2017deepmon,chen2015glimpse,xu2017accelerating} studies the problem of what to cache based on communication and computation redundancy in some specific applications, e.g., computer vision. Most researchers adopt collaborative caching amongst edge devices, especially in the network with dense users. They usually formulate the caching problem into an optimisation problem on the content replacement \cite{chen2017optimal,giatsoglou2017d2d,qiu2019popularity,malak2014optimal,peng2018optimal,ji2013optimal,ji2015throughput,chen2016cooperative,chen2017high,afshang2016fundamentals}, association policy \cite{golrezaei2012wireless,naderializadeh2014utilize,jarray2016effects,jarray2016effects,krishnan2017effect,bastug2014living,newman2010networks,bai2016caching}, and incentive mechanisms \cite{chen2016caching,taghizadeh2013distributed}.

Since the storage capacity of macro BSs, micro BSs, and edge devices is limited, the content replacement must be considered. Works on this problem focus on designing replacement policies to maximise the service quality, such as popularity based schemes \cite{blasco2014learning,bacstuug2013proactive}, and ML based schemes \cite{wang2018edge,bacstuug2013proactive}.



\subsection{Edge Training}
\begin{figure}[tp!]
    \begin{center}
    \includegraphics[width=0.99\columnwidth]{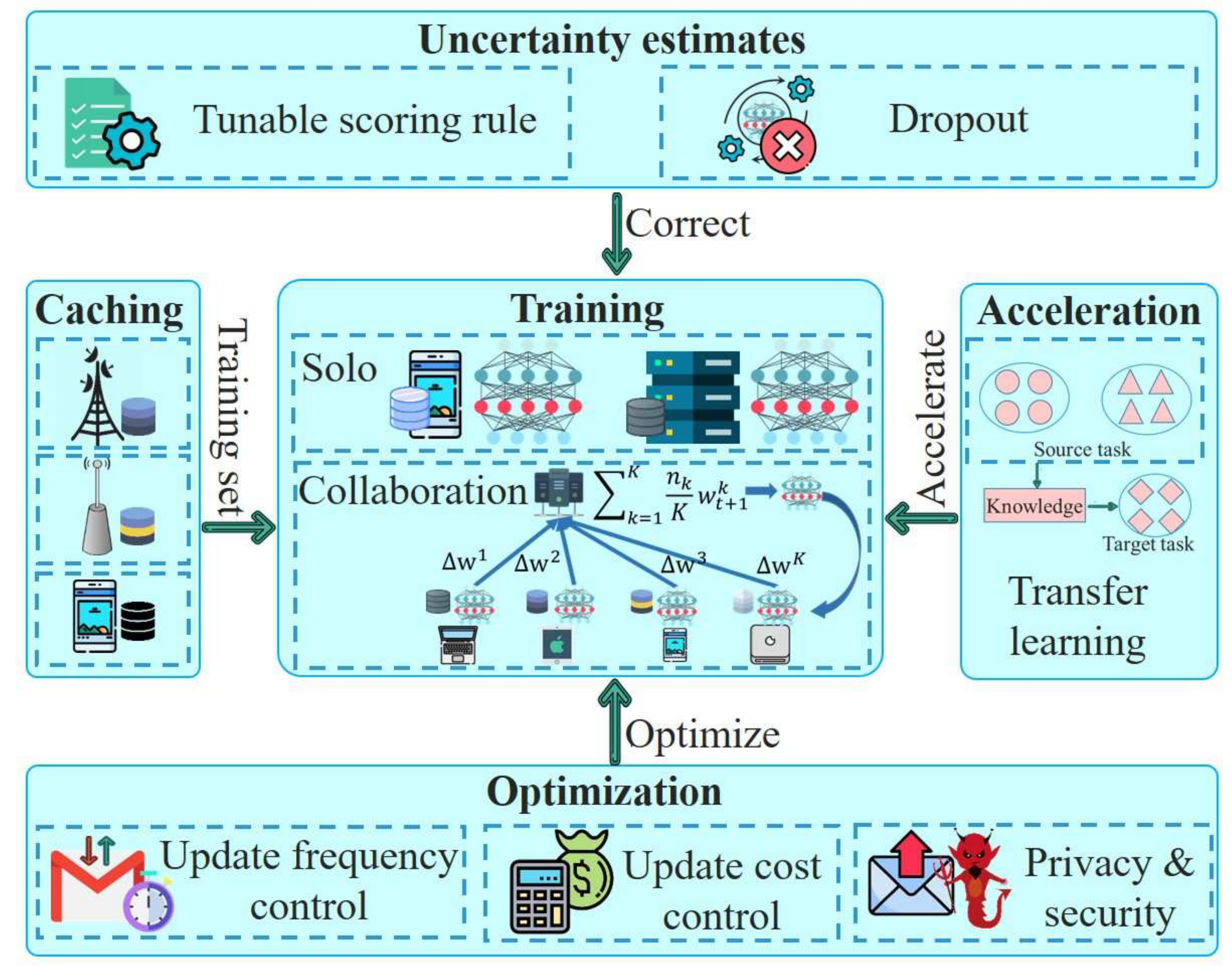}
    \end{center}
    \caption{The illustration of edge training. The model/algorithm is trained either on a single device (solo training), or by the collaboration of edge devices (collaborative training) with training sets cached at the edge. Acceleration module speeds up the training, whilst the optimisation module solves problems in training, e.g., update frequency, update cost, and privacy and security issues. Uncertainty estimates module controls the uncertainty in training.}
    \label{training}
\end{figure}

Edge training refers to a distributed learning procedure that learns the optimal values for all the weights and bias, or the hidden patterns based on the training set cached at the edge. For example, Google develops an intelligent input application, named G-board, which learns user's input habits with the user's input history and provides more precise prediction on the user's next input \cite{mcmahan2017federated}. The architecture of edge training is shown as Fig.~\ref{training}. Different from traditional centralised training procedures on powerful servers or computing clusters, edge training usually occurs on edge servers or edge devices, which are usually not as powerful as centralised servers or computing clusters. Hence, in addition to the problem of training set (caching), four key problems should be considered for edge training: ($\romannumeral 1$) how to train (the training architecture), ($\romannumeral 2$) how to make the training faster (acceleration), ($\romannumeral 3$) how to optimise the training procedure (optimisation), and ($\romannumeral 4$) how to estimate the uncertainty of the model output (uncertainty estimates). The structure of this section is organised as the left module in Fig.~\ref{overview}.

For the first problem, researchers design two training architectures: solo training \cite{chen2018exploring,lane2017squeezing,radu2018multimodal,lane2015early} and collaborative training \cite{mcmahan2017federated,li2016deepcham,huang2018task,valerio2017communication,xing2018enabling}. Solo training means training tasks are performed on a single device, without assistance from others, whilst collaborative training means that multiple devices cooperate to train a common model/algorithm. Since solo training has a higher requirement on the hardware, which is usually unavailable, most existing literature focuses on collaborative training architectures.

Different from centralised training paradigms, in which powerful CPUs and GPUs could guarantee a good result with a limited training time, edge training is much slower. Some researchers pay attention to the acceleration of edge training. Corresponding to training architecture, works on training acceleration are divided into two categories: acceleration for solo training \cite{chen2018exploring,valery2017cpu,valery2018low,miu2015bootstrapping,9073978,shahmohammadi2017smartwatch,flutura2018drinkwatch}, and collaborative training \cite{xing2018enabling,smith2017federated,3375312}.

Solo training is a closed system, in which only iterative computation on single devices is needed to get the optimal parameters or patterns. In contrast, collaborative training is based on the cooperation of multiple devices, which requires periodic communication for updating. Update frequency and update cost are two factors which affect the performance of communication efficiency and training result. Researchers on this area mainly focus on how to maintain the performance of the model/algorithm with lower update frequency \cite{konevcny2015federated,konevcny2016federated,konevcny2015federated,konevcny2016federated,mcmahan2016communication,zhao2018federated,8995775,8963610,wang2019adaptive,wang2017co,8952884,9026922,9014530,8950073,chen2016revisiting}, and update cost \cite{konevcny2016federated,konevcny2017stochastic,lin2017deep,hardy2017distributed,smith2017federated,caldas2018expanding}. In addition, the public nature of collaborative training is vulnerable to malicious users. There is also some literature which focuses on the privacy \cite{mcmahan2016communication,yang2019securing,bonawitz2017practical,liu2018secure,gentry2009fully,geyer2017differentially,9069945,dwork2011differential,abadi2016deep,mcmahan2017learning,zhuo2019federated,cheng2019secureboost} and security \cite{biggio2012poisoning,steinhardt2017certified,fung2018mitigating,douceur2002sybil,blanchard2017machine,chen2017distributed,yin2018byzantine,chen2017targeted,bagdasaryan2018backdoor,bhagoji2018analyzing,blanchard2017machine,yin2018byzantine} issues.

In DL training, the output results may be erroneously interpreted as model confidence. Estimating uncertainty is easy on traditional intelligence, whilst it is resource-consuming for edge training. Some literature \cite{gal2016dropout,yao2018rdeepsense} pays attention to this problem and proposes various kinds of solutions to reduce computation and energy consumption.

We also summarised some typical applications of edge training \cite{bonawitz2019towards,nishio2019client,konevcny2015federated,konevcny2016federated,mcmahan2016communication,hard2018federated,chen2019federated,mcmahan2017federated,yang2018applied,ramaswamy2019federated,ramaswamy2019federated,sheller2018multi,roy2019braintorrent,samarakoon2018distributed,nguyen2018diot} that adopt the above-mentioned solutions and approaches.

\subsection{Edge Inference}\label{edgeinference}
\begin{figure}[tp!]
    \begin{center}
    \includegraphics[width=0.99\columnwidth]{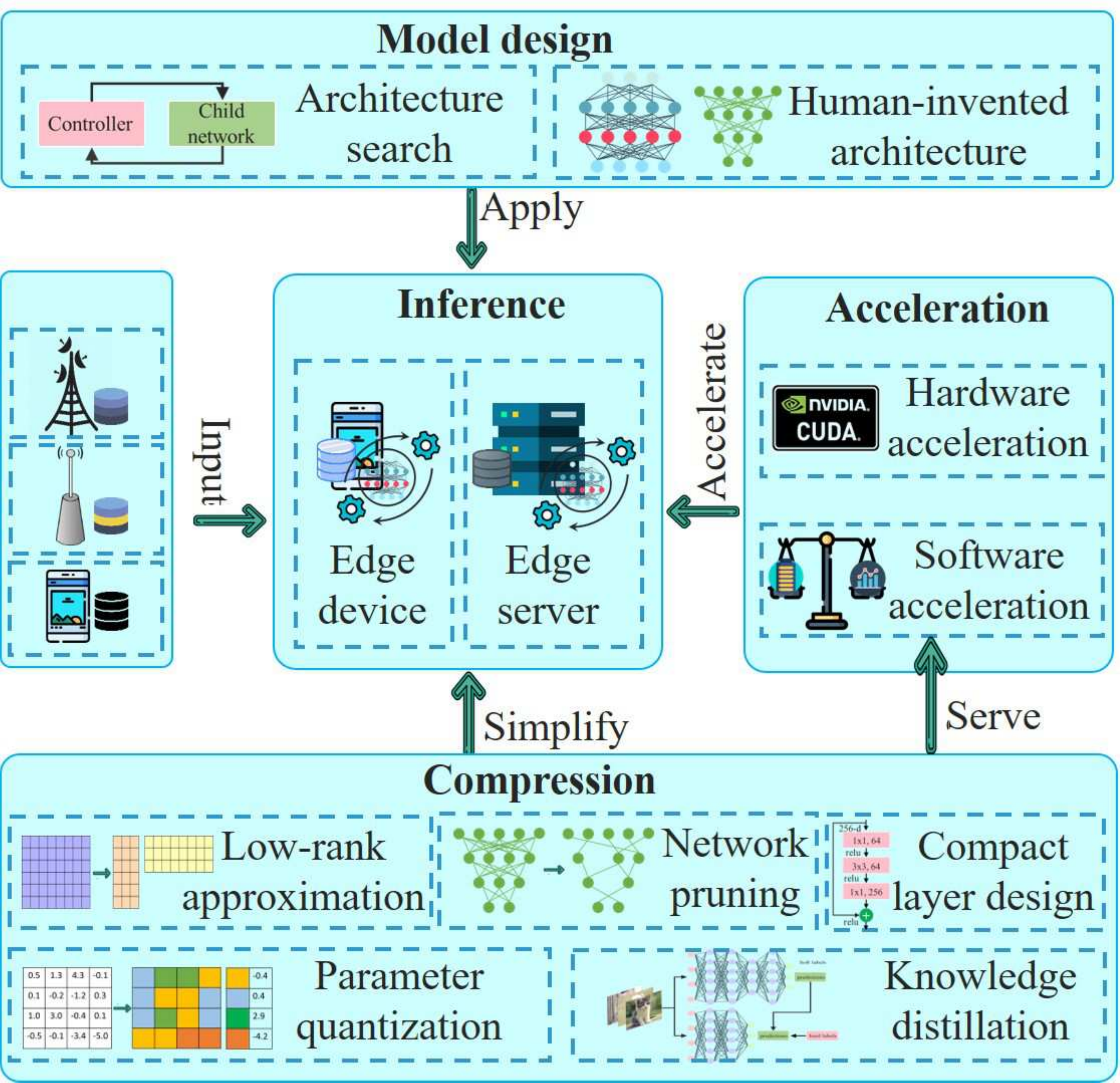}
    \end{center}
    \caption{The illustration of edge inference. AI models/algorithms are designed either by machines or humans. Models could be further compressed through compression technologies: low-rank approximation, network pruning, compact layer design, parameter quantisation, and knowledge distillation. Hardware and software solutions are used to accelerate the inference with input data.}
    \label{inference}
    \end{figure}

Edge inference is the stage where a trained model/algorithm is used to infer the testing instance by a forward pass to compute the output on edge devices and servers. For example, developers have designed a face verification application based DL, and employ on-device inference \cite{chen2018mobilefacenets,duong2018mobiface}, which achieves high accuracy and low computation cost. The architecture of edge inference is shown as Fig.~\ref{inference}. Most existing AI models are designed to be implemented on devices which have powerful CPUs and GPUs, this is not applicable in an edge environment. Hence, the critical problems of employing edge inference are: ($\romannumeral 1$) how to make models applicable for their deployment on edge devices or servers (design new models, or compress existing models), and ($\romannumeral 2$) how to accelerate edge inference to provide real-time responses. The structure of this section is organised as the right module in Fig.~\ref{overview}.

For the problem of how to make models applicable for the edge environment, researchers mainly focus on two research directions: design new models/algorithms that have less requirements on the hardware, naturally suitable for edge environments, and compress existing models to reduce unnecessary operation during inference. For the first direction, there are two ways to design new models: let machines themselves design optimal models, i.e., architecture search \cite{zoph2018learning,real2018regularized,cortes2017adanet,real2018regularized,tan2018mnasnet,liu2018darts,real2018regularized}; and human-invented architectures with the application of depth-wise separable convolution \cite{howard2017mobilenets,zhang2017hello,wofk2019fastdepth} and group convolution \cite{zhang2018shufflenet,qin2018merging}. We also summarise some typical applications based on these architectures, including face recognition \cite{chen2018mobilefacenets,duong2018mobiface,sandler2018mobilenetv2}, human activity recognition (HAR) \cite{bhattacharya2016smart,almaslukh2018robust,almaslukh2018,sundaramoorthy2018harnet,radu2016towards,cruciani2018automatic,bo2018detecting,yao2017deepsense,yao2018qualitydeepsense}, vehicle driving \cite{streiffer2017darnet,liu2015toward,bo2013you,yang2011detecting}, and audio sensing \cite{lane2015deepear,georgiev2017low}.
For the second direction, i.e., model compression, researchers focus on compressing existing models to obtain thinner and smaller models, which are more computation- and energy-efficient with negligible or even no loss on accuracy. There are five commonly used approaches on model compression: low-rank approximation \cite{jaderberg2014speeding,denton2014exploiting,maji2017adapt,kim2015compression,wang2016accelerating,bhattacharya2016sparsification}, knowledge distillation \cite{bucilua2006model,hinton2015distilling,romero2014fitnets,zagoruyko2016paying,sau2016deep,crowley2018moonshine,li2018deeprebirth,zhou2018rocket,lopes2017data}, compact layer design \cite{szegedy2015going,he2016deep,alex2019yolo,iandola2016squeezenet,szegedy2017inception,shafiee2017squishednets,yang2019cdeeparch,zhang2018dynamically,shen2018cs}, network pruning \cite{guo2017pruning,han2015learning,gordon2018morphnet,manessi2018automated,molchanov2016pruning,you2019gate,yang2017designing,yao2017deepiot,liu2017learning,chen2016deep,LUO2020107461,9097925,Oyedotun_2020_WACV,Singh_2020_WACV,9018278}, and parameter quantisation \cite{gong2014compressing,chen2015compressing,han2015deep,wu2016quantized,courbariaux2015,courbariaux2016binarized,rastegari2016xnor,lin2015neural,denton2014exploiting,vanhoucke2011improving,alvarez2016efficient,nasution2017faster,peng2017running,anwar2015fixed,langroudi2018deep,soudry2014expectation,esser2015backpropagation,9018278}. In addition, we also summarise some typical applications \cite{mathur2017deepeye,zeng2017mobiledeeppill,wang2018deepsearch,kim2017design,xu2017fast,liu2018demand} that are based on model compression.

Similar to edge training, edge devices and servers are not as powerful as centralised servers or computing clusters. Hence, edge inference is much slower. Some literature focuses on solving this problem by accelerating edge inference. There are two commonly used acceleration approaches: hardware acceleration and software acceleration. Literature on hardware acceleration \cite{alzantot2017rstensorflow,loukadakis2018accelerating,oskouei2015gpu,latifi2016cnndroid,tsung2016high,rizvi2016gpgpu,rizvi2017deep,rizvi2017optimized,rizvi2017general,cao2017mobirnn,cao2017mobirnn,guihot2012renderscript,motamedi2016fast,motamedi2018cappuccino,motamedi2017machine,huynh2016deepsense,huynh2017deepmon,taylor2017adaptive,rallapalli2016very,redmon2016you,bettoni2017convolutional,ma2017optimizing,park2018implementation,chen2018understanding,chen2016eyeriss,chen2019eyeriss,9034111} mainly focuses on the parallel computing which is available as hardware on devices, e.g., CPU, GPU, and DSP. Literature on software acceleration \cite{han2016mcdnn,georgiev2017accelerating,lane2016deepx,lane2016accelerated,lane2016dxtk,yang2018netadapt,ma2017deeprt,abtahi2018accelerating,xu2017accelerating,cavigelli2019cbinfer,huynh2017deepmon} focus on optimising resource management, pipeline design, and compilers, based on compressed models.

\subsection{Edge offloading}
\begin{figure}[tp!]
    \begin{center}
    \includegraphics[width=0.99\columnwidth]{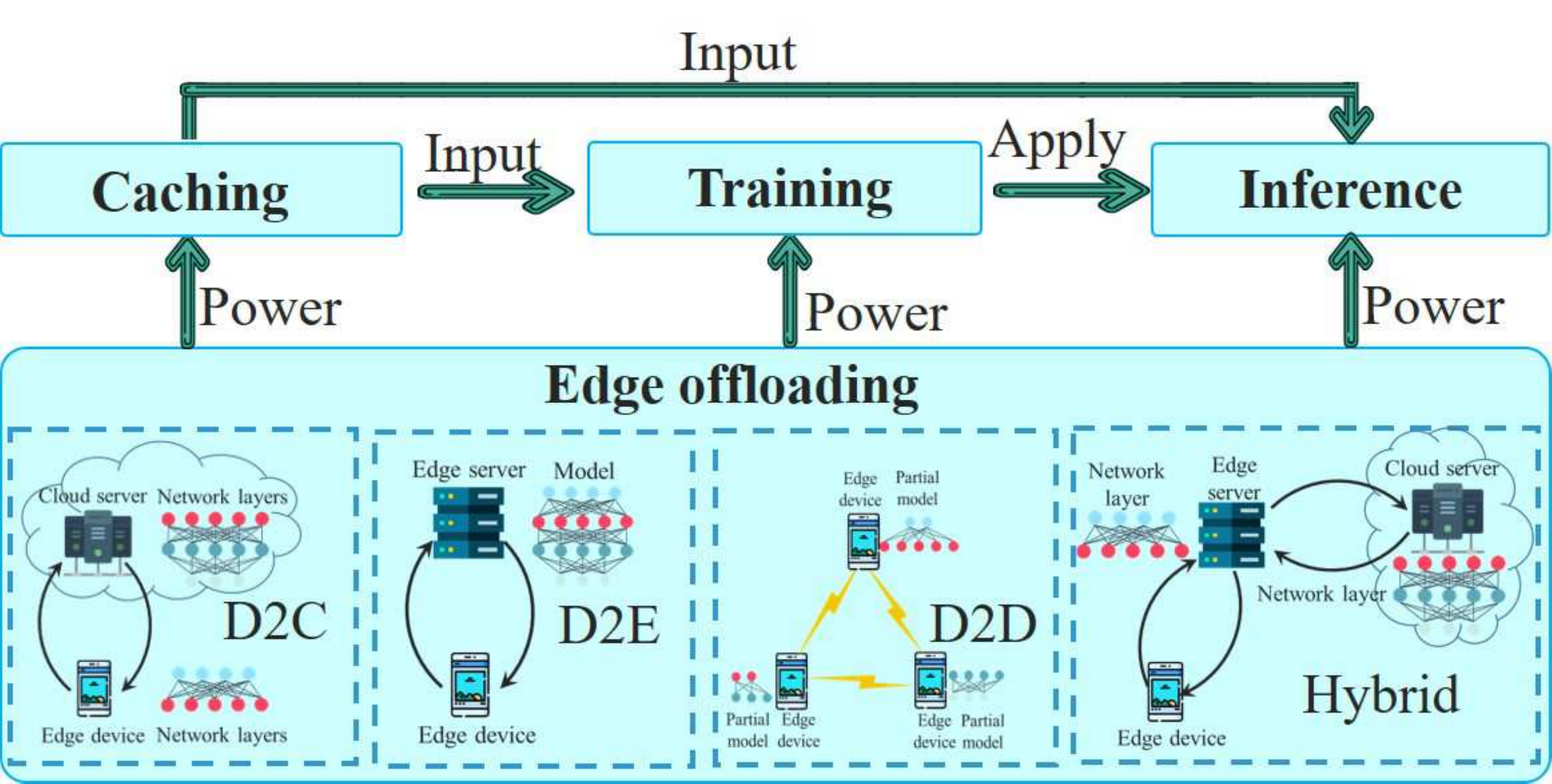}
    \end{center}
    \caption{The illustration of edge offloading. Edge offloading is located at the bottom layer in edge intelligence, which provides computing services for edge caching, edge training, and edge inference. The computing architecture includes D2C, D2E, D2D, and hybrid computing.}
    \label{computation}
    \end{figure}

As a necessary component of edge intelligence, edge offloading refers to a distributed computing paradigm, which provides computing service for edge caching, edge training, and edge inference. If a single edge device does not have enough resource for a specific edge intelligence application, it could offload application tasks to edge servers or other edge devices. The architecture of edge offloading is shown as Fig.~\ref{computation}. Edge offloading layer transparently provides computing services for the other three components of edge intelligence. In edge offloading, Offloading strategy is of utmost importance, which should give full play to the available resources in edge environment. The structure of this section is organised as the top module in Fig.~\ref{overview}.

Available computing resources are distributed in cloud servers, edge servers, and edge devices. Correspondingly, existing literature mainly focuses on four strategies: device-to-cloud (D2C) offloading, device-to-edge server (D2E) offloading, device-to-device (D2D) offloading, and hybrid offloading. Works on the D2C offloading strategy \cite{li2018learning,huang2017deep,eshratifar2018energy,kang2017neurosurgeon,osia2017hybrid,liu2017deeprotect,xu2019edgesanitizer,ananthanarayanan2017real,ali2018edge,naderiparizi2017glimpse,sanabria2018code,hanhirova2018latency,qi2017dnn,ran2017delivering,ran2018deepdecision,georgiev2016leo} prefer to leave pre-processing tasks on edge devices and offload the rest of the tasks to a cloud server, which could significantly reduce the amount of uploaded data and latency. Works on D2E offloading strategy \cite{ra2011odessa,streiffer2017eprivateeye,li2018edge,ko2018edge,tian2019lep,zhang2019task,jeong2018computation} also adopt such operation, which could further reduce latency and the dependency on cellular network. Most works on D2D offloading strategy \cite{xu2017enabling,liu2018edgeeye,song2018situ,yi2017lavea,hadidi2018musical,talagala2018eco,de2018dianne,fukushima2018microdeep,bach2017knowledge} focus on smart home scenarios, where IoT devices, smartwatches and smartphones collaboratively perform training/inference tasks. Hybrid offloading schemes \cite{morshed2017deep,teerapittayanon2017distributed,yousefpour2019guardians} have the strongest ability of adaptiveness, which makes the most of all the available resources.

We also summarise some typical applications that are based on these offloading strategies, including intelligent transportation \cite{ferdowsi2019deep}, smart industry \cite{li2018deep}, smart city \cite{tang2017incorporating}, and healthcare \cite{liu2017new} \cite{muhammed2018ubehealth}.

\subsection{Summary}
\begin{figure}[t]
    \begin{center}
    \includegraphics[width=\columnwidth]{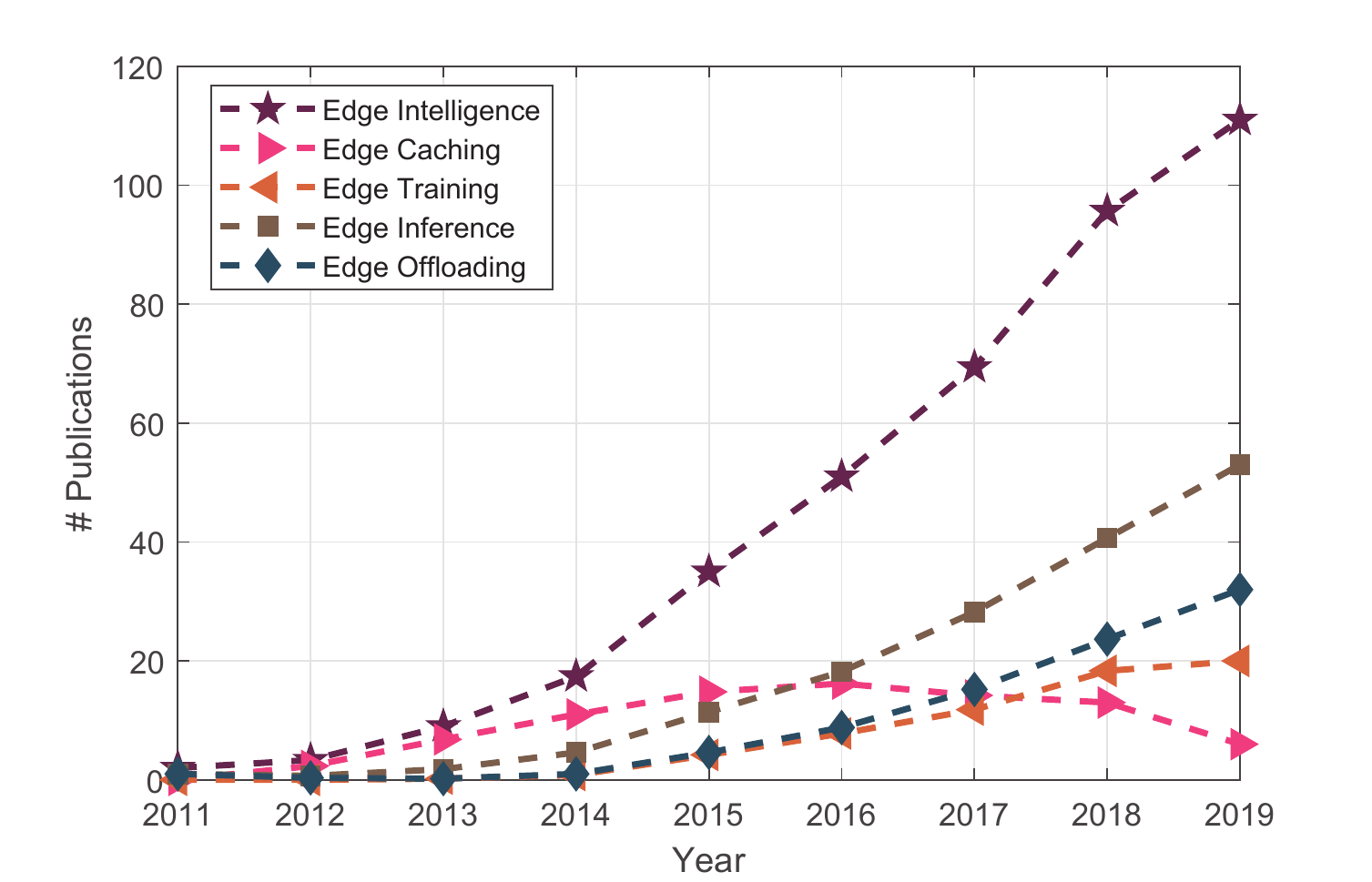}
    \end{center}
    \caption{Publication volume over time. These curves show the trend of publication volume in edge caching, edge training, edge computing, edge inference, and edge intelligence, respectively.}
    \label{trend}
    \end{figure}

In our survey, we identify four key components of edge intelligence, i.e. edge caching, edge training, edge inference, and edge offloading.
Edge intelligence shows an explosive developing trend with a huge amount of researcher have been carried out to investigate and realise edge intelligence over the past five years. We count the publication volume of edge intelligence, as shown in Fig.~\ref{trend}.

We see that this research area started from 2011, then grew at a slow pace before reaching 2014. Most strikingly, after 2014, there is a rapid rise in the publication volume of edge training, edge inference, and edge offloading. Meanwhile, the publication of edge caching is gradually winding down. Overall, the publication volume of edge intelligence is booming, which demonstrates a research field replete with activity. Such prosperity of this research filed owes to the following three reasons.

First, it is the booming development of intelligent techniques, e.g., deep learning and machine learning techniques that provides a theoretical foundation for the implementation of edge intelligence \cite{schwabacher2007survey,he2010survey,bahrammirzaee2010comparative}. Intelligent techniques achieve state-of-the-art performance on various fields, ranging from voice recognition, behaviour prediction, to automatic piloting. Benefiting from these achievements, our life has been dramatically changed. People hope to enjoy smart services anywhere and at any time. Meanwhile, most existing intelligent services are based on cloud computing, which brings inconvenience for users. For example, more and more people are using voice assistant on smartphone, e.g., MI AI and Apple Siri. However, such applications can not work without networks.

Second, the increasing big data distributed at the edge, which fuels the performance of edge intelligence \cite{zhang2017survey,singh2015survey,yi2015survey}. We have entered the era of IoT, where a giant amount of IoT devices collect sensory data from surrounding environment day and night and provide various kinds services for users. Uploading such giant amount of data to cloud data centres would consume significant bandwidth resources. Meanwhile, more and more people are concerned about privacy and security issues behind the uploaded data. Pushing intelligent frontiers is a promising solution to solve these problems and unleash the potential of big data at the edge.

Third, the maturing of edge computing systems \cite{tran2017collaborative,hu2015mobile} and peoples' increasing demand on smart life \cite{kim2017smart,su2011smart} facilitate the implementation of edge intelligence. Over the past few years, the theories of edge computing have moved towards application, and various kinds of applications have been developed to improve our life, e.g., augmented reality \cite{zhang2018jaguar,zhang2017cloudar,lin2016ubii}. At the same time, with the wide spreading of 5G networks, more and more IoT devices are implemented to construct smart cities. People are increasingly reliant on the convenient service provided from a smart life. Large efforts from both academia and industry are enacted to realise these demands.



%


%

\section{Edge Caching}\label{S3}

Initially, the concept of caching comes from computer systems. The cache was designed to fill the throughput gap between the main memory and registers \cite{hennessy2011computer} by exploring correlations of memory access patterns. Later, the caching idea was introduced in networks to fill the throughput gap between core networks and access networks. Nowadays, the cache is deployed in edge devices, like various base stations and end devices. By leveraging the spatiotemporal redundancy of communication and computation tasks, caching at the edge can significantly reduce transmission and computation latency and improve users' QoE~\cite{paschos2018role, li2018survey, wang2017survey}.

From existing studies, the critical issues of caching technologies in edge networks fall into three aspects: the cached content, caching places, and caching strategies. Next, we discuss and analyse relevant literature in edge caching in terms of the above three perspectives. The related subjects include the preliminary of caching, cache deployment, and cache replacement.

\subsection{Preliminary of Caching}\label{S3.1}
The critical idea of edge caching technologies is to explore the spatiotemporal redundancy of users' requests. The redundancy factor largely determines the feasibility of caching techniques. Generally, there are two categories, i.e., communication redundancy, and computation redundancy.

\subsubsection{Communication Redundancy}
Communication redundancy is caused by the repetitive access of popular multimedia files, such as audio, video, and webpages. Content with high popularity tends to be requested by mobile users many times. Thus, the network needs to transmit the content to these users over and over again. In this case, caching popular content at edge devices can eliminate enormous duplicated transmissions from core networks.

To better understand communication redundancy, many existing studies investigate content request patterns of mobile users. For example, Crane \emph{et al.}~\cite{crane2008robust} analyse 5 million videos on YouTube and regards the number of daily views as the popularity of videos. They then discover four temporal popularity patterns of videos and demonstrate the temporal communication redundancy of popular videos from the angle of content providers'. Meanwhile, Adar \emph{et al.} \cite{adar2008large} analyse five weeks of webpage interaction logs collected from over 612,000 users. They show that temporal revisitation also exists at the individual level. In \cite{traverso2013temporal}, Traverso \emph{et al.} characterise the popularity profiles of YouTube videos by using the access logs of 60,000 end-users. They propose a realistic arrival process model, i.e., the Shot Noise Model (SNM), to model the temporal revisitation of online videos. Moreover, Dernbach \emph{et al.} \cite{dernbach2016cache} exhibit the existence of regional movie preferences, i.e., the spatial communication redundancy of content by analysing the MovieLens dataset, which contains 6,000 user viewing logs. Consequently, the above studies apply large-scale and real-world datasets demonstrating communication redundancy from both the temporal and spatial dimensions. These studies support the feasibility of edge caching ideas and pave the way for employing edge caching technologies in real-world scenarios.

\subsubsection{Computation Redundancy}

Computation redundancy is caused by commonly used high computational complexity applications or AI models.
In the wave of AI, we are now surrounded by various intelligent edge devices such as smartphones, smart watches, and smart brands. These intelligent edge devices provide diverse applications to augment users' understanding of their surroundings \cite{GoogleGlass, Hololens}. For example, speech-recognition based AI assistants, e.g., Siri and Cortana, and song identification enabled music applications, have been widely used in peoples' daily lives. Such AI-based applications are of high computational complexity and cause high power consumption of the device~\cite{braud2020multipath, zhang2018jaguar, zhang2017cloudar}.

Meanwhile, some researchers have discovered the computation redundancy in AI-based applications. For example, in a park, nearby visitors may use their smartphones to recognise the same flowers and then search the information accordingly. In this case, there are a lot of unnecessary computations across devices. Therefore, if we offload such a painting recognition task to edges and cache the computation results, redundant computations can be further eliminated~\cite{lovagnini2018circe, guo2018foggycache}. In \cite{guo2018foggycache}, Guo \emph{et al.} crawl street views by using the Google Streetview API~\cite{GoogleStreetView} and builds an 'outdoor input image set'. They then find that around 83\% of the images exhibit redundancy, which leads to a large number of potential unnecessary computations for image reorganisation application. Also, they analyse NAVVIS \cite{huitl2012tumindoor}, an indoor view dataset, and observed that nearly 64\% of indoor images exhibit redundancy. To our knowledge, this work is the earliest one using real-world datasets to demonstrate the existence of computation redundancy.

\begin{figure}[tp!]
  \centering
  \includegraphics[width=1\columnwidth]{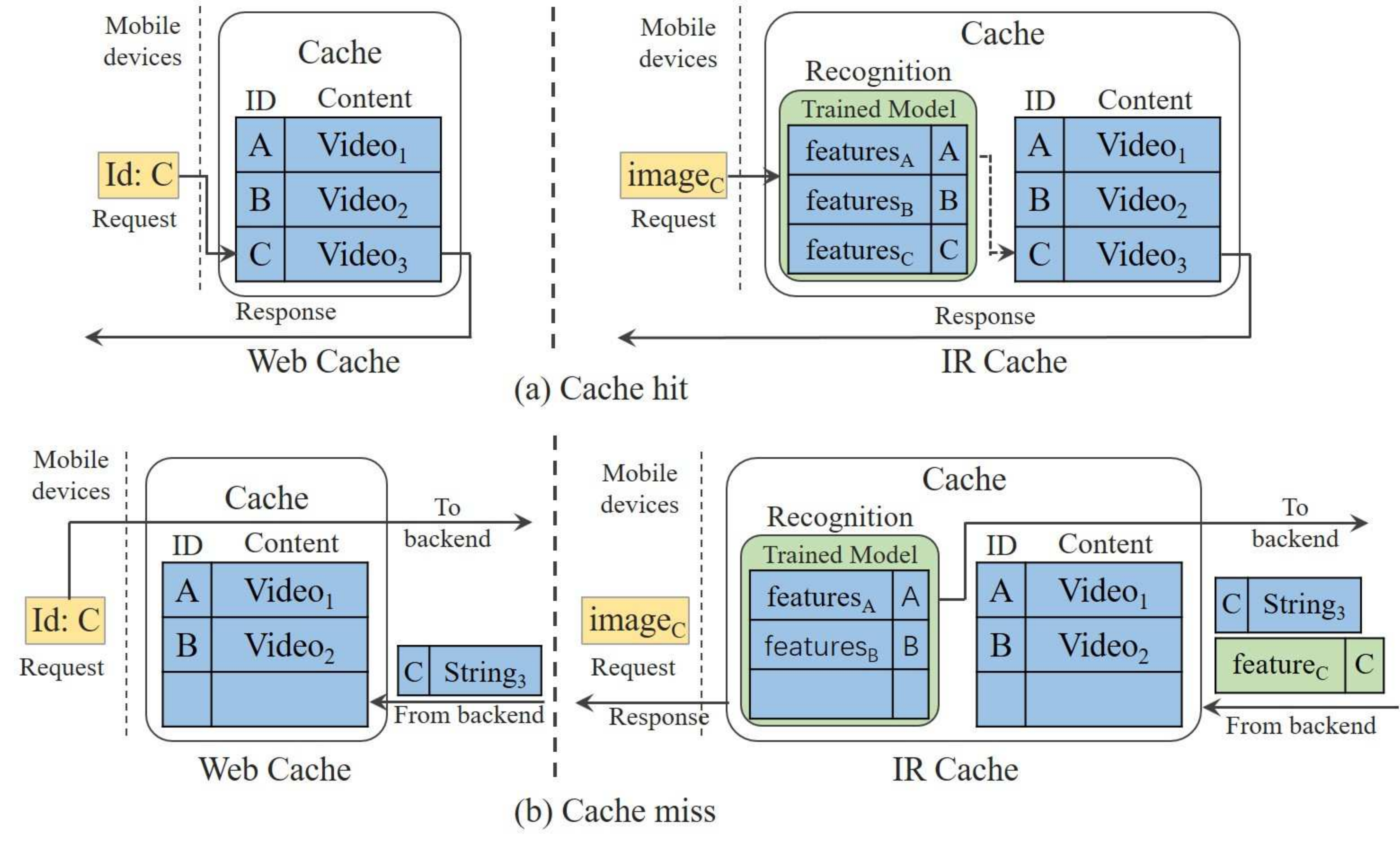}
   \caption{Illustration of image recognition (IR) cache, as well as the comparison with traditional web cache. The request/input for IR is an image. The system run the trained model to recognize the image, which will be labelled with an identifier. Then, the recognized image's identifier is used to find relevant content within cache. If there is a cache hit, content is returned. Otherwise, IR modelling is performed and the result is placed in the cache.}
   \label{fig:Cache1}
 \vspace*{-3mm}
\end{figure}

In the elimination of computation redundancy, an important step is to capture and quantify the similarity of users' requests. As shown in \figurename~\ref{fig:Cache1}, in the case of communication redundancy, a unique identifier can identify users' request content, e.g., Universal Resource Identifier, URI.
However, for computation redundancy, we first need to obtain the features of users' requests and then find the best match of the computation results according to the extracted features. It is notable that in computation redundancy, the cached content is computation results instead of requested files.

\subsection{Cache Deployment}\label{S3.2}
\begin{figure}[tp!]
  \centering
  \includegraphics[width=1.0\columnwidth]{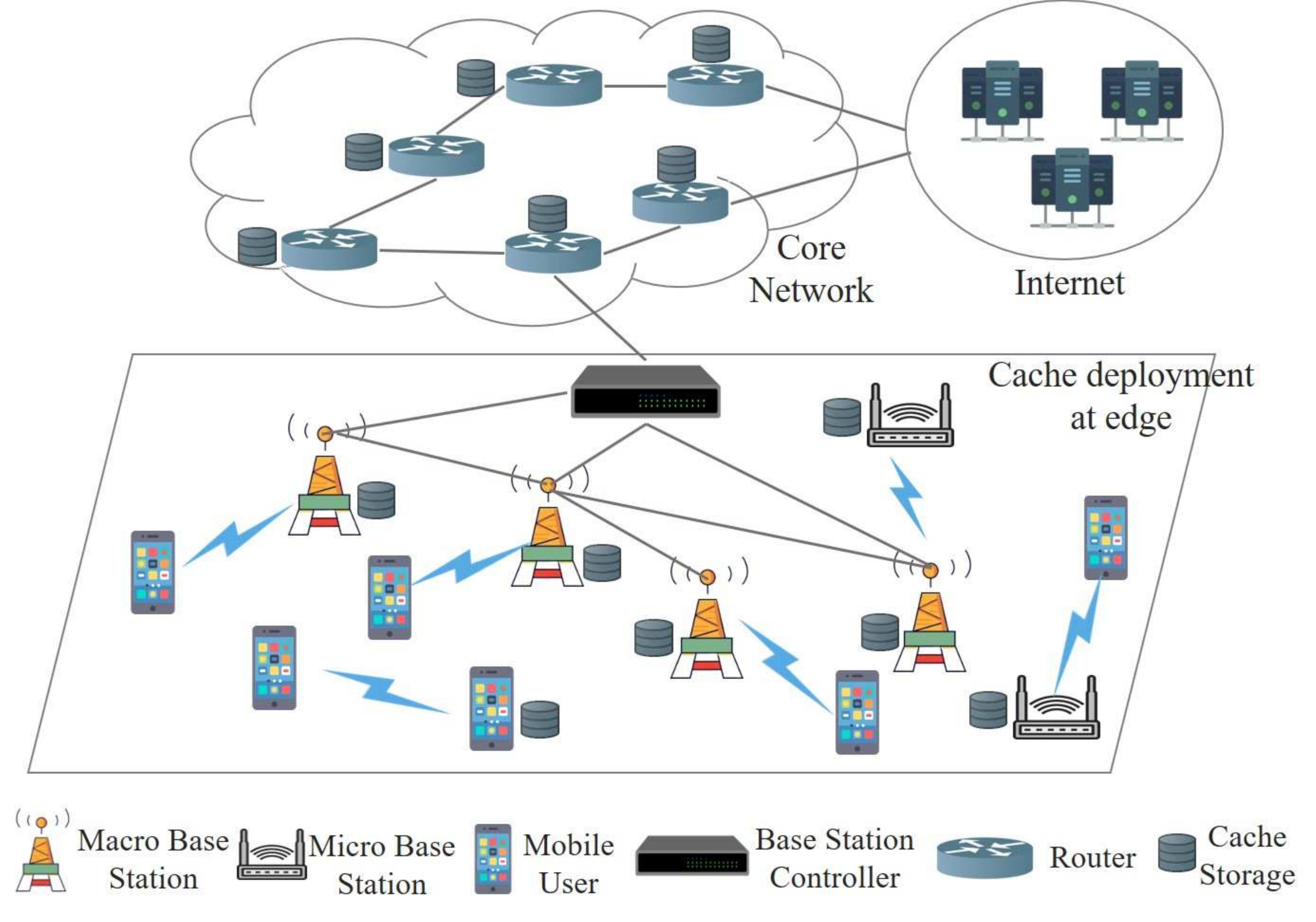}
   \caption{Cache deployment at the edge. There are three places to deploy caches: macro base stations, micro base stations, and end devices.}
   \label{fig:Cache2}
 \vspace*{-3mm}
\end{figure}

As shown in \figurename~\ref{fig:Cache2}, in edge networks, there are three main places to deploy cache units, i.e., macro base stations, small base stations, and end devices. Caching at different places have different characteristics, and we now provide a detailed discussion.

\subsubsection{Caching at Macro Base Stations}
The main purpose of deploying cache units in macro base stations (MBSs) is to relax the burden of backhaul~\cite{liu2016caching} by exploring the communication redundancy and cache machine learning models by reducing the computation redundancy. Caching popular files and models in MBSs, content could be directly fetched from MBSs instead of core networks. In this way, redundant transmission and computation are eliminated. Compared with other cache places in the edge networks, MBSs have the most extensive coverage range and the most massive cache spaces. The typical coverage radius of an MBS is nearly 500 meters \cite{li2016performance}. Due to its broad coverage range and vast cache spaces, MBS can serve more users. Thus, caching at MBSs can explore more of both communication redundancy and computation redundancy, and obtain a better cache hit probability. Also, since MBSs are deployed by operators, the topology structure of MBSs is stable and does not change over time. The comparison of different cache places is summarised in Table \ref{tab:CacheCompar1}.

\begin{table}[ht]
\small
\centering
\caption{Comparison of different cache places.}
\label{tab:CacheCompar1}
\begin{tabular}{|c|c|c|c|}
\hline
\bf{\tabincell{c}{Cache places}} & \bf{MBSs} & \bf{SBSs} & \bf{Devices}\\
\hline
\bf{\tabincell{c}{Coverage radius}} & 500m & 20 $\sim$ 200m & 10m\\
\hline
\bf{\tabincell{c}{Cache spaces}} & Large & Medium & Small\\
\hline
 \bf{Served users} & Massive & Small & Few\\
\hline
\bf{\tabincell{c}{Topology structure}} & Stable & \tabincell{c}{Change \\ slightly} & \tabincell{c}{Change \\ dramatically}\\
\hline
\bf{\tabincell{c}{Redundancy potential}}& High & Medium & Low\\
\hline
\bf{\tabincell{c}{Computational power}}& High & Medium & Low\\
\hline
\end{tabular}
\vspace{-2mm}
\end{table}

Some researchers first studied the most fundamental problem, i.e., what files should be cached at MBSs to improve end-users' QoE. One straightforward idea is to cache the most popular files. However, in \cite{blaszczyszyn2015optimal}, Blaszczyszyn \emph{et al.} find that always caching the most popular contents maybe not the optimal strategy. Specifically, they assume the distribution of MBSs follows a Poisson Poisson point process~\cite{andrews2011tractable}, and both users' requests and the popularity of content follow Zipf's law \cite{roadknight2000file}. By maximising users' cache hit ratio, they derived the optimal scheme, in which less popular contents are also cached. Furthermore, Ahlehagh and Dey~\cite{ahlehagh2014video} take user preference profiles into account, and the videos of user-preferred video categories have a high priority for caching. Alternatively, Chatzieleftheriou \emph{et al.}~\cite{chatzieleftheriou2017caching} explore the effect of content recommendations on the MBSs caching system. They discover that caching the recommended content can improve the cache hit ratio. Still, the recommendations of service providers may distort user preferences.


Apart from popular files, caching machine learning models at MBSs is another promising field. In 2016, Google proposed a novel machine learning framework, i.e., federated learning with the objective of training a global model, e.g., deep neural networks, across multiple devices \cite{konevcny2016federated}. In detail, the multiple devices train local models on local data samples. And they only exchange parameters, e.g., the weights of a deep neural network, between these local models and finally converge to a global model. Note that data is only kept in local devices and not exchanged across devices. Thus, federated learning can ensure data privacy and data secrecy and attract widespread attention from both industry~\cite{li2019abnormal} and academia~\cite{li2019federated}. Because MBSs can serve more mobile devices and have more powerful computation units, they are usually acted as the central server to orchestrate the different steps of the federated learning algorithm by caching the global model and collecting parameters from multiple devices~\cite{abad2019hierarchical,lim2019federated}.

To fully explore the communication and computation redundancy, MBSs are allowed to cooperatively cache content, including files, computation results, and AI-models. In other words, a mobile user can be neighbouring base stations~\cite{wang2014cache}. In \cite{peng2015backhaul}, Peng \emph{et al.} consider a collaborative caching network where all base stations are managed by a controller. They discover that contents with the highest popularity should be cached first in the case of long latency through backhaul network. Otherwise, caches should keep content diversity, i.e., to cache as much different content as possible. Also, Tuyen \emph{et al.}~\cite{tran2017collaborative} propose a collaborative computing framework across multiple MBSs. The computation workload at different base stations usually exhibits spatial diversity~\cite{xu2017understanding}. Therefore, offloading computation tasks to nearby idle MBSs and cache the computation results or trained-AI models cooperatively can facilitate the performance of mobile networks.

\subsubsection{Caching at Small Base Stations}
Small base stations (SBSs) are a set of low-powered mobile access points that have a range of 20 metres to 200 metres, e.g., microcell, picocell, and femtocell \cite{xiao2016dynamic}. By deploying small base stations on hot spots, mobile users will have a better quality of experience, such as high end-rate, due to the benefit from spatial spectrum reuse \cite{xiao2016analytical, xiao2016apollonius}. Therefore, densely deploying small base stations~\cite{lai2019clustering, xiao2016resource} is a promising approach in future mobile networks and also brings huge potential to reduce both communication and computation redundancy by caching at SBSs~\cite{hamidouche2016mean,zhao2016communications,liu2016cache}.

In \cite{bacstug2015cache}, Bacstug \emph{et al.} use a stochastic geometry method to model the caching network of small cells, where users' most popular files or computation results are stored on SBSs. They theoretically demonstrate that employing storage units in SBSs indeed brings gains in terms of the average delivery rate. Unlike \cite{bacstug2015cache} which considers a centralised control method, Chen \emph{et al.} \cite{chen2015cache} propose a distributed caching strategy where each SBS only considers local users' request pattern instead of global popularity. Each SBS maintains a content list by sampling requested files.


Compared with MBS, SBS has less redundancy potential due to its smaller coverage range, the number of served users, and even cache spaces. Thus, various technologies are adopted to explore caching benefits fully. A commonly used technique is the cooperative transmission. In SBS networks, the coverage area of SBSs is usually overlapped with each other, especially for the dense deployment scenario. In other words, a mobile user is able to receive content from multiple SBSs, making cooperative caching of multiple SBSs possible.  There are a lot of transmission technologies applied in SBS networks, such as multicast, beamforming, cooperative multi-point (CoMP), and so on.

In \cite{liao2017optimizing}, Liao \emph{et al.} investigate caching enabled SBS networks. By exploring the potential of multicast transmission from SBSs to mobile users, their approach could reduce the backhaul cost in SBS networks. Similarly, Poularakis \emph{et al.}~\cite{poularakis2014multicast} also delve into multicast opportunities in cache-enabled SBS networks. Different from \cite{liao2017optimizing}, they assume that both MBS and SBSs are able to use multicast. Each SBS can create multicast transmissions to end-users, while each MBS can provide multicast of popular content to SBSs within the coverage area. In terms of simulation results, the serving cost reduction up to 88\% compared to unicast transmissions.

\begin{figure}[tp!]
  \centering
  \includegraphics[width=1.0\columnwidth]{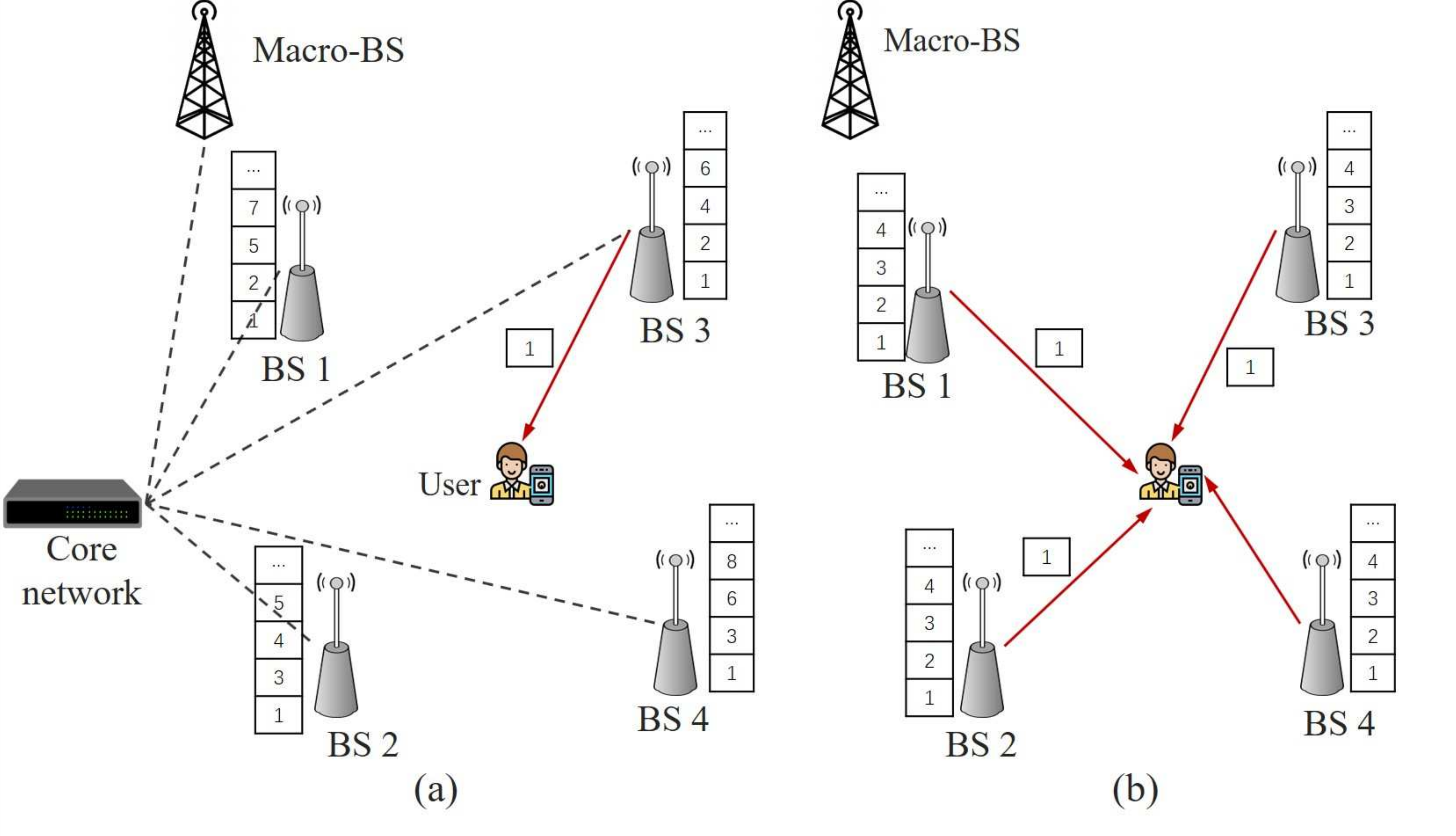}
   \caption{Cache model at SBS. (a) mobile use is associated with one SBS. (b) mobile use is associated with multiple SBS, which send content to the user with beamforming and CoMP.}
   \label{fig:Cache3}
\end{figure}

In SBS networks, since a mobile user has the potential to receive signals from multiple SBSs, there are two association cases, as shown in \figurename \ref{fig:Cache3}. In the first case, the mobile user is only associated with one SBS. Alternatively, the mobile user is associated with multiple SBSs. By using the beamforming \cite{dahrouj2010coordinated} and CoMP technology \cite{marsch2011coordinated}, the associated SBSs can jointly send content for downlink transmission.

In \cite{pantisano2014cache}, Pantisano \emph{et al.} consider the first association case and presents a cache-aware user association approach. They adjust users' association to improve the local cache hit ratio based on whether the associated base station contains files and AI models required by the user. The problem is formulated as a one-to-many matching game and they propose a distributed algorithm based on the deferred acceptance scheme to solve it.

Alternatively, some scholars focus on the second association case and explore the power of cooperative transmission in cache-enabled SBS networks \cite{shanmugam2013femtocaching,liu2017energy,ao2015distributed,ao2018fast,chen2017cooperative}. In \cite{shanmugam2013femtocaching}, Shanmugam \emph{et al.} study the problem of content deployment in SBS networks. They assume that mobile users could communicate with multiple cache units and formulate the optimisation problem to minimise the average downloading delay. Liu \emph{et al.}~\cite{liu2017energy} explore the potential of energy efficiency of cache-enabled SBS networks where all SBSs use CoMP to transmit cached content cooperatively. By maximising energy efficiency, the optimal transmit power of SBSs are worked out. In \cite{ao2015distributed} and \cite{ao2018fast}, Ao \emph{et al.} study a distributed caching strategy in SBS networks where CoMP technology is applied. In the CoMP enabled networks, caching strategy can bring two different gains. On the one hand, diverse contents could be cached in nearby BSs to maximise the cache hit. On the other hand, caching the same content in nearby BSs can let corresponding BSs transmit concurrently and bring multiplexing gain. By trading off both gains, they devised a near-optimal strategy to maximise the system throughput. Moreover, they find when content is of skewed popularity distribution, caching multiple copies of popular files and AI models yields larger caching gains. Further, Chen \emph{et al.} \cite{chen2017cooperative} consider a similar system. Unlike \cite{ao2015distributed} and \cite{ao2018fast}, where all nearby SBSs can employ CoMP, Chen \emph{et al.} first group SBSs into multiple disjointed clusters and only the SBSs in the same cluster are able to transmit content cooperatively. To trade off the parallel transmission and joint transmission, they divide the cache space into two parts. One is in charge of caching content with less popularity to improve content diversity, while the other is used to cache contents with the highest popularity. Then, they optimize the problem of space assignment.

\begin{figure}[tp!]
  \centering
  \includegraphics[width=1.0\columnwidth]{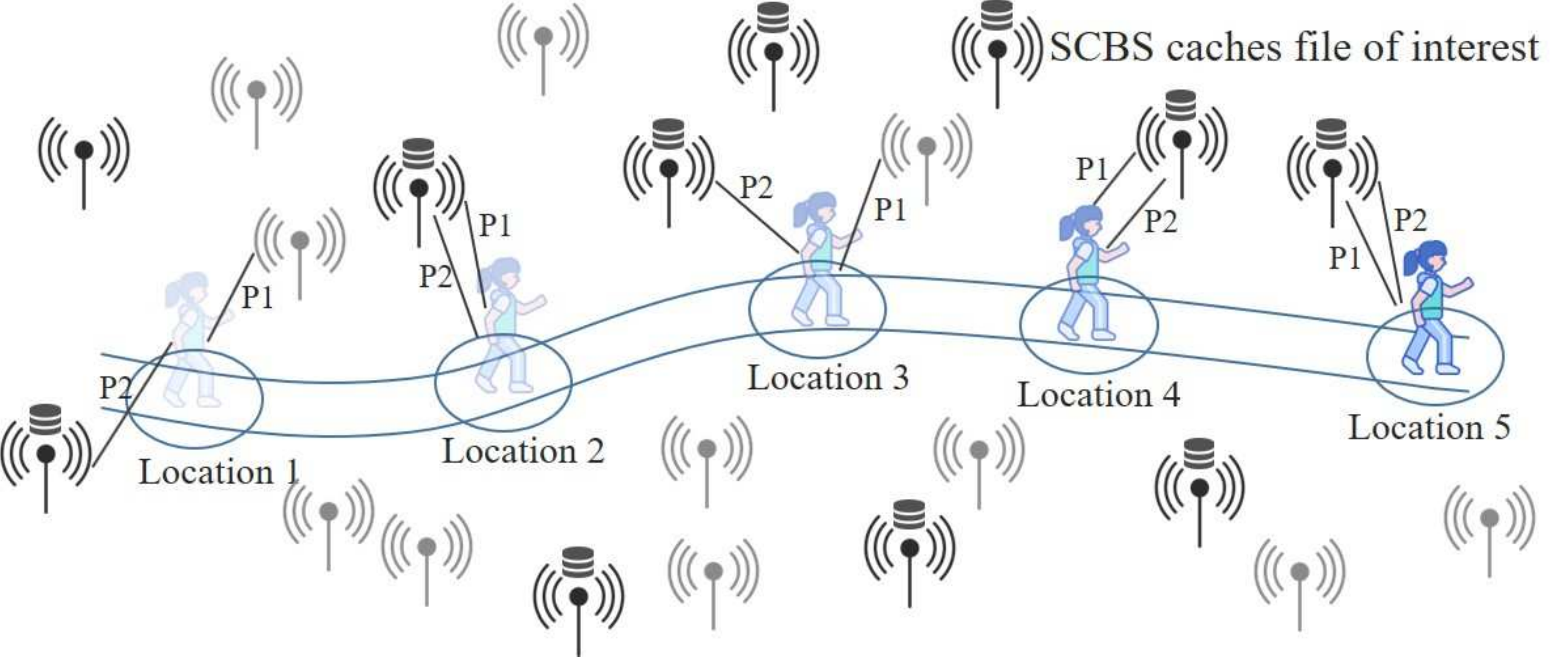}
   \caption{User gets file of interest from SCBS using policy 1, denoted as P1, and policy 2, denoted as P2, while moving from location 1 to location 5. P1 means connecting to the SCBS that provides the highest average received power. P2 means connecting to the nearest SCBS that could provide files of interest.}
   \label{fig:Cache4}
\end{figure}

Since the coverage range of each SBS is too small, mobile users will go through multiple SBSs within a short time, as shown in \figurename ~\ref{fig:Cache4}. This frequent handover behaviour will cause the degradation of caching performance. In \cite{krishnan2017effect}, Krishnan \emph{et al.} investigate the retransmission in cache-enabled SBS networks. Because of the frequent handover between different SBSs, sometimes, when none of the SBSs in the user's vicinity has cached the requested file or AI models, the file transmission will be interrupted. Nevertheless, the retransmission will be triggered when the requested file or AI models are cached at vicinity SBSs. By using stochastic geometry to analyse the cache hit probability, Krishnan \emph{et al.} find that SBSs should cache content diversely for mobile users. In \cite{guan2014mobicacher}, Guan \emph{et al.} assume that users' preferences for content and mobility patterns are known prior, and users' preferences remain constant over a short period.
They then formulate an optimisation problem with the objective of maximising the utility of caching and devise a heuristic caching strategy. In \cite{Poularakis2017Code} and \cite{Ozfatura2018Mobility}, the same caching system model is investigated where mobile users migrate between multiple SBSs. Due to the limited transmission time, users may not be able to download the complete requested files or parameters of AI models from the associated SBS, and the requests can be redirected to MBS. In \cite{Poularakis2017Code}, Poularakis \emph{et al.} use random walks to model user movements and formulated an optimisation problem based on the Markov chain aiming to maximize the probability of serving by SBSs. They further propose two caching strategies, i.e., a centralised solution for the small-scale system and a distributed solution for the large-scale system. Unlike \cite{Poularakis2017Code}, Ozfatura \emph{et al.} propose a distributed greedy algorithm to minimise the amount of data downloaded from MBS \cite{Ozfatura2018Mobility}. Requests with deadlines below a given threshold are responded by SBSs while other requests are served by MBS.

In SBS caching systems, predicting users' requests to improve the cache hit ratio is also a commonly used method, which falls into the field of artificial intelligence applications. By applying AI technology to historical users' request logs, we can profile user preference or content popularity patterns. Next, we can predict users' requests or content popularity, respectively \cite{chang2018learn}. In \cite{kader2015leveraging}, Kader \emph{et al.} design a big data platform and collects mobile traffic data from a telecom operator in Turkey. They then used collaborative filtering, a common machine learning method, to estimate content popularity. The simulation results demonstrate that the caching benefits are further explored with the help of content popularity prediction. Similar to \cite{kader2015leveraging}, in \cite{pantisano2015match}, Pantisano \emph{et al.} also apply collaborative filtering to predict content popularity. They then devised a user-SBS association scheme based on estimated popularity and the current cache composition to minimise the backhaul bandwidth allocation. In \cite{cheng2019localized}, Bastug \emph{et al.} focus on individual content request probability instead of global content request probability. They propose to use the Bayesian learning method to predict personal preferences and then incorporate this crucial information into the caching strategy. If we lack the historical data of user request logs, how can we predict content popularity? In \cite{bacstuug2015transfer}, Bastug \emph{et al.} investigate this open issue and proposes a transfer learning-based caching procedure. Specifically, they exploit contextual information, e.g., social networks, and referred to it as a source domain. Then the prior information in the source domain is incorporated in the target domain to estimate content popularity.

Also, SBSs are used to cache the data from end devices, like smartphones and IoT. In recent times, IoT devices are widely distributed in homes, streets, and even whole cities to allow users to monitor the ambient environment \cite{el2018edge, lin2016ubii}. By collecting and analysing the big data on IoT devices, a smarter physical world can be built. Considering the demand of real-time data analysis, caching and processing the data at the edge is a common and promising method. In \cite{quevedo2014case}, Quevedo \emph{et al.} introduce a caching system for IoT data and proof that the caching system could reduce the energy consumption of IoT sensors. In \cite{sharma2017live}, Sharma \emph{et al.} propose a collaborative edge and cloud data processing framework for IoT networks where SBSs are in charge of caching IoT data, extracting useful features and uploading features to cloud part.

Meanwhile, since SBSs are often deployed at hot points, the requested computation tasks from served users will exhibit spatiotemporal locality. Therefore, by caching computation results at SBSs, the redundant computation tasks can be eliminated. Drolia \emph{et al.}~\cite{drolia2017cachier} propose a caching strategy, Cachier, to cache the recognition results on edge servers by release repetitive recognition computation. Specifically, Cachier first extracts features of the requested recognition task, and then tries to match a similar object from the cache. If there is a cache hit, the corresponding computation results would be sent back to the mobile device. Otherwise, the request would be sent to the cloud. To identify similar recognition tasks, they used a Locality Sensitive Hashing (LSH) algorithm \cite{lv2007multi} to determine the best match. Furthermore, to overcome the unbalanced and time-varying distribution of users' requested tasks, Guo \emph{et al.} \cite{guo2018foggycache} design an Adaptive LSH-Homogenized kNN joint algorithm which outperforms LSH in terms of evaluation results.
Drolia \emph{et al.} further introduce a proactive caching strategy into their system by predicting the requirement of users and proactively caching parts of models on SBSs server for pre-processing to further reduce the latency~\cite{drolia2017precog}. Such a strategy is also used in \cite{venugopal2018shadow} to deal with unstructured data at SBSs.

Moreover, in some task-fickle scenarios, multiple different kinds of tasks, e.g., voice recognition and object recognition, are offloaded from devices to SBSs. By pre-caching multiple kinds of deep learning models at SBSs for different kinds of tasks, we can reduce the computation time and further improve users' QoE. Taylor \emph{et al.} propose an adaptive model selection scheme to select the best model for users \cite{taylor2018adaptive}. They use a supervised learning method to train a predictor model offline and then deploy it on an edge server. When a request arrives, the predictor will select an optimal model for the task. In \cite{zhao2018privacy}, Zhao \emph{et al.} propose a system, Zoo, to compose different models to provide a satisfactory service for users. Ogden \emph{et al.} propose a deep inference platform, MODI, to determine what model to cache and what model to use for specific tasks \cite{ogden2018modi}. There is a decision engine inside MODI, which aggregates previous results to decide what new models are required to cache.

\subsubsection{Caching at Devices}

Caching at devices exploits the available storage space of end equipment, like mobile phones and IoT devices. These devices can leverage the communication and computation redundancy locally. Furthermore, they can fetch the requested content or computation results from other devices in proximity through device-to-device (D2D) communication \cite{ji2016wireless, chen2016mitigating}.

\begin{figure}[tp!]
\begin{center}
\includegraphics[width=0.8\linewidth]{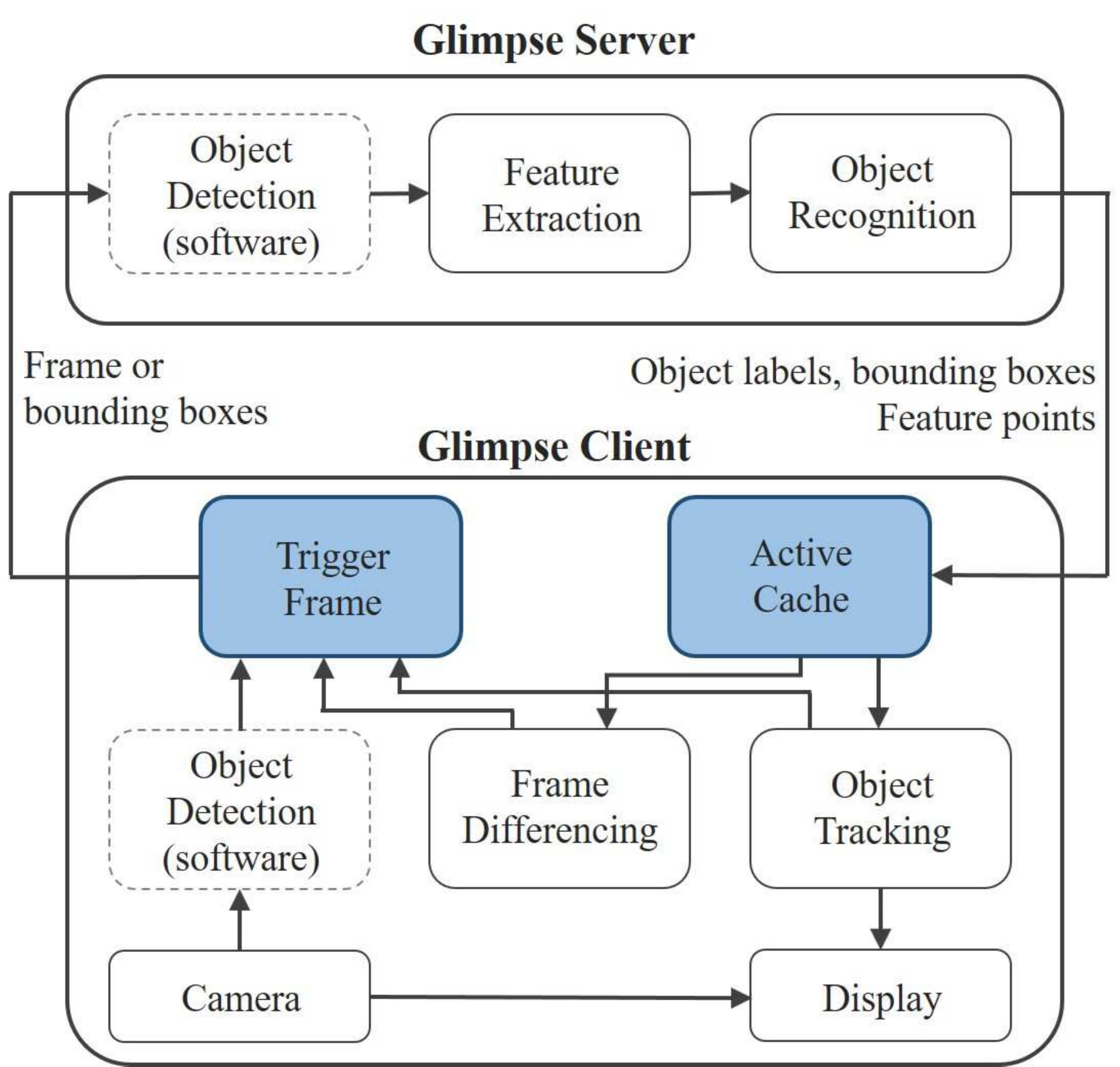}
\end{center}
\caption{The architecture of Glimpse. Edge device, i.e., glimpse client, only uploads trigger frames to the cloud to save bandwidth resources. Glimpse server transmits the recognition results and features back to edge devices. Edge devices deal with local frames with these features.}
\label{glimpse}
\end{figure}

First of all, end devices can explore the communication and computation redundancy locally. For instance, in some static continuous computer vision applications, such as monitoring, the captured consecutive images are similar to some extent. Therefore, the results of previous images could be reused for the latter inference. In \cite{xu2017accelerating,cavigelli2019cbinfer,huynh2017deepmon}, they cache the results of previous frames to reduce redundant computation and latency. In some mobile continuous computer vision applications, such as driving assistance, the system is required to provide high trackability. The system needs to recognise, locate, and label the tracked object, e.g., road signs, on the screen in real-time. The recognised object would repeatedly appear in multiple images for a period. Chen \emph{et al.} develop an active cache based continuous object recognition system, called Glimpse, to achieve real-time trackability \cite{chen2015glimpse}. The structure of Glimpse is shown as Fig.~\ref{glimpse}. Glimpse caches frames locally and only uploads trigger frames to the cloud server. Trigger frames refers to the frames, for which the recognition from the server is different from current local tracking. The cloud server sends back the recognised object, its labels, bounding boxes, and features, which would be cached locally on devices. Then the devices would track the object with the labels, bounding boxes, and features locally on captured frames. A similar approach is also adopted in CNNCache \cite{xu2017accelerating}.

On the other hand, compared with MBSs and SBSs, devices have very limited cache spaces and coverage range, due to the cost constraint of end devices and their low transmission power. Although these limitations seem to only take the small caching benefits for local devices, the situation will be changed when the networks are with dense users. The benefits of caching will be amplified with the number of users increases. In \cite{chen2016d2d}, Chen \emph{et al.} study the difference between caching at SBSs and devices where content is cached according to a joint probability distribution. By applying stochastic geometry, they derive the closed-form expression of hit probability and request density. Although the cache hit probability of device caching is always lower than SBS caching due to the small cache spaces, the request density of device caching is much higher than SBS caching. This is because, in device caching, more concurrent links are allowed, compared with the case of SBS caching, especially in the dense user scenario. Similarly, in \cite{gregori2016wireless}, Gregori \emph{et al.} investigate caching at both devices and SBSs as well. However, they do not compare these two different scenarios and only design joint transmission and caching policies for them to minimise energy consumption separately.

In the device caching system, the number of coexisting D2D links affect the device caching performance dramatically. The D2D links are the fundamental requirement for end-devices sharing files, models, and computation results and further reducing communication and computation redundancy amongst devices. First of all, the establishment of a D2D connection depends on content placement. In other words, when a device discovers that the requested content is placed in a nearby device within the D2D transmission range, the D2D link can be built, and the content will be transmitted directly. Therefore, there are a lot of scholars trying to maximise caching performance by optimising the content placement \cite{chen2017optimal, giatsoglou2017d2d, qiu2019popularity, malak2014optimal, peng2018optimal}. In \cite{malak2014optimal}, Malak \emph{et al.} model senders and receivers as members of Poisson Point Process and compute the probability of delivery in D2D networks. Considering the low transmission noise case, they find that the optimal content allocation could be approximately achieved by Benford's law when the path loss exponent equals 4. A similar system model is applied in \cite{peng2018optimal} but with a different performance metric. In \cite{peng2018optimal}, Peng \emph{et al.} analyse the outage probability of D2D transmission in cache-enabled D2D networks. They then obtain the optimal caching strategy by using a gradient descent method. In \cite{chen2017optimal}, Chen \emph{et al.} aim to maximise successful offloading probability. Different from \cite{peng2018optimal, malak2014optimal} which did not consider the time divided transmission, Chen \emph{et al.} divide time into multiple slots, and each transmitter independently chooses a time slot to transmit files. Employing the gradient descent method, they design a distributed caching policy. Unlike the above studies where each end-user applies the same caching strategy, Giatsoglou \emph{et al.} \cite{giatsoglou2017d2d} divide the $2K$ most popular contents into two groups of the same size. $K$ is the cache capacity of each user. Then randonly allocate these two groups to users, i.e., some users cache group A, whilst others cache group B.

Apart from content placement, association policy is also important to the establishment of D2D links. In \cite{golrezaei2012wireless}, Golrezaei \emph{et al.} optimise the collaboration distance for D2D communications with distributed caching, where the collaboration distance is the maximum allowable distance for D2D communication. They assume each user employs a simple randomised caching policy. In \cite{naderializadeh2014utilize}, Naderializadeh \emph{et al.} propose a greedy association method, i.e., greedy closest-source policy. In this association policy, starting from the first user, each user chooses the closest user with the desired file forming a D2D pair. They assume that each file is randomly cached in devices and derive a lower bound on the cached content reuse.

Generally, end devices are controlled by end consumers who are able to decide whether to cache and share content or to not do so. Therefore, the incentive mechanism is introduced in D2D networks to encourage users to exploit the storage space of their equipment and share cached content, e.g., files, AI models and computation results, with other users. In \cite{chen2016caching}, Chen \emph{et al.} propose an incentive mechanism where the base station rewards the users who shares content with others via D2D communications. Since the base station will determine the reward to minimise its total cost while users would like to maximise their reward by choosing the caching policy, Chen \emph{et al.} model this conflict as a Stackelberg game and proposed an iterative gradient algorithm to obtain the Stackelberg Equilibrium. In \cite{taghizadeh2013distributed}, Taghizadeh \emph{et al.} consider a similar case where content providers pay the download cost to encourage users to download and share content. However, they do not model the conflict and merely design the caching strategy to minimise content provisioning costs.

\begin{figure}[tp!]
  \centering
  \includegraphics[width=1.0\columnwidth]{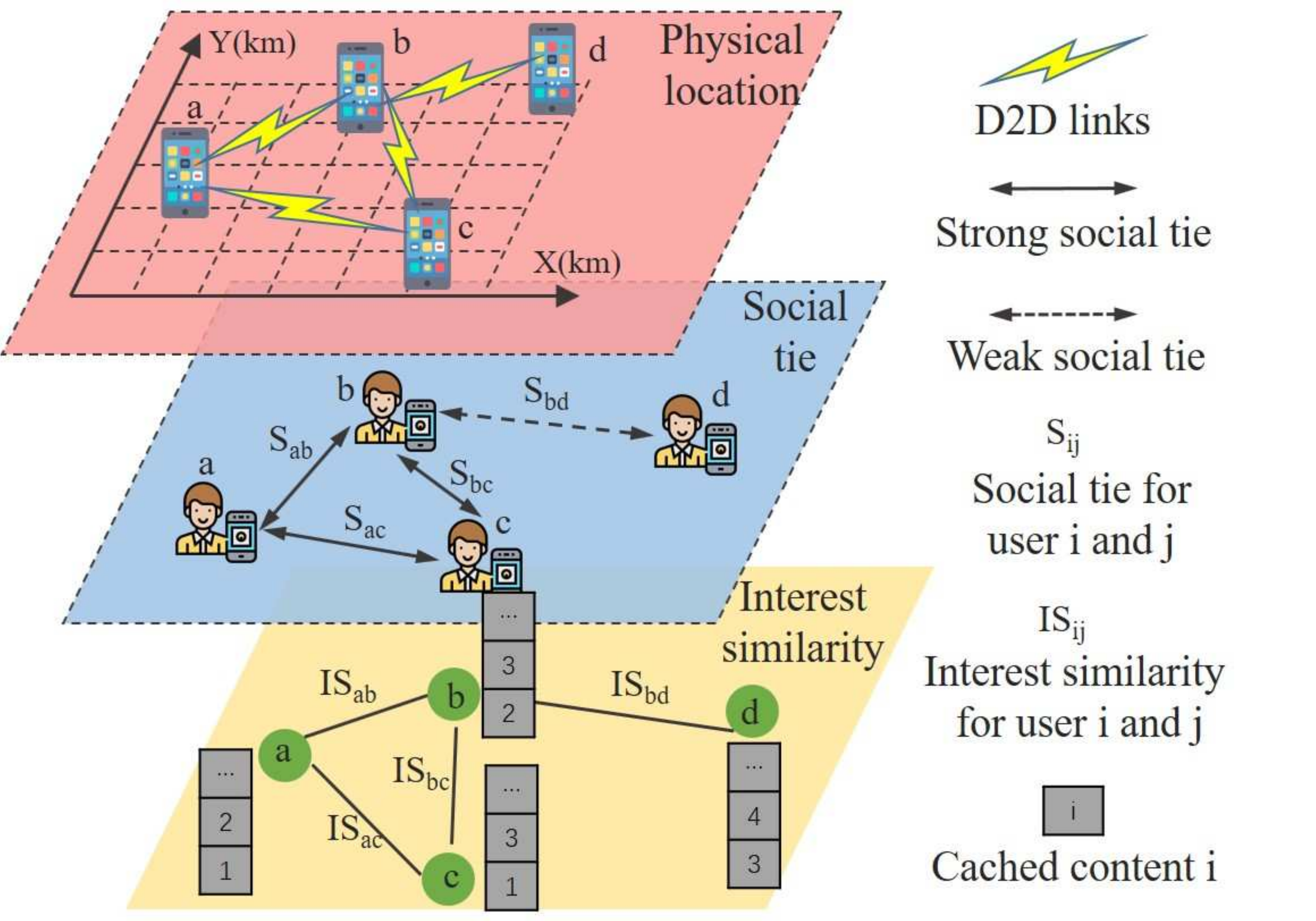}
   \caption{Social-aware caching at edge devices. In location based framework, users close to each other could exchange cached content, .e.g., user a, b, and c. In social tie based framework, user could also exchange cached content with others if they have strong social tie. In interest based framework, interest similarity is used to estimate the social tie among users.}
   \label{fig:Cache5}
 \vspace*{-3mm}
\end{figure}

Since end devices are bound up with users and affected by user attributes to some extent, some researchers focus on exploring the knowledge of user attributes, like social ties and interests, to assist device caching as shown in \figurename~\ref{fig:Cache5}. In \cite{bastug2014living}, Bastug \emph{et al.} propose to let influential users cache content, such that these users could disseminate the cached contents to others through their social ties. The influential users are determined by their social networks. First, a social graph is built based on past action history of users' encounters and file requests. Then, the influence of users is measured in terms of the centrality metric \cite{newman2010networks}. Apart from social ties, Bai \emph{et al.} \cite{bai2016caching} consider users' interests as well. They use the hypergraph to model the relationships among social ties, common interests, and spectrum resources and design an optimal caching strategy to maximise the cache hit ratio.

\begin{table*}[htp]
\footnotesize
\centering
\caption{Comparison of different cache deployment strategies.}
\label{tab:CacheCompar2}
\vspace{-0.1in}
\begin{tabular}{c c c c c c}
\toprule
\bf{Ref.} & \bf{Cache places}  & \bf{Performance metrics} & \bf{Mathematical tools} & \bf{Control methods} & \bf{\tabincell{c}{Transmission Cooperativity}}\\
\hline
\cite{blaszczyszyn2015optimal} & MBSs & Hit probability & \tabincell{c}{Stochastic geometry} & Centralised & Non-cooperative\\
\hline
\cite{chatzieleftheriou2017caching} & MBSs & Cache hit ratio & Optimisation & Centralised & Non-cooperative\\
\hline
\tabincell{c}{\cite{liu2014will} \\\cite{liu2016energy}}& MBSs & Energy efficiency & \tabincell{c}{Stochastic  geometry} & Centralised & Non-cooperative\\
\hline
\cite{ahlehagh2014video} & MBSs & \tabincell{c}{The number of \\ concurrent videos} & Optimisation & Distributed & Non-cooperative\\
\hline
\cite{peng2015backhaul} & MBSs & \tabincell{c}{The average \\download delay} & Optimisation & Centralised & Cooperative\\
\hline
\cite{gu2013proactive} & MBSs & Storage space & Optimisation & Centralised & Cooperative\\
\hline
\cite{khreishah2015collaborative} & MBSs & \tabincell{c}{Aggregate \\operational cost} & Optimisation & Centralised & Cooperative \\
\hline
\cite{ostovari2016efficient} & MBSs & \tabincell{c}{The aggregated \\caching and download cost} & Optimisation & Centralised & Cooperative \\
\hline
\cite{wang2016mobility} & MBSs & \tabincell{c}{Cache \\failure probability} & Optimisation & Centralised & Cooperative \\
\hline
\cite{bacstug2015cache} & SBSs & \tabincell{c}{Outage probability \\ Content delivery rate} & \tabincell{c}{Contract theory} & Centralised & Non-cooperative \\
\hline
\cite{chen2015cache} & SBSs & \tabincell{c}{The cache \\service probability} & \tabincell{c}{Stochastic geometry} & Distributed & Non-cooperative \\
\hline
\cite{li2018contract} & SBSs & \tabincell{c}{The profit \\of NSP and VPs} & \tabincell{c}{Stochastic geometry} & Centralised & Non-cooperative\\
\hline
\cite{poularakis2014exploiting} & SBSs & \tabincell{c}{The number of \\ requests served by SBSs} & Optimisation & Centralised & Non-cooperative \\
\hline
\cite{liao2017optimizing} & SBSs & Backhaul cost & Optimisation & Centralised & Non-cooperative \\
\hline
\cite{poularakis2014multicast} & SBSs & Servicing cost & Optimisation & Centralised & Non-cooperative \\
\hline
\cite{pantisano2014cache} & SBSs & Caching utility & Matching theory & Distributed &  Non-cooperative \\
\hline
\cite{shanmugam2013femtocaching} & SBSs & \tabincell{c}{Downloading \\time of files} & Optimisation & Centralised & Cooperative \\
\hline
\cite{liu2017energy} & SBSs & \tabincell{c}{Energy efficiency} & Optimisation & Centralised & Cooperative \\
\hline
\tabincell{c}{\cite{ao2015distributed}\\ \cite{ao2018fast}} & SBSs & System throughput & Optimisation & Distributed & Cooperative\\
\hline
\cite{chen2017cooperative} & SBSs & \tabincell{c}{Cache hit probability \\ and energy efficiency} & \tabincell{c}{Stochastic geometry} & Centralised & Cooperative \\
\hline
\cite{krishnan2017effect} & SBSs & \tabincell{c}{Cache \\hit probability} & \tabincell{c}{Stochastic geometry} & Centralised & Non-cooperative \\
\hline
\cite{guan2014mobicacher} & SBSs & Maximise caching utility & Optimisation & Centralised & Non-cooperative \\
\hline
\cite{Poularakis2017Code} & SBSs & \tabincell{c}{The probability of \\ response from MBS} & Optimisation  & \tabincell{c}{Centralised \& Distributed} & Non-cooperative \\
\hline
\cite{Ozfatura2018Mobility} & SBSs & \tabincell{c}{The amount of \\ data downloaded from MBS} & Optimisation & Distributed & Non-cooperative \\
\hline
\cite{kader2015leveraging} & SBSs & \tabincell{c}{Backhaul load} & \tabincell{c}{Machine learning} & Centralised & Non-cooperative \\
\hline
\cite{pantisano2015match} & SBSs & \tabincell{c}{Backhaul\\ bandwidth allocation} & \tabincell{c}{Machine learning} & Distributed & Non-cooperative \\
\hline
\cite{cheng2019localized} & SBSs & \tabincell{c}{System throughput} & \tabincell{c}{Machine learning} & Centralised & Non-cooperative  \\
\hline
\cite{bacstuug2015transfer} & SBSs & \tabincell{c}{Backhaul \\ offloading gains} & \tabincell{c}{Machine learning} & Centralised & Non-cooperative \\
\hline
\cite{chen2016d2d} & Devices & \tabincell{c}{Cache hit ratio\\ Density of cache-served requests} & \tabincell{c}{Stochastic  Geometry} & Distributed & Non-cooperative  \\
\hline
\cite{gregori2016wireless} & Devices & \tabincell{c}{Energy consumption} & \tabincell{c}{Optimisation} & Distributed & Non-cooperative \\
\hline
\cite{malak2014optimal} & Devices &  \tabincell{c}{The probability \\ of successful content delivery} & \tabincell{c}{Stochastic Geometry} & Distributed & Non-cooperative \\
\hline
\cite{peng2018optimal} & Devices & \tabincell{c}{Outage probability} & Optimisation & Distributed & Non-cooperative \\
\hline
\cite{chen2017optimal} & Devices & \tabincell{c}{offloading probability} & Optimisation & Distributed & Non-cooperative\\
\hline
\cite{giatsoglou2017d2d} & Devices & \tabincell{c}{offloading gain} & \tabincell{c}{Stochastic Geometry} & Centralised & Non-cooperative\\
\hline
\cite{golrezaei2012wireless} & Devices & \tabincell{c}{number \\ of D2D links} & \tabincell{c}{Optimisation} & Distributed & Non-cooperative \\
\hline
\cite{naderializadeh2014utilize} & Devices & Spectral reuse & Optimisation & Distributed & Non-cooperative\\
\hline
\cite{ji2013optimal} \cite{ji2015throughput} & Devices & System throughput & Optimisation & Centralised & Non-cooperative\\
\hline
\cite{chen2016cooperative} \cite{chen2017high} & Devices & Network throughput & Optimisation & Centralised & Cooperative \\
\hline
\cite{afshang2016fundamentals} & Devices & Coverage probability & \tabincell{c}{Stochastic Geometry} & Centralised & Non-cooperative \\
\hline
\cite{jarray2016effects} & Devices & Service success probability & \tabincell{c}{Stochastic Geometry} & Distributed & Non-cooperative \\
\hline
\cite{krishnan2017effectnew} & Devices & Coverage probability & \tabincell{c}{Stochastic Geometry} & Distributed & Non-cooperative \\
\hline
\cite{chen2016caching} & Devices & Caching reward & \tabincell{c}{Game theory} & Distributed & Non-cooperative \\
\hline
\cite{taghizadeh2013distributed} & Devices & Content provisioning costs & \tabincell{c}{Optimisation} & Distributed & Non-cooperative \\
\hline
\cite{bastug2014living} & Devices & Backhaul costs & \tabincell{c}{Graph Theory} & Centralised & Non-cooperative \\
\hline
\cite{bai2016caching} & Devices & Cache hit ratio & \tabincell{c}{Graph Theory} & Centralised & Non-cooperative \\
\bottomrule
\end{tabular}
\end{table*}

\subsection{Cache Replacement}\label{S3.3}
In practice, the request distribution of content varies with time, and new content is constantly being created. Hence, it is critical to update caches at intervals. The cache update process generally takes place when new content is delivered and needs to be cached, but all cache units are occupied. Hence, some cached old content needs to be replaced. Therefore, the cache update process is called cache replacement as well.

Several conventional cache replacement strategies have been proposed, such as first-in first-out (FIFO), least frequently used (LFU), least recently used (LRU), and their variants \cite{ioannou2016survey}. FIFO evicts the content in terms of cached time without any regard to how often or how many times it was accessed before. LFU keeps the most frequently requested content, while LRU keeps the most recently accessed content. However, these replacement strategies merely consider content request features in a short time window and may not obtain the global optimal solutions.

Another popular method is to replace the content based on its popularity. In \cite{blasco2014learning}, Blasco \emph{et al.} divide time into periods and within each period there is a cache replacement phase. During each cache replacement phase, the content of the lowest popularity is discarded. Apart from historical popularity, Bacstuug \emph{et al.} \cite{bacstuug2013proactive} take the future content popularity into consideration as well. They propose a proactive popularity caching (PropCaching) method to estimate content popularity and then determine which content should be evicted.

Mathematically, the cache replacement problem could be formulated as a Markov Decision Process (MDP) \cite{bacstuug2013proactive,wang2018edge}. The MDP model can be represented into a tuple $(\bm{S}, \bm{A}, R(s, a))$. $\bm{S}$ refers to the set of possible states for caches. $\bm{A}$ is the set of eviction actions. $R(s,a)$ is the reward function that determines the reward when cache performers action $a$ in the state $s$. The reward is usually modelled as the cache hit or the changes in transmission cost. In \cite{bacstuug2013proactive}, Bacstuug \emph{et al.} obtain the cache replacement actions based on Q-learning. In \cite{wang2018edge}, the method is upgraded. Wang \emph{et al.} apply deep reinforcement learning to solve it.


\section{Edge Training}\label{S4}



The standard learning approach requires centralising training data on one machine, whilst edge training relies on distributed training data on edge devices and edge servers, which is more secure and robust for data processing. The main idea of edge training is to perform learning tasks where the data is generated or collected with edge computing resources. It is not necessary to send users' personal data to a central server, which effectively solves the privacy problem and saves network bandwidth.

Training data could be solved through edge caching. We discuss how to train an AI model in edge environments in this section. Since the computing capacity on edge devices and edge servers is not as powerful as central servers, the training style changes correspondingly in the edge environment. The major change is the distributed training architecture, which must take the data allocation, computing capability, and network into full consideration. New challenges and problems, e.g., training efficiency, communication efficiency, privacy and security issues, and uncertainty estimates, come along with the new architecture. Next, we discuss these problems in more detail.

\subsection{Training architecture}
Training architecture depends on the computing capacity of edge devices and edge servers. If one edge device/server is powerful enough, it could adopt the same training architecture as a centralised server, i.e., training on a single device. Otherwise, cooperation with other devices is necessary. Hence, there are two kinds of training architectures: solo training, i.e., perform training tasks on a single edge device/server, and collaborative training, i.e., few devices and servers work collaboratively to perform training tasks.

\subsubsection{Solo training}
Early researchers mainly focus on verifying the feasibility of directly training deep learning models on mobile platforms. Chen \emph{et al.} find that the size of neural network and the memory resource are two key factors that affect training efficiency \cite{chen2018exploring}. For a specific device, training efficiency could be improved significantly by optimising the model. Subsequently, Lane \emph{et al.} successfully implement a constrained deep learning model on smartphones for activity recognition and audio sensing \cite{lane2017squeezing}. The demonstration achieves a better performance than shallow models, which demonstrates that ordinary smart devices are qualified for simple deep learning models. Similar verification is also done on wearable devices \cite{radu2018multimodal} and embedded devices \cite{lane2015early}.

\subsubsection{Collaborative training}
The most common collaborative training architecture is the master-slave architecture. Federated learning \cite{mcmahan2017federated} is a typical example, in which a server employs multiple devices and allocates training tasks for them. Li \emph{et al.} develop a mobile object recognition framework, named DeepCham, which collaboratively trains adaptation models \cite{li2016deepcham}. The DeepCham framework consists of one master, i.e., edge server and multiple workers, i.e., mobile devices. There is a training instance generation pipeline on workers that recognises objects in a particular mobile visual domain. The master trains the model using the training instance generated by workers. Huang \emph{et al.} consider a more complex framework with additional scheduling from the cloud \cite{huang2018task}. Workers with training instances first uploads a profile of the training instance and requests to the cloud server. Then, the cloud server appoints an available edge server to perform the model training.

Peer-to-peer is another collaborative training architecture, in which participants are equal. Valerio \emph{et al.} adopt such training architecture for data analysis \cite{valerio2017communication}. Specifically, participants first perform partial analytic tasks separately with their own data. Then, participants exchange partial models and refine them accordingly. The authors use an activity recognition model and a pattern recognition model to verify the proposed architecture and find that the trained model could achieve similar performance with the model trained by a centralised server. Similar training architecture is also used in \cite{xing2018enabling} to enable knowledge transferring amongst edge devices.

\subsection{Training Acceleration}
Training a model, especially deep neural networks, is often too computationally intensive, which may result in low training efficiency on edge devices, due to their limited computing capability. Hence, some researchers focus on how to accelerate the training at edge. Table \ref{trainacceleration} summaries existing literature on training acceleration.


Chen \emph{et al.} find that the size of a neural network is an important factor that affects the training time \cite{chen2018exploring}. Some efforts \cite{valery2017cpu,valery2018low} investigate transfer learning to speed up the training. In transfer learning, learned features on previous models could be used by other models, which could significantly reduce the learning time. Valery \emph{et al.} propose to transfer features learned by the trained model to local models, which would be re-trained with the local training instances \cite{valery2017cpu}. Meanwhile, they exploit the shared memory of the edge devices to enable the collaboration between CPU and GPU. This approach could reduce the required memory and increase computing capacity. Subsequently, the authors further accelerate the training procedure by compressing the model by replacing float-point with 8-bit fixed point \cite{valery2018low}.

In some specific scenarios, interactive machine learning (iML) \cite{amershi2014power,ware2001interactive} could accelerate the training. iML engages users in generating classifiers. Users iteratively supply information to the learning system and observe the output to improve the subsequent iterations. Hence, model updates are more fast and focused. For example, Amazon often asks users targeted questions about their preferences for products. Their preferences are promptly incorporated into a learning system for recommendation services. Some efforts \cite{miu2015bootstrapping,shahmohammadi2017smartwatch} adopt such approach in model training on edge devices. Shahmohammadi \emph{et al.} apply iML on human activity recognition, and find that only few training instances are enough to achieve a satisfactory recognition accuracy \cite{shahmohammadi2017smartwatch}. Based on such theory, Flutura \emph{et al.} develop DrinkWatch to recognise drink activities based on sensors on smartwatch \cite{flutura2018drinkwatch}.

\begin{figure}[tp!]
\begin{center}
\includegraphics[width=\linewidth]{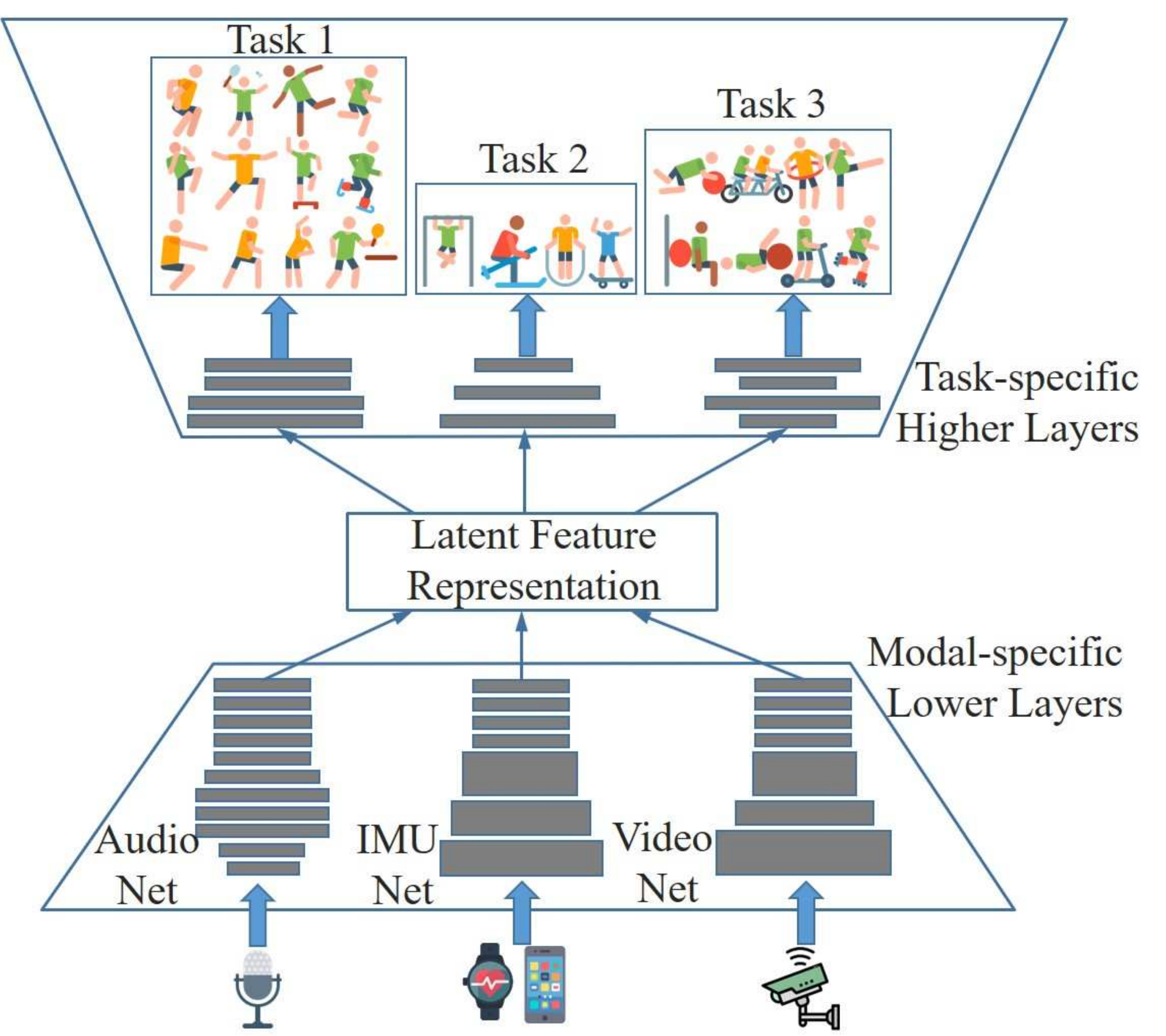}
\end{center}
\caption{Illustration of the hourglass model. The lower part represents lower layers of specific sensing models. The latent feature representation part is the common layer. Lower layers project their data into this layer for knowledge transfer. The upper part represents task-specific higher layers, which are trained for specific recognition tasks.}
\label{hourglass}
\end{figure}

In a collaborative training paradigm, edge devices are enabled to learn from each other to increase learning efficiency. Xing \emph{et al.} propose a framework, called RecycleML, which uses cross modal transfer to speed up the training of neural networks on mobile platforms across different sensing modalities in the scenario that the labelled data is insufficient \cite{xing2018enabling}. They design an hourglass model for knowledge transfer for multiple edge devices, as shown in Fig.~\ref{hourglass}. The bottom part denotes lower layers of multiple specific models, e.g., AudioNet, IMUNet, and VideoNet. The middle part represents the common layers of these specific models. These models project their data into the common layer for knowledge transfer. The upper part represents the task-specific layers of different models, which are trained in a targeted fashion. Experiments show that the framework achieves 50x speedup for the training. Federated learning could be also applied to accelerate the training of models on distributed edge devices. Smith \emph{et al.} propose a systems-aware framework to optimise the setting of federated learning (e.g., update cost and stragglers) and to speed up the training \cite{smith2017federated}.

\begin{table*}[ht]
    \footnotesize
    \centering
    \caption{Literature summary of model acceleration in training.}
    \label{trainacceleration}
    \begin{tabular}{c c c c c c c}
    \toprule
    \bf{Ref.} & \bf{Model}  & \bf{Approach} & \bf{Learning method} & \bf{Object} & \bf{Performance} \\
    \hline
    \cite{chen2018exploring}&DNN& Hardware acceleration  & Transfer learning & Review training factors & N/A\\
    \hline
    \cite{valery2017cpu} &CNN& Hardware acceleration  & Transfer learning & Alleviate memory constraint & Faster than Caffe-OpenCL trained\\
    \hline
    \cite{valery2018low}&CNN& \makecell{Hardware acceleration\\ parameter quantisation}  & Transfer learning & Alleviate memory constraint & Faster than Caffe-OpenCL trained\\
    \hline
    \cite{9073978}&DNN& Analog memory  & Transfer learning & Better energy-efficiency & Close to software baseline of 97.9\\
    \hline
    \cite{shahmohammadi2017smartwatch}&\makecell{RF, ET, NB\\ LR, SVM} & Human annotation  & Incremental learning & Investigate iML for HAR & 93.3\% accuracy \\
    \hline
    \cite{flutura2018drinkwatch}&Naive Bayes & Human annotation  & Incremental learning & Reduce limitations in learning & 6-8 hours to train a model\\
    \hline
    \cite{xing2018enabling}&CNN& Software acceleration  & Transfer learning & Reduce required labelled data & $50\times$ faster\\
    \hline
    \cite{smith2017federated}&Statistical model& Software acceleration  & Federated learning & Address statistical challenges& Outperform global, local manners\\

    \hline
    \cite{3375312}& GCN & \makecell{Software-hardware \\ Co-optimization}  & Supervised learning & \makecell{Accelerate GCN training \\ on heterogeneous platform}& An order of magnitude faster\\

    \bottomrule
    \end{tabular}
    \end{table*}

\subsection{Training optimisation}

Training optimisation refers to optimising the training process to achieve some given objectives, e.g., energy consumption, accuracy, privacy-preservation, security-preservation, etc. Since solo training is similar to training on a centralised server to a large extent, existing work mainly focuses on collaborative training. Federated learning is the most typical collaborative training architecture, and almost all literature on collaborative training is relevant to this topic.

\begin{figure}
\begin{center}
\includegraphics[width=0.85\linewidth]{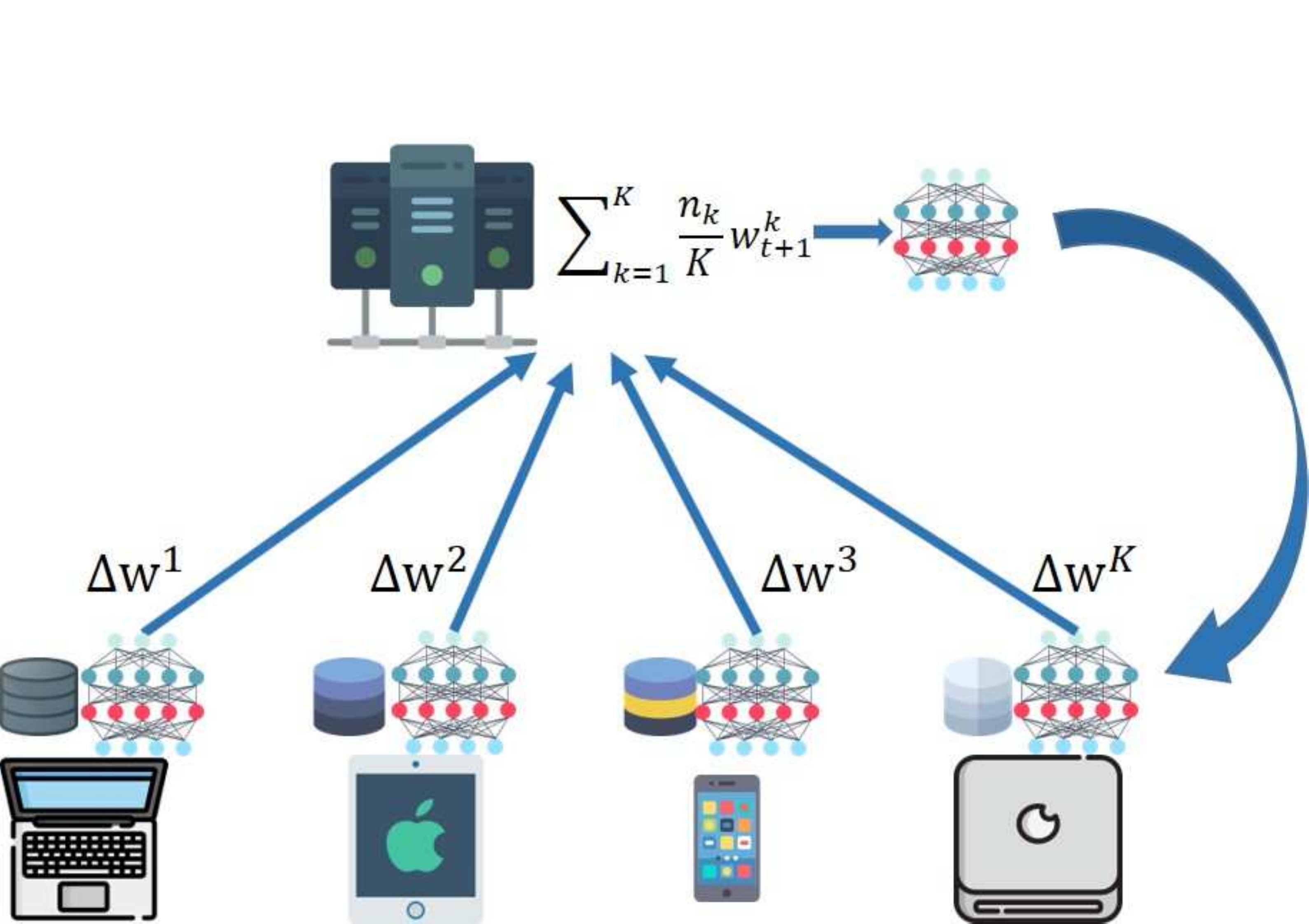}
\end{center}
\caption{The illustration of federated learning. Each training participant trains the shared model with cached data. After training, an update, i.e., $\Delta$w will be uploaded to the central server. All received updates from training participants would be aggregated to update the shared model. Then, the new shared model would be sent to all edge devices for the next round of learning.}
\label{federatedlearning}
\end{figure}

Federated learning is a kind of distributed learning \cite{predd2006distributed,valerio2017communication,lavassani2018combining}, which allows training sets and models to be located in different, non-centralised positions, and learning can occur independent of time and places. This training architecture is first proposed by Google, which allows smartphones to collaboratively learn a shared model with their local training data, instead of uploading all data to a central cloud server \cite{mcmahan2017federated}. The learning process of federated learning is shown as Fig.~\ref{federatedlearning}. There is a untrained shared model On the central server, which will be allocated training participants for training. Training participants, i.e., edge devices train the model with the local data. After local learning, changes of the model are summarised as a small focused update, which will be sent to the central server through encrypted communication. The central server averages received changes from all mobile devices and updates the shared model with the averaged result. Then, mobile devices download the update for their local model and repeats the procedure to continuously improve the shared model. In this learning procedure, only the encrypted changes are uploaded to the cloud and the training data of each mobile user remains on mobile devices. Transfer learning and edge computing are combined to learn a smarter model for mobile users. In addition, since learning occurs locally, federated learning could effectively protect user privacy, when compared with a traditional centralised learning approach.

Typical edge devices in federated learning are smartphones with unreliable and slow network connections. Moreover, due to the unknown mobility, these devices may be intermittently available for working. Hence, the communication efficiency between smartphones and the central server is of the utmost importance to the training. Specifically, there are factors affecting the communication efficiency: communication frequency and communication cost. In addition, the update from edge devices is vulnerable to malicious users. Hence privacy and security issues should also be considered. We discuss these problems in detail next. Table \ref{trainingoptimization} summarises literature on training optimisation.

\subsubsection{Communication Frequency}

In federated learning, communication between edge devices and the cloud server is the most important operation, which uploads the updates from edge devices to the could server and downloads the aggregated update from the shared model to local models. Due to the possible unreliable network condition of edge devices, minimising the number of update rounds, i.e., communication frequency between edge devices and cloud server is necessary. Jakub \emph{et al.} are the first to deploy federated learning framework and propose the setting for federated optimisation \cite{konevcny2015federated}. In \cite{konevcny2016federated}, the authors characterise the training data as massively distributed (data points are stored across massive edge devices), non-IID (training set on devices may be drawn from different distributions), and unbalanced (different devices have different number of training samples). In each round, each device sends an encrypted update to the central server. Then they propose a federated stochastic variance reduced gradient (FSVRG) algorithm to optimise the federated learning. They find that the central shared model could be trained with a small number of communication rounds.

McMahan \emph{et al.} propose a federated averaging algorithm (FedAvg) to optimise federated learning in the same scenario with \cite{konevcny2015federated,konevcny2016federated} and further evaluate the framework with five models and four datasets to proof the robustness of the framework \cite{mcmahan2016communication}. Although FedAvg could reduce the number of communication rounds for certain datasets, Zhao \emph{et al.} find that using this algorithm to train CNN models with highly skewed non-IID dataset would result in the significant reduction of the accuracy \cite{zhao2018federated}. They find the accuracy reduction results from the weight divergence, which refers to the difference of learned weights between two training processes with the same weight initialisation. Earth mover's distance (EMD) between the distribution over classes on each mobile device and the distribution of population are used to quantify the weight divergence. They then propose to extract a subset of data, which is shared by all edge devices to increase the accuracy.

Strategies that reduce the number of updates should be on the premise of not compromising the accuracy of the shared model. Wang \emph{et al.} propose a control algorithm to determine the optimal number of global aggregations to maximise the efficiency of local resources \cite{wang2019adaptive}. They first analyse the convergence bound of SGD based federated learning. Then, they propose an algorithm to adjust the aggregation frequency in real-time to minimise the resource consumption on edge devices, with the joint consideration of data distribution, model characteristics, and system dynamics.

Above-mentioned works adopt a synchronous updating method, where in each updating round, updates from edge devices are first uploaded to the central server, and then aggregated to update the shared model. Then the central server allocates aggregated updates to each edge device. Some researchers think that it is difficult to synchronise the process. On one hand, edge devices have significantly heterogeneous computing resources, and the local model are trained asynchronously on each edge device. On the other hand, the connection between edge devices and the central server is not stable. Edge devices may be intermittently available, or response with a long latency due to the poor connection. Wang \emph{et al.} propose an asynchronous updating algorithm, called CO-OP through introducing an age filter \cite{wang2017co}. The shared model and downloaded model by each edge device would be labelled with ages. For each edge device, if the training is finished, it would upload its update to the central server. Only when the update is neither obsolete nor too frequent, it will be aggregated to the shared model. However, most works adopt synchronous approaches in federated learning, due to its effectiveness \cite{mcmahan2016communication,chen2016revisiting}.

\subsubsection{Communication cost}
In addition to communication frequency, communication cost is another factor that affects the communication efficiency between edge devices and the central server. Reducing the communication cost could significantly save bandwidth and improve communication efficiency. Konevcny \emph{et al.} propose and proof that the communication cost could be lessened through structured and sketched updates \cite{konevcny2016federated,konevcny2017stochastic}. The structured update means learning an update from a restricted space that could be parametrised with few variables through using low rank and random mask structure, while sketched update refers to compressing the update of the full model through quantisation, random rotations and sub-sampling.

Lin \emph{et al.} find most of the gradient update between edge devices and the central server is redundant in SGD based federated learning \cite{lin2017deep}. Compressing the gradient could solve the redundancy problem and reduce the update size. However, compression methods, such as gradient quantisation and gradient sparsification would lead to the decreased accuracy. They propose a deep gradient compression (DGC) method to avoid the loss of accuracy, which use momentum correction and local gradient clipping on top of the gradient sparsification. Hardy \emph{et al.} also try to compress the gradient and propose a compression algorithm, called AdaComp \cite{hardy2017distributed}. The basic idea of AdaComp is compute staleness on each parameter and remove a large part of update conflicts.

Smith \emph{et al.} propose to combine multi-task learning and federated learning together, which train multiple relative models simultaneously \cite{smith2017federated}. It is quite cost-effective for a single model, during the training. They develop an optimisation algorithm, named MOCHA, for federated setting, which allows personalisation through learning separate but related models for each participant via multi-task learning. They also prove the theoretical convergence of this algorithm. However, this algorithm is inapplicable for non-convex problems.

Different from the client-to-server federated learning communication in \cite{konevcny2016federated,lin2017deep,hardy2017distributed}, Caldas \emph{et al.} propose to compress the update from the perspective of server-to-client exchange and propose Federated Dropout to reduce the update size \cite{caldas2018expanding}. In client-to-server paradigm, edge devices download the full model from the server, while in a server-to-client paradigm, each edge device only downloads a sub-model, which is a subset of the global shared model. This approach both reduces the update size and the computation on edge devices.

\subsubsection{Privacy and security issues}
After receiving updates from edge devices, the central server needs to aggregate these updates and construct an update for the shared global model. Currently, most deep learning models rely on variants of stochastic gradient descent (SGD) for optimisation. FedAvg, proposed in \cite{mcmahan2016communication}, is a simple but effective algorithm to aggregate SGD from each edge device through weighted averaging. Generally, the update from each edge device contains significantly less information of the users' local data. However, it is still possible to learn the individual information of a user from the update \cite{yang2019securing,9048613}. If the updates from users are inspected by malicious hackers, participant edge users' privacy would be threatened. Bonawitz \emph{et al.} propose Secure Aggregation to aggregate the updates from all edge devices, which makes the participant’ updates un-inspectable by the central server \cite{bonawitz2017practical}. Specifically, each edge device uploads a masked update, i.e., parameter vector to the server, and then the server accumulates a sum of the masked update vectors. As long as there is enough edge devices, the masks would be counteracted. Then, the server would be able to unmask the aggregated update. During the aggregation, all individual updates are non-inspectable. The server can only access the aggregated unmasked update, which effectively protect participants' privacy. Liu \emph{et al.} introduce homomorphic encryption to federated learning for privacy protection \cite{liu2018secure}. Homomorphic encryption \cite{gentry2009fully} is an encryption approach that allows computation on ciphertexts and generates an encrypted result, which, after decryption, is the same with the result achieved through direct computation on the plain text. The central server could directly aggregate the encrypted updates from participants.

Geyer \emph{et al.} propose an algorithm to hide the contribution of participants at the clients' based on differential privacy \cite{geyer2017differentially}. Similar to differential privacy-preserving traditional approaches \cite{dwork2011differential,abadi2016deep}, the authors add a carefully calibrated amount of noise to the updates from edge devices in federated learning. The approach ensures that attackers could not find whether an edge device participated during the training. A similar differential privacy mechanisms are also adopted in federated learning based recurrent language model and federated reinforcement learning in \cite{mcmahan2017learning} and \cite{zhuo2019federated}.

\begin{figure}
\begin{center}
\includegraphics[width=0.7\linewidth]{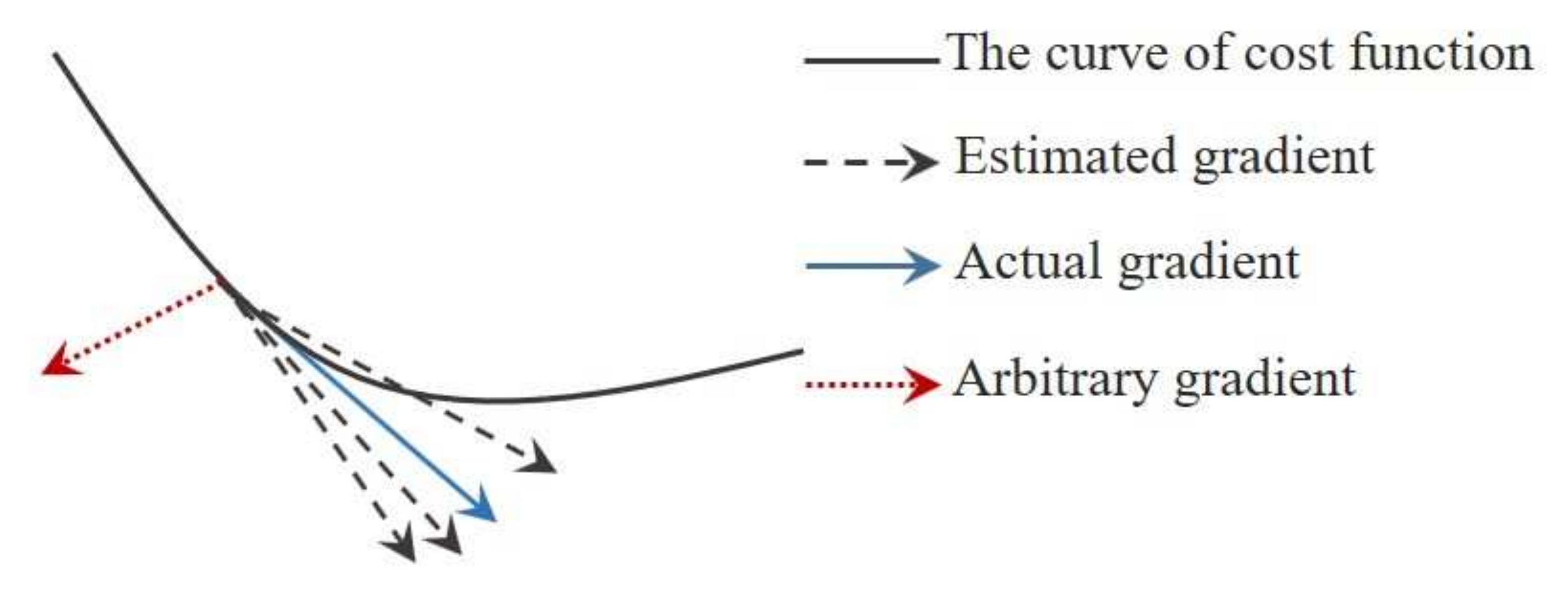}
\end{center}
\caption{The impact of data-poisoning attack. The black dashed arrows refers to the gradient estimates computed by honest participants, which are distributed around the actual gradient. The red dotted arrow indicates the arbitrary gradient computed by malicious participants, which hampers the convergence of the training.}
\label{data-poisoning}
\end{figure}

In federated learning, the participants could observe intermediate model states and contribute arbitrary updates to the global shared model. All aforementioned research assumes that the participants in federated learning are un-malicious, which provides a real training set and uploads the update based on the training set. However, if some of the participants are malicious, who uploads erroneous updates to the central server, the training process fails. In some cases, the attack would result in large economic losses. For example, in a backdoor attacked face recognition based authentication system, attackers could mislead systems to identify them as a person who can access a building through impersonation. According to their attack patterns, attacks could be classified into two categories: data-poisoning and model-poisoning attacks. Data-poisoning means compromising the behaviour and performance of the model through changing the training set, e.g., accuracy, whilst model-poisoning only change the model's behaviour on specific inputs, without impacting the performance on other inputs. The impact of data-poisoning attack is shown as Fig.~\ref{data-poisoning}.

The work in \cite{biggio2012poisoning} tests the impact of a data-poisoning attack on SVM through injecting specially crafted training data, and find that the SVM's test error increases with the attack. Steinhardt \emph{et al.} construct the approximate upper bound of the attack loss on SVM and provides a solution to eliminate the impact of the attack \cite{steinhardt2017certified}. In particular, they first remove outliers residing outside a feasible bound, and then minimise the margin-based loss on the rest data.

Fung \emph{et al.} evaluate the impact of sybil-based data-poisoning attack on federated learning and propose a defense scheme, FoolsGold, to solve the problem \cite{fung2018mitigating}. A sybil-based attack \cite{douceur2002sybil} means that a participant edge device has a wrong training dataset, in which the data is the same with other participants whilst its label is wrong. For example, in digit recognition, the digit `1' is labelled with `7'. They find that attackers may overpower other honest participants by poisoning the model with sufficient sybils. The proposed defense system, FoolGold, is based on contribution similarity. Since sybils share a common objective, their updates appear more similar than honest participants. FoolGold eliminates the impact of sybil-based attacks through reducing the learning rate of participants that repeatedly upload the same updates.

Blanchard \emph{et al.} evaluate the Byzantine resilience of SGD in federated learning \cite{blanchard2017machine}. Byzantine refers to arbitrary failures in federated learning, such as erroneous data and software bugs. They find that linear gradient aggregation has no tolerance for even one Byzantine failure. Then they propose a Krum algorithm for aggregation with the tolerance of $f$ Byzantines out of $n$ participants. Specifically, the central server computes pairwise distances amongst all updates from edge devices, and takes the sum of $n-f-2$ closest distance for all updates. The update with the minimum sum would be used to update the global shared model. However, all updates from edge devices are inspectable during computation, which may result in the risk of privacy disclosure. Chen \emph{et al.} propose to use the geometric median of gradients as the update in federated learning \cite{chen2017distributed}. This approach could tolerate $q$ Byzantine failures up to $2q(1+\epsilon)\leq m$, in which $q$ is the number of Byzantine failures, $m$ refers to the headcount of participants, and $\epsilon \large$ is a small constant. This approach groups all participants into mini-batches. However, Yin \emph{et al.} find that the approach fails if there is one Byzantine in each mini-batch \cite{yin2018byzantine}. They then propose a coordinate-wise median based approach to deal with the problem.

\begin{figure}
\begin{center}
\includegraphics[width=\linewidth]{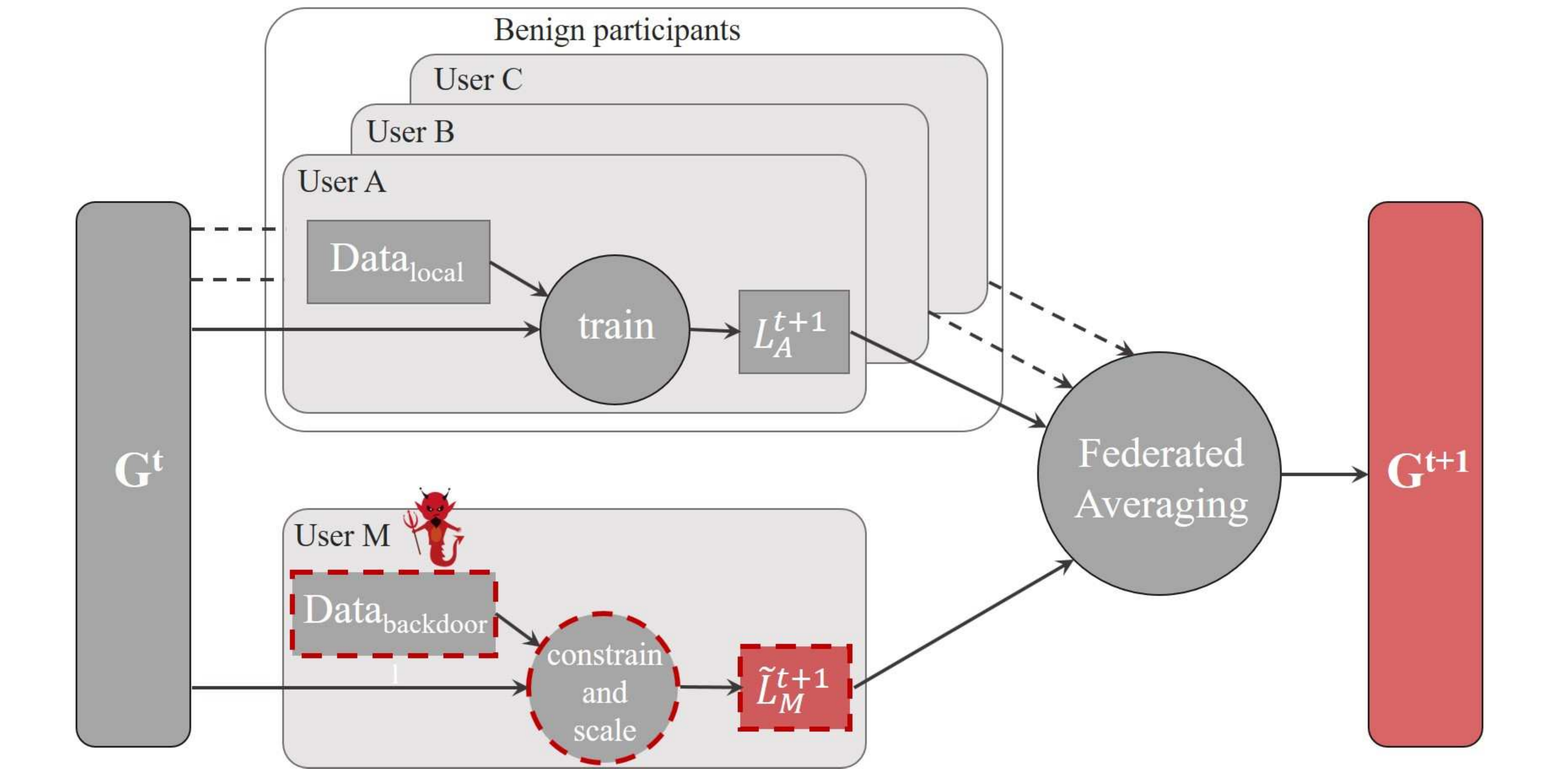}
\end{center}
\caption{Overview of model-poisoning based attack. Attackers train the backdoor model with local data. Then, attackers scale up the weight of the update to guarantee that the backdoor model would not be cancelled out by other updates.}
\label{backdoor}
\end{figure}

In fact, data-poisoning based attacks on federated learning is low in efficiency in the condition of small numbers of malicious participants. Because there are usually thousands of edge devices participating in the training in federated learning. The arbitrary update would be offset by averaging aggregation. In contrast, model-poisoning based attacks are more effective. Attackers directly poison the global shared model, instead of the updates from thousands of participants. Attackers introduce hidden backdoor functionality in the global shared model. Then, attackers use key, i.e., input with attacker-chosen features to trigger the backdoor. The model-poisoning based attack is shown as Fig.~\ref{backdoor}. Works on model-poisoning mainly focus on the problem of how backdoor functionality is injected in federated learning. Hence, we will focus on this direction as well.

Chen \emph{et al.} evaluate the feasibility of conducting a backdoor in deep learning through adding few poisoning samples into the training set \cite{chen2017targeted}. They find that only 5 poisoning samples out of 600,000 training samples are enough to create a backdoor. Bagdasaryan \emph{et al.} propose a model replacement technique to open a backdoor to the global shared model \cite{bagdasaryan2018backdoor}. As we aforementioned, the central server computes an update through averaging aggregation on updates from thousands of participants. The model replacement method scales up the weights of the `backdoored' update to ensure that the backdoor survives the averaging aggregation. This is a single-round attack. Hence, such attack usually occurs during the last round update of federated learning. Different from \cite{bagdasaryan2018backdoor}, Bhagoji \emph{et al.} propose to poison the shared model even when it is far from convergence, i.e., the last round update \cite{bhagoji2018analyzing}. To prevent that, the malicious update is offset by updates from other participants, they propose a explicit boosting mechanism to negate the aggregation effect. They evaluate the attack technique against some famous attack-tolerant algorithms, i.e., Krum algorithm \cite{blanchard2017machine} and coordinate-wise median algorithm \cite{yin2018byzantine}, and find that the attack is still effective.

\subsection{Uncertainty Estimates}
Standard deep learning method for classification and regression could not capture model uncertainty. For example, in model for classification, obtained results may be erroneously interpreted as model confidence. Such problems exist as well in edge intelligence. Efficient and accurate assessment of the deep learning output is of crucial importance, since the erroneous output may lead to undesirable economy loss or safety consequence in practical applications.

In principle, the uncertainty could be estimated through extensive tests. \cite{gal2016dropout} propose a theoretical framework that casts dropout training in DNNs as approximate Bayesian inference in deep Gaussian processes. The framework could be used to model uncertainty with dropout neural networks through extracting information from models. However, this process is computation intensive, which is not applicable on mobile devices. This approach is based on sampling, which requires sufficient output samples for estimation. Hence, the main challenge to estimate uncertainty on mobile devices is the computational overhead. Based on the theory proposed in \cite{gal2016dropout}, Yao \emph{et al.}  propose RDeepSence, which integrates scoring rules as training criterion that measures the quality of the uncertainty estimation to reduce energy and time consumption \cite{yao2018rdeepsense}. RDeepSence requires to re-train the model to estimate uncertainty. The authors further propose ApDeepSence, which replaces the sampling operations with layer-wise distribution approximations following closed-form representations \cite{yao2018apdeepsense}.

\subsection{Applications}

\begin{figure*}
\begin{center}
\includegraphics[width=0.75\linewidth]{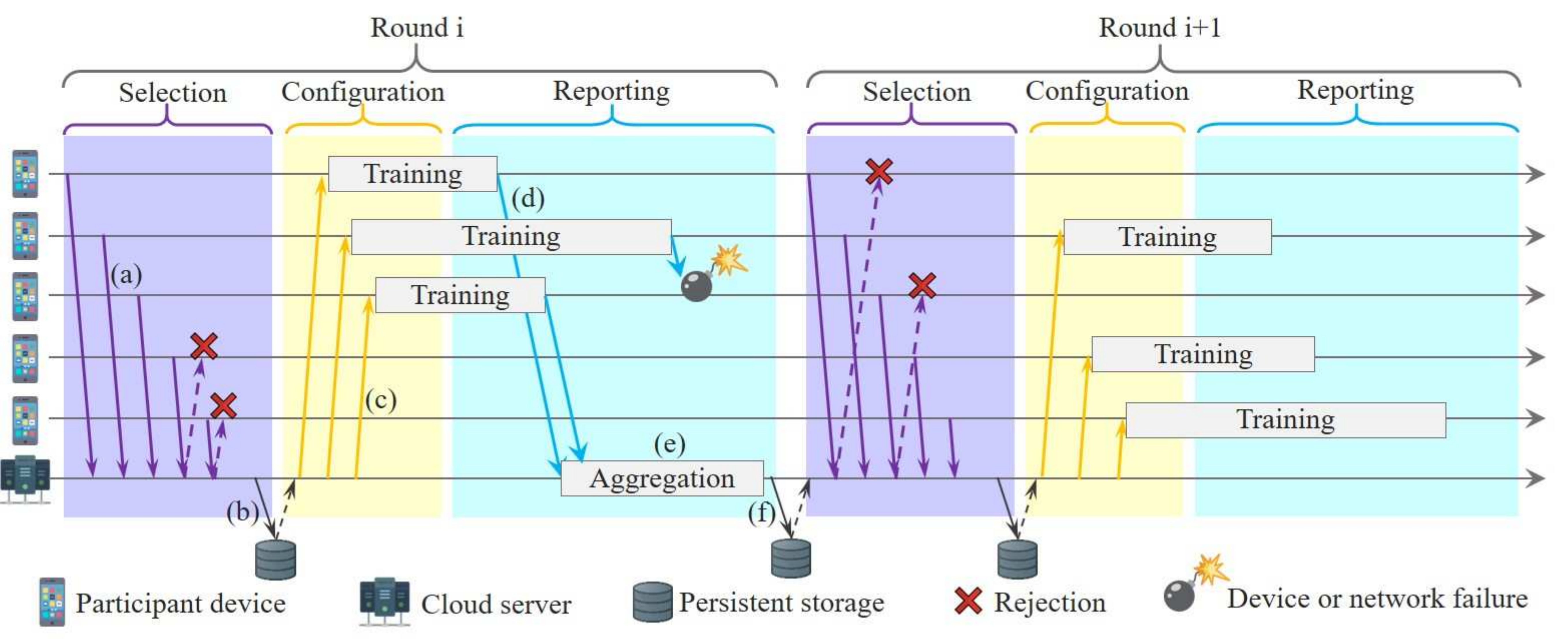}
\end{center}
\caption{The illustration of a TensorFlow-based federated learning system. (a) edge devices register to participate in federated training. Un-selected devices would be suggested to participate in the next round. (b) server reads the checkpoint of the model from storage. (c) server sends a shared model to each selected edge device. (d) edge devices train the model with local data and uploads their updates. (e) All received updates are aggregated. (f) the server save the checkpoint of the model.}
\label{federatedprotocol}
\end{figure*}

Bonawitz \emph{et al.} develop a scalable product system for federated learning on mobile devices, based on TensorFlow \cite{bonawitz2019towards}. In this system, each updating round consists of three phases: client selection, configuration, and reporting, as shown in Fig.~\ref{federatedprotocol}. In the client selection phase, eligible edge devices, e.g., devices with sufficient energy and computing resources, periodically send messages to the server to report the liveness. The server selects a subset among them according to a given objective. In the configuration phase, the server sends a shared model to each selected edge device. In the reporting phase, each edge device reports the update to the server, which would be aggregated to update the shared model. This protocol presents a framework of federated learning, which could adopt multiple strategies and algorithms in each phase. For example, the client selection algorithm proposed in \cite{nishio2019client} could be used in the client selection phase. The communication strategy in \cite{konevcny2015federated,konevcny2016federated} could be used for updating, and the FedAvg algorithm in \cite{mcmahan2016communication} is adopted as an aggregation approach.

Researchers from Google have been continuously working on improving the service of Gboard with federated learning. Gboard consists of two parts: text typing and a search engine. The text typing module is used to recognise users' input, whilst the search engine provides user relevant suggestions according to their input. For example, when you type `let's eat', Gboard may display the information about nearby restaurants. Hard \emph{et al.} train a RNN language model using a federated learning approach to improve the prediction accuracy of the next-word for Gboard \cite{hard2018federated}. They compare the training result with traditional training methods on a central server. Federated learning achieves comparable accuracy with the central training approach. Chen \emph{et al.} use federated learning to train a character-level RNN to predict high-frequent words on Gboard \cite{chen2019federated}. The approach achieves 90.56\% precision on a publicly-available corpus. McMahan \emph{et al.} undertake the first step to apply federated learning to enhance the search engine of Gboard \cite{mcmahan2017federated}. When users search with Gboard, information about the current context and whether the clicked suggestion would be stored locally. Federated learning processes this information to improve the recommendation model. Yang \emph{et al.} further improve the recommendation accuracy by introducing an additional \emph{triggering model} \cite{yang2018applied}. Similarly, there are some works \cite{ramaswamy2019federated,ramaswamy2019federated} focusing on emoji prediction on mobile keyboards.

Federated learning has great potential in the medical imaging domain, where patient information is highly sensitive. Sheller \emph{et al.} train a brain tumour segmentation model with data from multi-institution by applying federated learning \cite{sheller2018multi}. The encrypted model is first sent to data owners , i.e., institutions, then the data owners decode, train, encrypt and upload the model back to the central aggregator. Roy \emph{et al.} further develop an architecture of federated learning that uses peer-to-peer communications to replace the central aggregator towards medical applications \cite{roy2019braintorrent}.

Samarakoon \emph{et al.} apply federated learning in vehicular networks to jointly allocate power and resources for ultra reliable low latency communication \cite{samarakoon2018distributed}. Vehicles train and upload their local models to the roadside unit (RSU), and RSU feeds back the global model to vehicles. Vehicles could use the model to estimate the queue length in city. Based on the queue information, the traffic system could reduce the queue length and optimise the resource allocation.

Nguyen \emph{et al.} develop DIoT, a self-learning system to detect infected IoT devices by malware botnet in smart home environments \cite{nguyen2018diot}. IoT devices connect to the Internet through a gateway. They design two models for IoT device identification and anomaly detection. These two models are trained through the federated learning approach.

\begin{table*}[ht]
    \footnotesize
    \centering
    \caption{Literature summary of training optimisation.}
    \label{trainingoptimization}
    \begin{tabular}{c c c c c c c}
    \toprule
    \bf{Ref.}  & \bf{Problem} & \bf{Solution} & \bf{Dataset}  & \bf{Performance} \\
    \hline
    \cite{konevcny2015federated} & Communication efficiency & FSVRG & Google+ posts  & Less rounds\\
    \hline
    \cite{konevcny2016federated} & Communication efficiency & FSVRG & Google+ posts  & Less rounds\\
    \hline
    \cite{mcmahan2016communication}& Communication efficiency & FedAvg &MNIST, CIFAR-10, KWS  & $10-100\times$ less rounds \\
    \hline
    \cite{zhao2018federated}& Communication efficiency & FedAvg,  data sharing&MNIST, CIFAR-10, KWS  & 30\% higher accuracy\\
    \hline

    \cite{8995775}& Uncoordinated communication & \makecell{Incentive mechanism \\ Admission control}&N/A  & 22\% gain in reward\\
    \hline

    \cite{8963610}& Incentive mechanism & Deep reinforcement learning& MNIST  & Lower communication cost\\
    \hline
    \cite{wang2019adaptive}& Communication frequency & Aggregation control &MNIST  & Near to the optimum\\
    \hline
    \cite{wang2017co}& Communication frequency & CO-OP&MNIST & 80\% accuracy\\
        \hline
    \cite{8952884}& Communication bandwidth &  Beamforming design &CIFAR-10 & Lower training loss, higher accuracy\\
    \hline
    \cite{9026922}& Noisy communication &  Successive convex approximation &MNIST & Approach to centralized method\\
    \hline

    \cite{9014530}& Wireless fading channel &  D-DSGD, CA-DSGD &MNIST & Converges faster, higher accuracy\\
    \hline
    \cite{8950073}& Single point of failure & Server-less aggregation &Real-world sensing data & One order of magnitude less rounds \\
    \hline
    \cite{konevcny2016federated}& Communication cost & \makecell{Structured update \\ sketched update} &CIFAR-10, Reddit & 85\% accuracy\\
    \hline
    \cite{lin2017deep}& Communication cost & DGC&\makecell{ ImageNet, Penn Treebank\\ Cifar10, Librispeech Corpus}   & $270-600\times$ smaller update size\\
    \hline
    \cite{hardy2017distributed}& Communication cost & \makecell{Compression\\staleness mitigation }&MNIST & $191\times$ smaller update size\\
    \hline
    \cite{smith2017federated}&\makecell{ Multi-task learning\\ Communication cost }& MOCHA & \makecell{Human Activity Recognition\\ GLEAM, Vehicle Sensor} & Lowest prediction error\\
    \hline
    \cite{caldas2018expanding}& Communication cost & Federated Dropout&MNIST, EMNIST, CIFAR-10 & $28\times$ smaller update size\\
    \hline
    \cite{bonawitz2017practical}& Information revealing & Secure Aggregation&N/A & $1.98\times$ expansion for $2^{14}$ users\\
    \hline
    \cite{liu2018secure}& Privacy protection & Homomorphic encryption&NUS-WIDE, Default-Credit & Little accuracy drop\\
    \hline
    \cite{geyer2017differentially}& Privacy protection & Differentially privacy &non-IID MNIST & Privacy maintained \\
    \hline

    \cite{9069945}& Privacy protection & \makecell{Differentially privacy \\ K-client random scheduling} & MNIST & Privacy maintained \\
    \hline
    \cite{zhuo2019federated}& Privacy protection & Gaussian differential&WHS, CT, WHG  &  F1 score 10\% - 20\% higher\\
    \hline
    \cite{cheng2019secureboost}& Privacy protection & SecureBoost&Credit 1, Credit 2  & Higher accuracy, F1-score\\
    \hline
    \cite{mcmahan2017learning}& Privacy protection & Differentially privacy &Reddit posts & Similar to un-noised models\\

    \hline
    \cite{fung2018mitigating}& Sybil-based attack & FoolGold&MNIST, VGGFace2 & Attacking rate \textless 1\% \\
    \hline
    \cite{blanchard2017machine}& Byzantine failure & Krum aggregation&MNIST, Spambase& Toleratable for 45\% Byzantines \\
    \hline
    \cite{chen2017distributed}& Byzantine failure & Batch gradients median& N/A & $2q(1+\epsilon)\leq m$ Byzantines\\
    \hline
    \cite{yin2018byzantine}& Byzantine failure & Coordinate-wise median & N/A & Optimal statistical error rate\\
    \hline

    \cite{bhagoji2018analyzing}& Backdoor attack & Explicit boosting  & Fashion-MNIST, Adult Census & 100\% backdoor accuracy \\

    \bottomrule
    \end{tabular}
    \end{table*}

\section{Edge inference}
The exponential growth of network size and the associated increase in computing resources requirement have been become a clear trend. Edge inference, as an essential component of edge intelligence, is usually performed locally on edge devices, in which the performance, i.e., execution time, accuracy, energy efficiency, etc. would be bounded by technology scaling. Moreover, we see an increasingly widening gap between the computation requirement and the available computation capacity provided by the hardware architecture \cite{nguyen2018diot}. In this section, we discuss various frameworks and approaches that contribute to bridging the gap.

\subsection{Model Design}
Modern neural network models are becoming increasingly larger, deeper and slower, they also require more computation resources \cite{hu2018squeeze,zoph2018learning,kong2017science}, which makes it quite difficult to directly run high performance models on edge devices with limited computing resources, e.g., mobile devices, IoT terminals and embedded devices. Guo \emph{emph} evaluate the performance of DNN on edge device and find inference on edge devices costs up to two orders of magnitude greater energy and response time than central server \cite{guo2018cloud}. Many recent works have focused on designing lightweight neural network models, which could be performed on edge devices with less requirements on the hardware. According to the approaches of model design, existing literature could be divided into two categories: architecture search, and human-invented architecture. The former is to let machine automatically design the optimal architecture, while the latter is to design architectures by human.

\subsubsection{Architecture Search}
Designing neural network architectures is quite time-consuming, which requires substantial effort of human experts. One possible research direction is to use AI to enable machine search for the optimal architecture automatically. In fact, some automatically searched architectures, e.g., NASNet \cite{zoph2018learning}, AmoebaNet \cite{real2018regularized}, and Adanet \cite{cortes2017adanet}, could achieve competitive even much better performance in classification and recognition. However, these architectures are extremely hardware-consuming. For example, it requires 3150 GPU days of evolution to search for the optimal architecture for CIFAR-10 \cite{real2018regularized}. Mingxing \emph{et al.} adopt reinforcement learning to design mobile CNNs, called MnasNet, which could balance accuracy and inference latency \cite{tan2018mnasnet}. Different from \cite{zoph2018learning,real2018regularized,cortes2017adanet}, in which only few kinds of cells are stacked, MnasNet cuts down per-cell search space and allow cells to be different. There are more $5 \times 5$ depthwise convolutions in MnasNet, which makes MnasNet more resource-efficient compared with models that only adopt $3 \times 3$ kernels.

Recently, a new research breakthrough of differentiable architecture search (DARTS) \cite{liu2018darts} could significantly reduce dependence on hardware. Only four GPU days are required to achieve the same performance as \cite{real2018regularized}. DARTS is based on continuous relaxation of the architecture representation and uses gradient descent for architecture searching. DARTS could be used for both convolutional and recurrent architectures.

Architecture search is hot research area and has a wide application future. Most literature on this area is not specially for edge intelligence. Hence, we will not further discuss on this field. Readers interested in this field could refer to \cite{elsken2018neural,wistuba2019survey}.

\subsubsection{Human-invented Architecture}
Although architecture search shows good ability in model design, its requirement on hardware holds most researchers back. Existing literature mainly focuses on human-invented architecture. Howard \emph{et al.} use depth-wise separable convolutions to construct a lightweight deep neural network, MobileNets, for mobile and embedded devices \cite{howard2017mobilenets}.
In MobileNets, a convolution filter is factorised into a depth-wise and a point-wise convolution filter. The drawback of depth-wise convolution is that it only filters input channels. Depth-wise separable convolution, which combines depth-wise convolution and $1 \times 1$ point-wise convolution could overcome this drawback. MobileNet uses $3 \times 3$ depth-wise separable convolutions, which only requires 8-9 times less computation than standard ones. Moreover, depth-wise and point-wise convolutions could also be applied to implement keyword spotting (KWS) models \cite{zhang2017hello} and depth estimation \cite{wofk2019fastdepth} on edge devices.

\begin{figure}[tp!]
\begin{center}
\includegraphics[width=\linewidth]{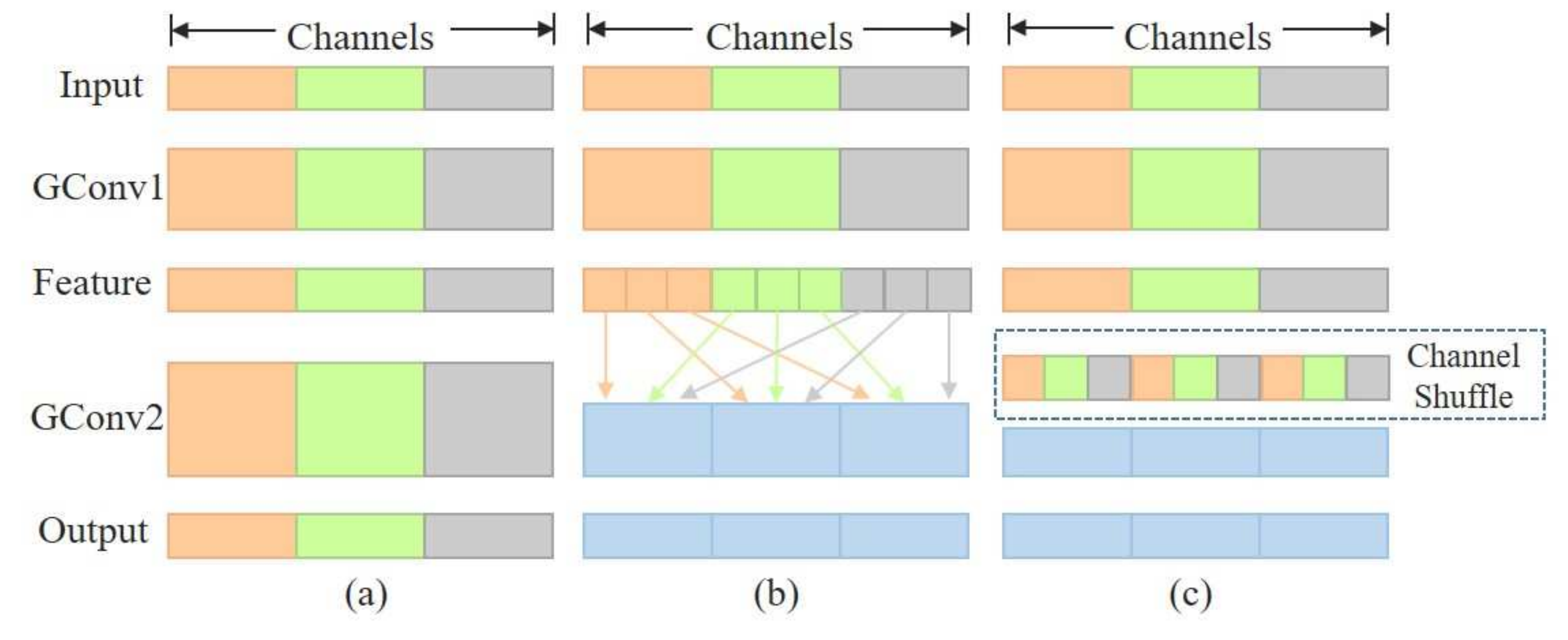}
\end{center}
\caption{Illustration of channel shuffle. GConv refers to group convolution. (a) Two stacked convolution layers. Each output channel is related with an input channel of the same group. (b) GConv2 takes data from different groups to make full relations with other channels. (c) The implementation of channel shuffle, which achieves the same effect with (b).}
\label{shufflenet}
\end{figure}

Group convolution is another way to reduce computation cost for model designing. Due to the costly dense $1 \times 1$ convolutions, some basic architectures, e.g., Xception \cite{chollet2017xception} and ResNeXt \cite{xie2017aggregated} cannot be used on resource-constrained devices. Zhang \emph{et al.} propose to reduce the computation complexity of $1 \times 1$ convolutions with pointwise group convolution \cite{zhang2018shufflenet}. However, there is a side effect brought on by group convolution, i.e., outputs of one channel are only derived from a small part of the input channels. The authors then propose to use a \emph{channel shuffle} operation to enable information exchanging among channels, as shown in Fig~\ref{shufflenet}.

Depth-wise convolution and group convolution are usually based on `sparsely-connected' convolutions, which may hamper inter-group information exchange and degrades model performance. Qin \emph{et al.} propose to solve the problem with merging and evolution operations \cite{qin2018merging}. In merging operation, features of the same location among different channels are merged to generate a new feature map. Evolution operation extracts the information of location from the new feature map and combines extracted information with the original network. Therefore, information is shared by all channels, so that the information loss problem of inter-groups is effectively solved.

\subsubsection{Applications}
A large number of models have been designed for various applications, including face recognition \cite{chen2018mobilefacenets,duong2018mobiface,sandler2018mobilenetv2}, human activity recognition (HAR) \cite{bhattacharya2016smart,almaslukh2018robust,almaslukh2018,sundaramoorthy2018harnet,radu2016towards,cruciani2018automatic,bo2018detecting,yao2017deepsense,yao2018qualitydeepsense}, vehicle driving \cite{streiffer2017darnet,liu2015toward,bo2013you,yang2011detecting}, and audio sensing \cite{lane2015deepear,georgiev2017low}. We introduce such applications next.

Face verification is increasingly attracting interests in both academic and industrial areas, and it is widely used in device unlocking \cite{fan2020emgauth} and mobile payments \cite{wen2015face}. Particularly, some applications, such as smartphone unlocking need to run locally with high accuracy and speed, which is challenging for traditional big CNN models due to constrained resources on mobile devices. Sheng \emph{et al.} present a compact but efficient CNN model, MobileFaceNets, which uses less than 1 million parameters and achieves similar performance to the latest big models of hundreds MB size \cite{chen2018mobilefacenets}. MobileFaceNets uses a global depth-wise convolution filter to replace the global average pooling filter and carefully design a class of face feature. Chi \emph{et al.} further lighten the weight of MobileFaceNets and presents MobiFace \cite{duong2018mobiface}. They adopt the Residual Bottleneck block \cite{sandler2018mobilenetv2} with expansion layers. Fast downsampling is also used to quickly reduce the dimensions of layers over $14 \times 14$. These two adopted strategies could maximise the information embedded in feature vectors and keep low computation cost.


\begin{table*}[htp]
    \footnotesize
    \centering
    \caption{Comparison of different HAR applications.}
    \label{HAR-comparison}
    \begin{tabular}{c c c c c c }
    \toprule
    \bf{Ref.}  & \bf{Model} & \bf{ML method} & \bf{Objective} & \bf{Dataset}\\
    \hline
     \cite{bhattacharya2016smart}  & RBM & Unsupervised Learning & Energy-efficiency, higher accurate & Opportunity dataset\\
     \hline
     \cite{almaslukh2018robust}  & CNN & Deep Learning & Improve accuracy & UCI \& WISDM \\
     \hline
     \cite{almaslukh2018}  & CNN & Deep Learning & Improve accuracy & RealWorld HAR\\
     \hline
     \cite{sundaramoorthy2018harnet}  & LSTM & Incremental learning & Minimise resource consumption & Heterogeneity Dataset \\
     \hline
     \cite{radu2016towards}  & CNN & Multimodal Deep Learning & Integrate sensor data & Opportunity dataset \\
     \hline
     \cite{cruciani2018automatic}  & Heuristic function & Supervised learning & Automatic labelling & 38 day-long dataset \\
     \hline
     \cite{bo2018detecting}  &\makecell{Random forest\\  Naive bayes \\ decision tree} & Ensemble learning & Detect label errors & CIMON \\
     \hline
     \cite{yao2017deepsense}  & CNN \& RNN & Supervised learning & Reduce data noise & Opportunity dataset \\
     \hline
     \cite{yao2018qualitydeepsense}  & CNN \& RNN & Supervised Learning & Heterogeneous sensing quality & Opportunity dataset \\
    \bottomrule
    \end{tabular}
\end{table*}

Edge intelligence could be used to extract contextual information from sensor data and facilitate the research on Human Activity Recognition (HAR). HAR refers to the problem of recognising when, where, and what a person is doing \cite{plotz2018deep}, which could be potentially used in many applications, e.g., healthcare, fitness tracking, and activity monitoring \cite{guan2017ensembles,shoaib2017resource}.
Table \ref{HAR-comparison} compares existing HAR technologies, regarding to their frameworks, models, ML methods, and objects.
The challenges of HAR on edge platforms could summarised as follows.
\begin{itemize}
    \item Commonly used classifiers for HAR, e.g., naive Bayes, SVM, DNN, are usually computation-intensive, especially when multiple sensors are involved.
    \item HAR requires to support near-real-time user experience in many applications.
    \item Very limited amount of labelled data is available for training HAR models.
    \item The data collected by on-device sensor includes noise and ambiguity.
\end{itemize}

Sourav \emph{et al.} investigate how to deploy Restricted Boltzmann Machines (RBM)-based HAR models on smartwatch platforms, i.e., the Qualcomm Snapdragon 400 \cite{bhattacharya2016smart}. They first test the complexity of a model that a smartwatch can afford. Experiments show that although a simple RBM-based activity recognition algorithm could achieve satisfactory accuracy, the resource consumption on a smartwatch platform is unacceptably high. They further develop pipelines of feature representation and RBM layer activation functions. The RBM model could effectively reduce energy consumption on smartwatches. Bandar \emph{et al.} introduce time domain statistical features in CNN to improve the recognition accuracy \cite{almaslukh2018robust}. In addition, to reduce the over-fitting problem of their model, they propose a data augmentation method, which applies a label-preserving transformation on raw data to create new data. The work is extended with extracting position features in \cite{almaslukh2018}.

Although deep learning could automatically extract features by exploring hidden correlations within and between data, pre-trained models sometimes cannot achieve the expected performance due to the diversities of devices and users, e.g., the heterogeneity of sensor types and user behaviour \cite{stisen2015smart}. Prahalathan \emph{et al.} propose to use on-device incremental learning to provide a better service for users \cite{sundaramoorthy2018harnet}. Incremental learning \cite{rosenfeld2018incremental} refers to a secondary training for a pre-trained model, which constrains newly learned filters to be linear combinations of existing ones. The re-trained model on mobile devices could provide personalised recognition for users.

Collecting fine-grained datasets for HAR training is challenging, due to a variety of available sensors, e.g., different sampling rated and data generation models. Valentin \emph{et al.} propose to use RBM architecture to integrate sensor data from multiple sensors \cite{radu2016towards}. Each sensor input is processed by a single stacked restricted Boltzmann machine in RBM model. Afterwards, all outputted results are merged for activity recognition by another stacked restricted Boltzmann machine. Supervised machine learning is a most commonly utilised approach for activity recognition, which requires a large amount of labelled data. Manually labelling requires extremely large amounts of effort. Federico \emph{et al.} propose a knowledge-driven automatic labelling method to deal with the data annotation problem \cite{cruciani2018automatic}. GPS data and step count information are used to generate weak labels for the collected raw data. However, such an automatic annotation approach may create labelling errors, which impacts the quality of the collected data. There are three types of labelling errors, including inaccurate timestamps, mislabelling, and multi-action labels. Multi-action labels means that individuals perform multiple different actions during the same label. Xiao \emph{et al.} solve the last two labelling errors through an ensemble of four stratified trained classifiers of different strategies, i.e., random forest, naive bayes, and decision tree \cite{bo2018detecting}.

The data collected by on-device sensors maybe noisy and it is hard to eliminate \cite{ang2007nonlinear,stisen2015smart}. For example, in movement tracking application on mobile devices, the travelled distance is computed with the sensory data, e.g., acceleration, speed, and time. However, the sensory data maybe noisy, which will result in estimation errors. Yao \emph{et al.} develop DeepSense, which could directly extracts robust noise features of sensor data in a unified manner \cite{yao2017deepsense}. DeepSense combines CNN and RNN together to learn the noise model. In particular, the CNN in DeepSense learns the interaction among sensor modalities, while the RNN learn the temporal relationship among them based on the output of the CNN. The authors further propose QualityDeepSense with the consideration of the heterogeneous sensing quality \cite{yao2018qualitydeepsense}. QualityDeepSense hierarchically adds sensor-temporal attention modules into DeepSense to measures the quality of input sensory data. Based on the measurement, QualityDeepSense selects the input with more valuable information to provide better predictions.

\begin{figure}[tp!]
\begin{center}
\includegraphics[width=\linewidth]{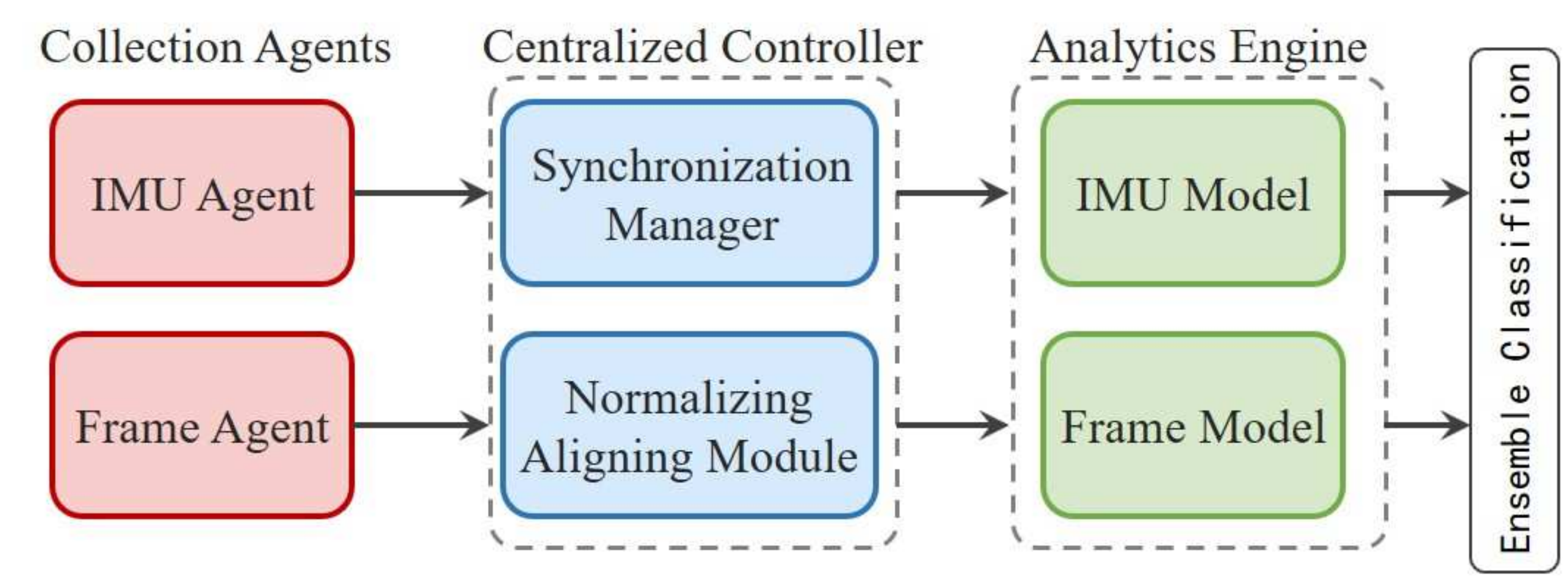}
\end{center}
\caption{Architecture of DarNet. IMU agent runs on IoT devices and frame agent runs on mobile devices. A centralised controller collects and pre-processes data for the analytic engine.}
\label{darnet}
\end{figure}

Distracted driving is a key problem, as it potentially leads to traffic accidents \cite{klauer2014distracted}. Some researchers address this problem by implementing DL models on smartphones to detect distracted driving behaviour in real-time. Christopher \emph{et al.} design DarNet, a deep learning based system to analyse driving behaviours and to detect distracted driving \cite{streiffer2017darnet}. There are two modules in the system: data collection and analytic engine. There is a centralised controller in the data collection component, which collects two kinds of data, i.e., IMU data from drivers' phones and images from IoT sensors. The analytic engine uses CNN to process image data, and RNN for sensor data, respectively. The outputs of these two models are combined through an ensemble-based learning approach to enable near real-time distracted driving activity detection. Fig. \ref{darnet} presents the architecture of DarNet. In addition to CNN and RNN models, there are also other models could be used to detect unsafe driving behaviours, such as SVM \cite{liu2015toward}, HMM \cite{bo2013you}, and decision tree \cite{yang2011detecting}.

\begin{figure}[tp!]
\begin{center}
\includegraphics[width=\linewidth]{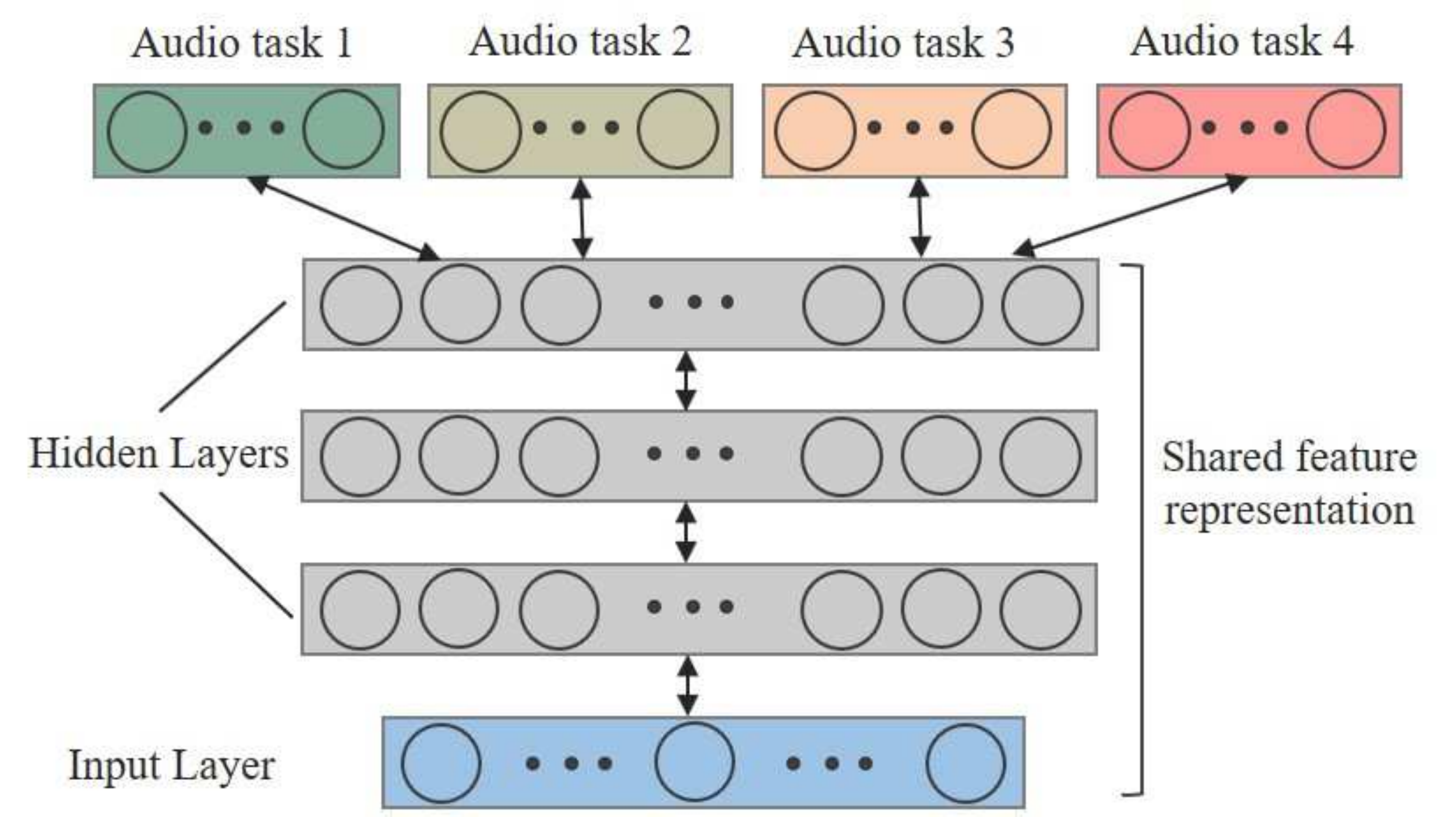}
\end{center}
\caption{Illustration of the multi-task audio sensing network.}
\label{multi-task}
\end{figure}

Audio sensing has become an essential component for many applications, such as speech recognition \cite{graves2013speech}, emotion detection \cite{rabbi2011passive}, and smart homes \cite{gebhart2017google}. However, directly running audio sensing models, even just the inference, would introduce a heavy burden on the hardware, such as digital signal processing (DSP) and battery. Nicholas \emph{et al.} develop DeepEar, a DNN based audio sensing prototype for the smartphone platform \cite{lane2015deepear}, including four coupled DNNs of stacked RBMs that collectively perform sensing tasks. These four DNNs share the same bottom layers, and each of them is responsible for a specific task, for example, emotion detection, and tone recognition.  Experiments show that only 6\% of the battery is enough to work through a day with the compromise of 3\% accuracy drop. Petko \emph{et al.} further improve the accuracy and reduces the energy consumption through applying multi-task learning and training shared deep layers \cite{georgiev2017low}. The architecture of multi-task learning is shown as Fig.~\ref{multi-task}, in which the input and hidden layers are shared for audio analysis tasks. Each task has a distinct classifier. Moreover, the shared representation is more scalable than DeepEar, since there is no limitation in the integration of tasks.

\subsection{Model Compression}\label{modelcompression}
Although neural networks are quite powerful in various promising applications, the increasing size of neural networks, both in depth and width, results in the considerable consumption of storage, memory and computing powers, which makes it challenging to run neural networks on edge devices. Moreover, statistic shows that the gaps between computational complexity and energy efficiency of deep neural networks and the hardware capacity are growing \cite{xu2018scaling}. It has been proved that neural networks are typically over-parameterised, which makes deep learning models redundant \cite{denil2013predicting}. To implement neural networks on powerless edge devices, large amounts of effort try to compress the models. Model compression aims to lighten the model, improve energy efficiency, and speed up the inference on resource-constraint edge devices, without lowering the accuracy. According to their approaches, we classify these works into five categories: low-rank approximation/matrix factorisation, knowledge distillation, compact layer, parameter quantising, and network pruning. Table~\ref{compression-comparison} summarises literature on model compression.

\begin{table*}[htp]
    \footnotesize
    \centering
    \caption{Literature summary of model compression.}
    \label{compression-comparison}
    \begin{tabular}{c c c c c c c}
    \toprule
    \bf{Ref.} & \bf{Model}  & \bf{Approach}  & \bf{Object} & \bf{Performance} &\bf{Type}\\
    \hline
    \cite{bucilua2006model} &NN& Knowledge distillation  & Less resource requirement &  Faster  & Lossless\\
     \hline
    \cite{hinton2015distilling}&NN& Knowledge distillation  & Compress model &  80\% improvement& Lossless\\
     \hline
    \cite{romero2014fitnets}&NN& Knowledge distillation  & Generate thinner model &  More accurate and smaller& Improved\\
     \hline
    \cite{zagoruyko2016paying}&CNN& \makecell{Knowledge distillation\\ Attention}  & \makecell{Improve performance \\with shallow model}&  1.1\% top-1 better& Improved\\
     \hline
     \cite{sau2016deep}&CNN& \makecell{Knowledge distillation\\ Regularisation}  & Reduce storage& $33.28\times$ smaller& Improved\\
     \hline
    \cite{crowley2018moonshine}&CNN& Knowledge distillation  & Less memory& 40\%smaller& Lossless\\
     \hline
    \cite{li2018deeprebirth}&GooLeNet& Knowledge distillation  & Less memory, acceleration& $3\times$ faster, $2.5\times$ less memory& 0.4\% drop\\
     \hline
     \cite{zhou2018rocket}&CNN& Knowledge distillation  &Improve training efficiency & $6.4\times$ smaller, $3\times$ faster& Lossy\\
     \hline
    \cite{lopes2017data}&CNN& Knowledge distillation  & Reconstruct training set& 50\%smaller& Lossy\\
     \hline
    \cite{jaderberg2014speeding}& CNN & Low-rank approximation  & Reduce runtime &  $4.5\times$ faster & Lossy\\
     \hline
    \cite{denton2014exploiting}& CNN & Low-rank approximation  & Reduce computation &  $2\times$ faster &Lossy\\
     \hline
    \cite{maji2017adapt}& CNN & Low-rank approximation  & Reduce computation &  $9\times$ speedup &Lossless\\
     \hline
    \cite{kim2015compression}& CNN & Low-rank approximation & Reduce energy consumption &  $4.26\times$ energy reduction &Lossy\\
     \hline
    \cite{wang2016accelerating}& CNN & \makecell{Low-rank approximation\\ Group sparsity}  & Reduce computation &  $5.91\times$ faster &Improved\\
    \hline
    \cite{bhattacharya2016sparsification}& \makecell{DNN\\ CNN} & \makecell{Low-rank approximation\\ Kernel separation}  & Use less resource & \makecell{$11.3\times$ less memory\\ $13.3\times$ faster} & Lossless \\
    \hline
     \cite{szegedy2015going} & CNN & Compact layer design  & Use less resources &  $3-10\times$ faster & \\
     \hline
     \cite{he2016deep} & ResNet & Compact layer design  & Training acceleration& 28\% relative improvement & Improved\\
     \hline
     \cite{alex2019yolo}& YOLO & Compact layer design  & Reduce model complexity & $15.1\times$ smaller, 34\% faster &Improved\\
     \hline
     \cite{iandola2016squeezenet} & CNN & Compact layer design  & Reduce parameters & $50\times$ fewer parameter &\\
     \hline
     \cite{szegedy2017inception} & CNN & Compact layer design  & Accelerates training &3.08\% top-5 error &\\
     \hline
    \cite{shafiee2017squishednets} & CNN & \makecell{Compact layer design \\ Task decomposition}  & \makecell{Utilise storage to trade \\for computing resources} &$5.17\times$ smaller & Improved\\
     \hline
     \cite{yang2019cdeeparch}& CNN & Compact layer design  & Simplify SqueezeNet &0.89MB total parameter & Lossy\\
     \hline
    \cite{zhang2018dynamically} & RNN & Compact layer design  & Improve compression rate &$7.9\times$ smaller & Lossy\\
     \hline
    \cite{shen2018cs} & CNN & Compressive sensing  & Training efficiency &6x faster & Improved\\
     \hline
    \cite{guo2017pruning}& NIN & Network pruning  & On-device customisation &$1.24\times$ faster & 3\% Lossy\\
    \hline
    \cite{han2015learning} & VGG-16 & Network pruning  & Reduce storage &$13\times$ fewer & Lossless\\
    \hline
    \cite{gordon2018morphnet}& DNN & Network pruning  & Higher energy efficiency &$20\times$ faster & Improved\\
    \hline
    \cite{manessi2018automated} & CNN & Network pruning  & Reduce iterations &33\% fewer & Lossy\\
    \hline
    \cite{molchanov2016pruning}& CNN & Network pruning  & Speed up inference &$10\times$ faster  & Lossy\\
    \hline
    \cite{you2019gate}& CNN & Global filter pruning  & Accelerate CNN &70\% FLOPs reduction  & Lossless\\
    \hline
    \cite{yang2017designing}& CNN & Network pruning  & Energy-efficiency &$3.7\times$ reduction on energy  & Lossy\\
    \hline

    \cite{yao2017deepiot} & RNN & Network pruning  & Reduce model size &\makecell{98.9\% smaller, 94.5\% faster\\ 95.7\% energy saved}  & Lossless\\
    \hline

    \cite{liu2017learning} & CNN & Network pruning  & Reduce memory footprint & $5\times$ less computation & Lossless\\
    \hline
    \cite{chen2016deep}& CNN & \makecell{Network pruning\\ Data reuse}  & Maximise data reusability &$1.43\times$ faster, 34\% smaller & Lossless\\
    \hline
    \cite{LUO2020107461}& CNN & Channel pruning & Speed up CNN inference &2\% higher top-1 accuracy & Improved\\
    \hline
    \cite{9097925}& CNN & Progressive Channel Pruning & Effective pruning framework &Up to 44.5\% FLOPs & Lossy\\
    \hline
    \cite{Oyedotun_2020_WACV}& DNN & Debiased elastic group LASSO & Structured Compression of DNN &Several folder smaller & Lossless\\
    \hline
    \cite{Singh_2020_WACV}& CNN & Filter correlations & Minimal information loss &96.4\% FLOPs pruning, 0.95\% error & Lossless\\
    \hline
    \cite{gong2014compressing} & CNN & Vector quantisation & Compress required storage &$16-24\times$ smaller & Lossy\\
     \hline
    \cite{chen2015compressing} & NN & Hash function  & Reduce model size &$8\times$ fewer & Lossy \\

     \hline
    \cite{han2015deep} & VGG-16 & \makecell{Parameter quantisation\\ Network pruning\\ Huffman coding} & Compress model &$49\times$ smaller & Lossless \\
     \hline
    \cite{wu2016quantized}& CNN & Parameter quantisation  & Compress model &$20\times$ smaller, $6\times$ faster & Lossy \\
     \hline
    \cite{courbariaux2015} & DNN & BinaryConnect  & Compress model  &State-of-the-art & Improved \\
     \hline

    \cite{courbariaux2016binarized} & DNN & Network Binarisation  & Speed up training &State-of-the-art & Improved \\
     \hline
    \cite{rastegari2016xnor} & DNN & Network Binarisation &Reduce model size &$32\times$ smaller, $58\times$ faster & Lossy \\
     \hline
    \cite{lin2015neural} & DNN & \makecell{Parameter quantisation\\ Binary Connect } & \makecell{Compress model\\ speed up training} &Better than standard SGD & Improved \\
     \hline
    \cite{denton2014exploiting} & DNN & \makecell{Parameter quantisation\\ Binary Connect}  & \makecell{Compress model\\ speed up training} &$2-3\times$ faster, $5-10\times$ smaller & Lossy \\
    \hline
    \cite{vanhoucke2011improving} & HMM & Parameter quantisation &Speed up training &$10\times$ speedup at most & Lossless \\
     \hline
     \cite{alvarez2016efficient} & LSTM & Quantisation aware training  &Recover accuracy loss  &4\% loss recovered & 8.1 \% Lossy \\
     \hline
     \cite{nasution2017faster} & \makecell{Faster \\R-CNN} & Parameter quantisation  &Reduce model size &$4.16\times$ smaller & Improved \\
    \hline
    \cite{peng2017running}& CNN & Parameter quantisation  &Save energy &4.45fps, 6.48 watts & Lossy \\
     \hline
    \cite{anwar2015fixed} & CNN & Parameter quantisation  &Reduce computation &1/10 memory shrinks & Improved \\
     \hline
    \cite{langroudi2018deep}& CNN & Posit number system  &Reduce model size &36.4\% memory shrinks & Lossy \\
     \hline
    \cite{soudry2014expectation} & MNN & Network Binarisation &Improve energy efficiency &State-of-the-art & Improved \\
     \hline
    \cite{esser2015backpropagation} & NN & Network Binarisation &Improve energy efficiency &State-of-the-art & Improved \\
    \hline
    \cite{9018278} & CNN & Non-parametric Bayesian &Improve quantisation efficiency &Better than RL methods & Lossy \\
    \bottomrule
    \end{tabular}
    \end{table*}

\subsubsection{Low-rank Approximation}

\begin{figure*}[tp!]
\begin{center}
\includegraphics[width=0.8\linewidth]{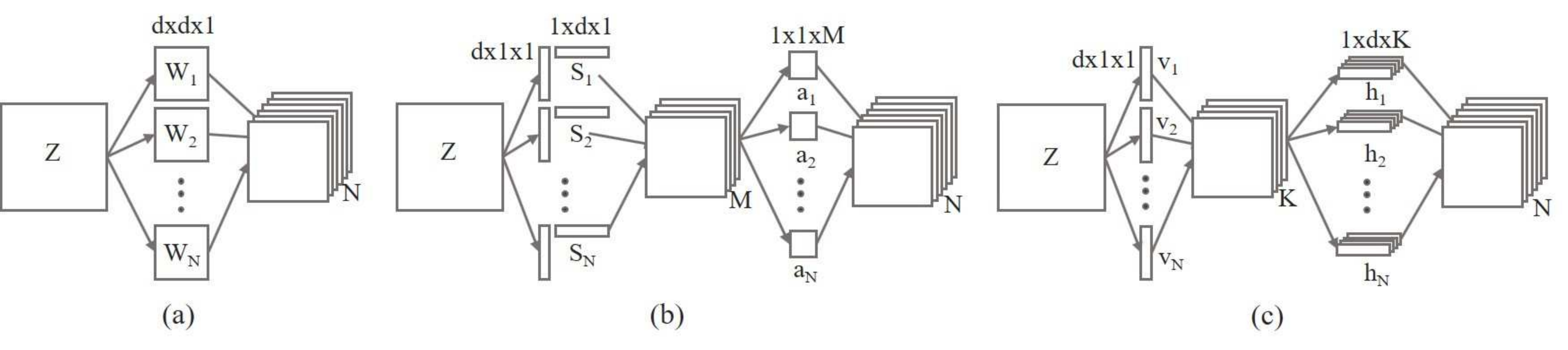}
\end{center}
\caption{The decomposition and approximation of a CNN. (a) The original operation of a convolutional layer acting on a single-channel input. (b) The approximation of the first scheme. (c) The approximation of the second scheme.}
\label{speeding}
\end{figure*}

The main idea of low-rank approximation is to use the multiplication of low-rank convolutional kernals to replace kernals of high dimension. This is based on the fact that a matrix could be decomposed into the multiplication of multiple matrices of smaller size. For example, there is a weight matrix \emph{W} of $m \times k$ dimension. The matrix \emph{W} could be decomposed into two matrices, i.e., \emph{X} $(m \times d)$ and \emph{Y} $(d \times k)$, and $W=UV$. The computational complexity of matrix \emph{W} is $O(m \times k)$, while the complexity for the decomposed two matrices is $O(m \times d ~+~ d \times k)$. Obviously, the approach could effectively reduce the model size and computation, as long as $d$ is small enough.

Jaderberg \emph{et al.} decompose the matrix of convolution layer $d \times d$ into the multiplication of two matrices $d \times 1$ and $1 \times d$ compress the CNNs \cite{jaderberg2014speeding}. The authors also propose two schemes to approximate the original filter. Fig.~\ref{speeding} presents the compression process. Fig.~\ref{speeding}(a) shows a convolutional layer acting on a single-channel input. The convolutional layer consists of $N$ filters. For the first scheme, they use the linear combination of $M$ ($M<N$) filters to approximate the operation of $N$ filters. For the second scheme, they factorise each convolutional layer into a sequence of two regular convolutional layers but with rectangular filters. The approach achieves a 4.5x acceleration with 1\% drop in accuracy. This work is a rank-1 approximation. Maji \emph{et al.} apply this rank-1 approximation on compressing CNN models on IoT devices, which achieves 9x acceleration of the inference \cite{maji2017adapt}. Denton \emph{et al.} explore the approximation of rank-$k$ \cite{denton2014exploiting}. They use monochromatic and biclustering to approximate the original convolutional layer.

Kim \emph{et al.} propose a whole network compression scheme with the consideration of entire convolutional and fully connected layers \cite{kim2015compression}. The scheme consists of three steps: rank selection, low-rank tensor decomposition, and fine-tuning. In particular, they first determine the rank of each layer through a global analytic solution of variational Bayesian matrix factorisation (VBMF). Then they apply Tucker decomposition to decompose the convolutional layer matrix into three components of dimension $1 \times 1$, $D \times D$ ($D$ is usually 3 or 5), and $1 \times 1$, which differs from SVD in \cite{denton2014exploiting}. The approach achieves a $4.26\times$ reduction in energy consumption. We note that the component of spatial size $w \times h$ still requires a large amount of computation. Wang \emph{et al.} propose a Block Term Decomposition (BTD) to further reduce the computation in operating the network, which is based on low-rank approximation and group sparsity \cite{wang2016accelerating}. They decompose the original weight matrix into the sum of few low multilinear rank weight matrices, which could approximately replace the original weight matrix. After fine-tuning, the compressed network achieves $5.91\times$ acceleration on mobile devices with the network, and a less than 1\% increase on the top-5 error.

Through optimising the parameter space of fully connected layers, weight factorisation could significantly reduce the memory requirement of DNN models and speed up the inference. However, the effect of the approach for CNN maybe not good, because there is a large amount of convolutional operations in CNN \cite{lane2015early}. To solve the problem, Bhattacharya \emph{et al.} propose a convolution kernel separation method, which optimises the convolution filters to significantly reduce convolution operations \cite{bhattacharya2016sparsification}. The authors verify the effectiveness of the proposed approach on various mobile platforms with popular models, e.g., audio classification and image recognition.

\subsubsection{Knowledge Distillation}

Knowledge distillation is based on transfer learning, which trains a neural network of smaller size with the distilled knowledge from a larger model. The large and complex model is called teacher model, whilst the compact model is referred as student model, which takes the benefit of transferring knowledge from the teacher network.

\begin{figure}[tp!]
\begin{center}
\includegraphics[width=0.8\linewidth]{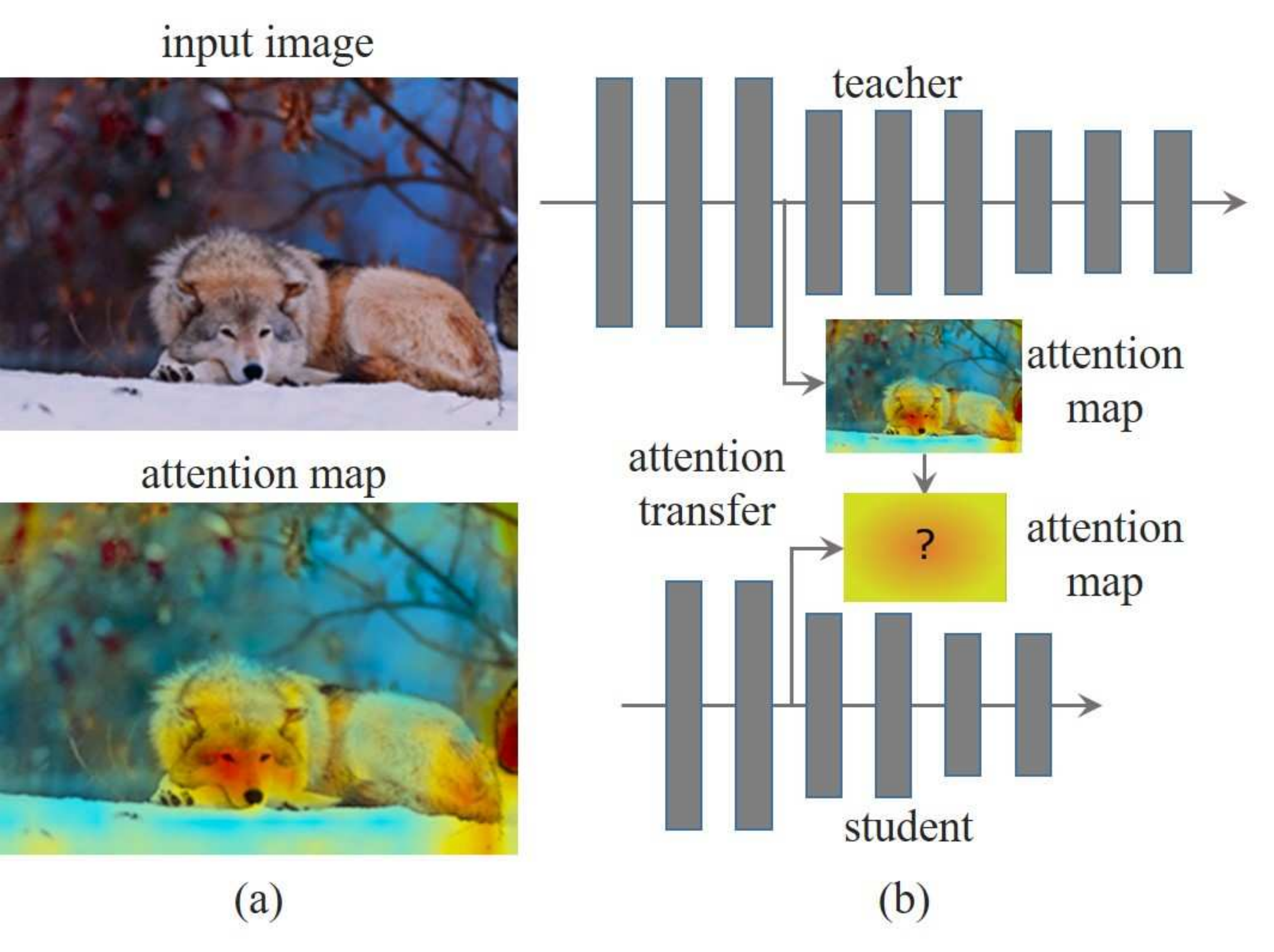}
\end{center}
\caption{The application of attention mechanism in teacher-student paradigm transfer learning. (a) The left image is an input and the right image is the corresponding spatial attention map of a CNN model which shows which feature affects the classification decision. (b) Schematic representation of attention transfer. The attention map of the teacher model is used to supervise the training of the student model.}
\label{attention}
\end{figure}

Bucilua \emph{et al.} take the first step towards compressing models with knowledge distillation \cite{buc2006model}. They first use a function learned by a high performing model to label pseudo data. Afterwards, the labelled pseudo data is utilised to train a compact but expressive model. The output of the compact model is compatible with the original high performing model. This work is limited to shallow models. The concept of knowledge distillation is first proposed in \cite{hinton2015distilling}. Hinton \emph{et al.} first train a large and complex neural model, which is an ensemble of multiple models. This complex model is the teacher model. Then they design a small and simple student model to learn its knowledge. Specifically, they collect a transfer dataset as the input of the teacher model. The data could be unlabelled data or the original training set of the teacher model. The temperature in softmax is raised to a high value in the teacher model, e.g., 20. Since the soft target of the teacher model is the mean result of multiple components of the teacher model, the training instances are more informative. Therefore, the student model could be trained on much less data than the teacher model. The authors prove the effectiveness on MNIST and speech recognition tasks. Sau \emph{et al.} propose to supervise the training of the student model with multiple teacher models, with the consideration that the distilled knowledge from a single teacher may be limited \cite{sau2016deep}. They also introduce a noise-based regulariser to improve the health in the performance of the student model.


Romero \emph{et al.} propose FitNet, which extends \cite{hinton2015distilling} to create a deeper and lighter student model \cite{romero2014fitnets}. Deeper models could better characterise the essence of the data. Both the output of the teacher model and the intermediate representations are used as hints to speed up training of the student model, as well as improve its performance. Opposite to \cite{romero2014fitnets}, Zagoruyko \emph{et al.} prove that shallow neural networks could also significantly improve the performance of a student model by properly defining attention \cite{zagoruyko2016paying}. Attention is considered as a set of spatial maps that the network focuses the most on in the input to decide the output decision. These maps could be represented as convolutional layers in the network. In the teacher-student paradigm, the spatial attention maps are used to supervise the student model, as shown in Fig.~\ref{attention}.

\begin{figure}[tp!]
\begin{center}
\includegraphics[width=0.9\linewidth]{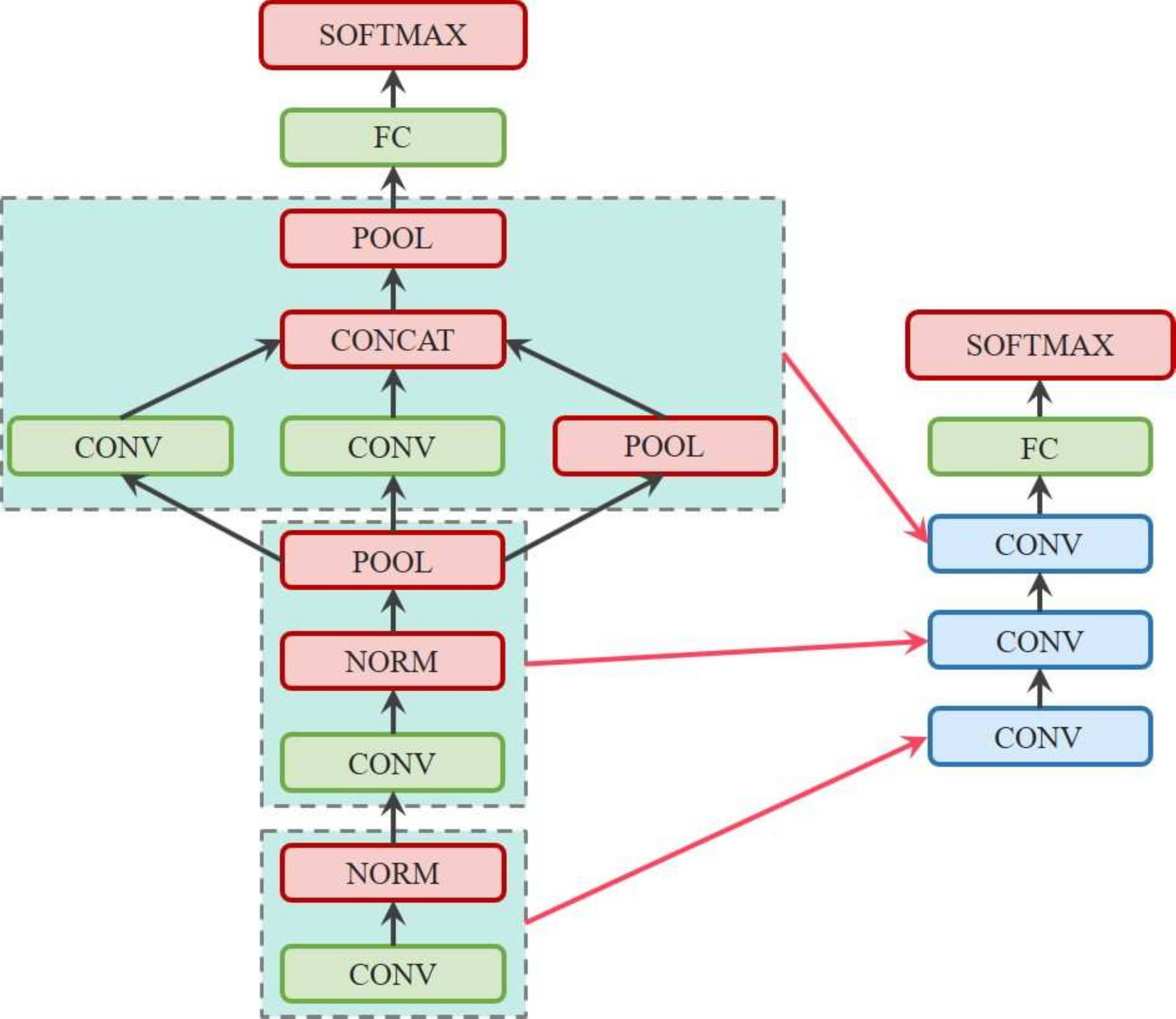}
\end{center}
\caption{The illustration of DeepRebirth. The upper model is the teacher model, while the lower is the student model. The highly correlated convolutional layer and non-convolutional layer are merged and become the new convolutional layer of the student model.}
\label{deeprebirth}
\end{figure}

There are also some efforts focusing on how to design the student model. Crowley \emph{et al.} propose to obtain the student model through replacing the convolutional layers of the teacher model with cheaper alternatives \cite{crowley2018moonshine}. The new generated student model is then trained under the supervision of the teacher model. Li \emph{et al.} design a framework, named DeepRebirth to merge the consecutive layers without weights, such as pooling and normalisation and convolutional layers vertically or horizontally to compress the model \cite{li2018deeprebirth}. The newly generated student model learns parameters through layer-wise fine-tuning to minimise the accuracy loss. Fig.~\ref{deeprebirth} presents the framework of DeepRebirth. After compression, GoogLeNet achieves 3x acceleration and 2.5x reduction in runtime memory.

\begin{figure}[tp!]
\begin{center}
\includegraphics[width=0.9\linewidth]{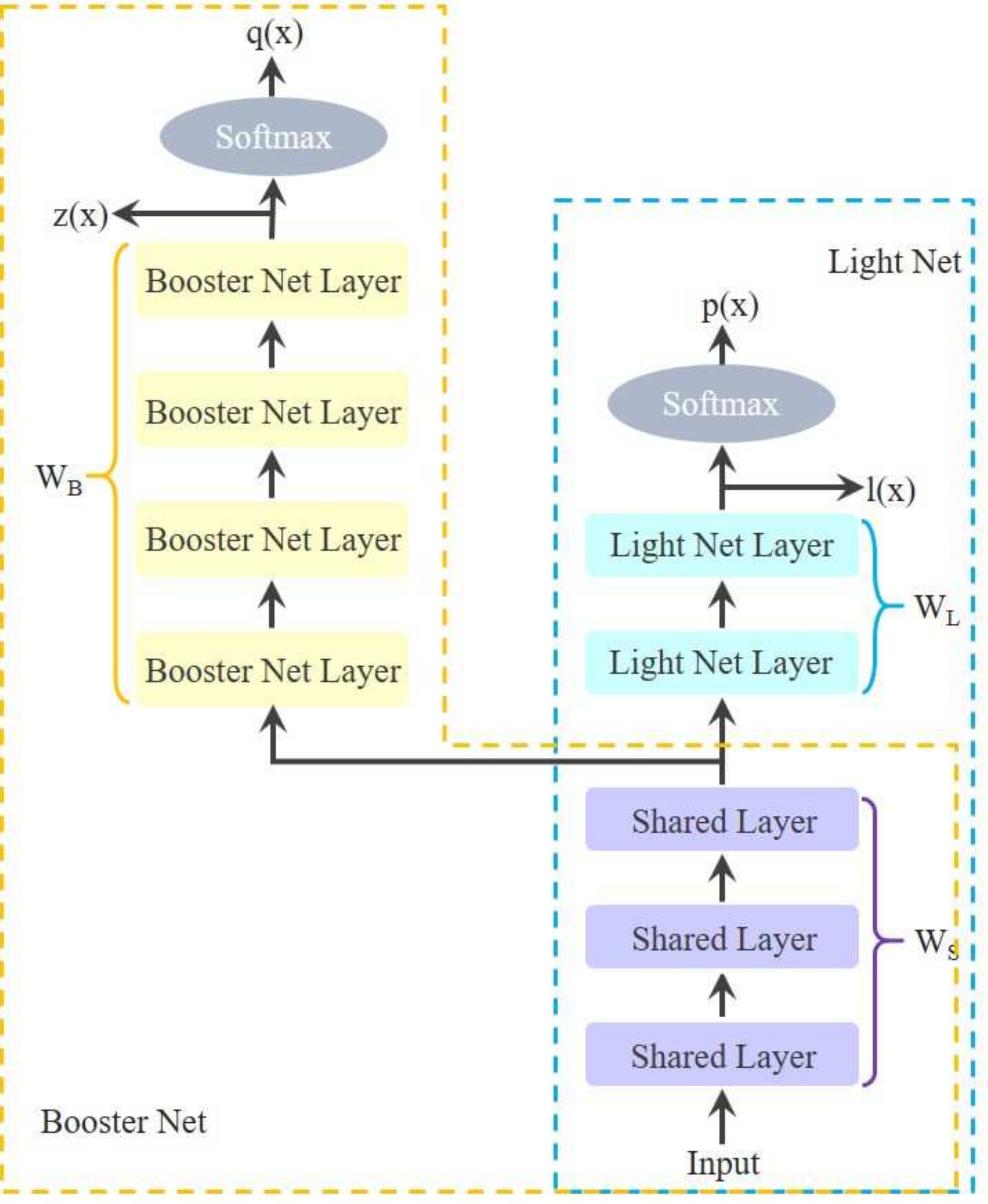}
\end{center}
\caption{The structure of Rocket Launching. $\boldsymbol{W_S}$, $\boldsymbol{W_L}$, and $\boldsymbol{W_B}$ denotes parameters. $z(x)$ and $l(x)$ represent the weighted sum before the softmax activation. $p(x)$ and $q(x)$ are outputs. Yellow layers are shared by the teacher and student.}
\label{rocket}
\end{figure}

The teacher model is pre-trained in most relevant works. Nevertheless, the teacher model and the student model could be trained in parallel to save time. Zhou \emph{et al.} propose a compression scheme, named Rocket Launching to exploit the simultaneous training of the teacher and student model \cite{zhou2018rocket}. During the training, the student model keeps acquiring knowledge learnt by the teacher model through the optimisation of the hint loss. The student model learns both the difference between its output and its target, and the possible path towards the final target learnt by the teacher model. Fig.~\ref{rocket} presents the structure of this framework.

When the teacher model is trained on a dataset concerning with privacy or safety, it is then difficult to train the student model. Lopes \emph{et al.} propose an approach to distill the learned knowledge of the teacher model without accessing the original dataset, which only needs some extra metadata \cite{lopes2017data}. They first reconstruct the original dataset with the metadata of the teacher model. This step could find the images that best match these given by the network. Then they remove the noise of the image to approximate the activation records through gradients, which could partially reconstruct the original training set of the teacher model.

\subsubsection{Compact layer design}
In deep neural networks, if weights end up to be close to 0, the computation is wasted. A fundamental way to solve this problem is to design compact layers in neural networks, which could effectively reduce the consumption of resources, i.e., memories and computation power. Christian \emph{et al.} propose to introduce sparsity and replace the fully connected layers in GoogLeNet \cite{szegedy2015going}. Residual-Net replaces the fully connected layers with global average pooling to reduce the resource requirements \cite{he2016deep}. Both GoogLeNet and Residual-Net achieve the best performance on multiple benchmarks.

\begin{figure}[tp!]
\begin{center}
\includegraphics[width=\linewidth]{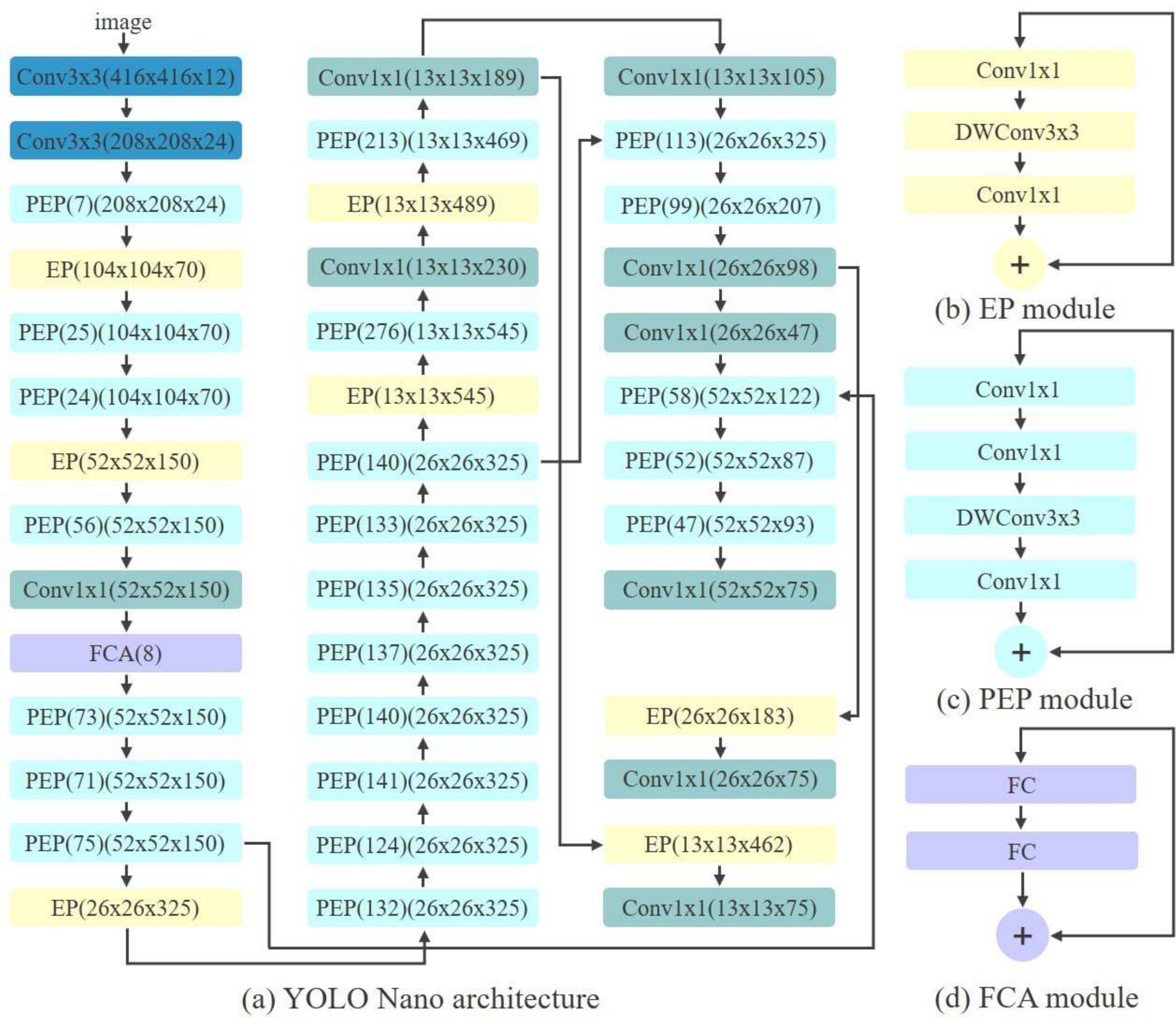}
\end{center}
\caption{The architecture of the YOLO Nano network. PEP(x) refers to x channels in PEP, while FCA(x) represents the reduction ratio of x.}
\label{YOLO}
\end{figure}

Alex \emph{et al.} propose a compact and lightweight CNN model, named YOLO Nano for image recognition \cite{alex2019yolo}. YOLO Nano is a highly customised model with module-level macro- and micro-architecture. Fig.~\ref{YOLO} shows the network architecture of YOLO Nano. There are three modules in YOLO Nano: expansion-projection (EP) macro-architecture, residual projection-expansion-projection (PEP) macro-architecture, and a fully-connected attention (FCA) module. PEP could reduce the architectural and computational complexity whilst preserving model expressiveness. FCA enables better utilisation of available network capacity.

Replacing a big convolution with multiple compact layers could effectively reduce the number of parameters and further reduce computations. Iandola \emph{et al.} propose to compress CNN models with three strategies \cite{iandola2016squeezenet}. First, decomposing $3 \times 3$ convolution into $1 \times 1$ convolutions, since it has much fewer parameters. Second, cut down input channels in $3 \times 3$ convolutions. Third, downsample late to produce big feature maps. The larger feature maps could lead to higher classification accuracy. The first two strategies are used to decrease the quantity of parameters in CNN models and the last one is used to maximise the accuracy of the model. Based on three above mentioned strategies, the authors design SqueezeNet, which can achieve $50\times$ reduction in the number of parameters, whilst remaining the same accuracy as the complete AlexNet. Similar approaches ares also used in \cite{szegedy2017inception}. Shafiee \emph{et al.} modify SqueezeNet for applications with fewer target classes and they propose SqueezeNet v1.1, which could be deployed on edge devices \cite{shafiee2017squishednets}. Yang \emph{et al.} propose to decompose a recognition task into two simple sub-tasks: context recognition and target recognition, and further design a compact model, namely cDeepArch \cite{yang2019cdeeparch}. This approach uses storage resource to trade for computing resources.

Shen \emph{et al.} introduce Compressive Sensing (CS) to jointly modify the input layer and reduce nodes of each layer for CNN models \cite{shen2018cs}. CS \cite{baraniuk2007compressive} could be used to reduce the dimensionality of the original signal while preserving most of its information. The authors use CS to jointly reduce the dimensions of the input layer whilst extracting most features. The compressed input layer also enables the reduction of the number of parameters.

Besides the above-mentioned works about CNNs, Zhang \emph{et al.} propose a dynamically hierarchy revolution (DirNet) to compress RNNs \cite{zhang2018dynamically}. In particular, they mine dictionary atoms from original networks to adjust the compression rate with the consideration of different redundancy degrees amongst layers. They then adaptively change the sparsity across the hierarchical layers.

\subsubsection{Network pruning}
\begin{figure}[tp!]
\begin{center}
\includegraphics[width=\linewidth]{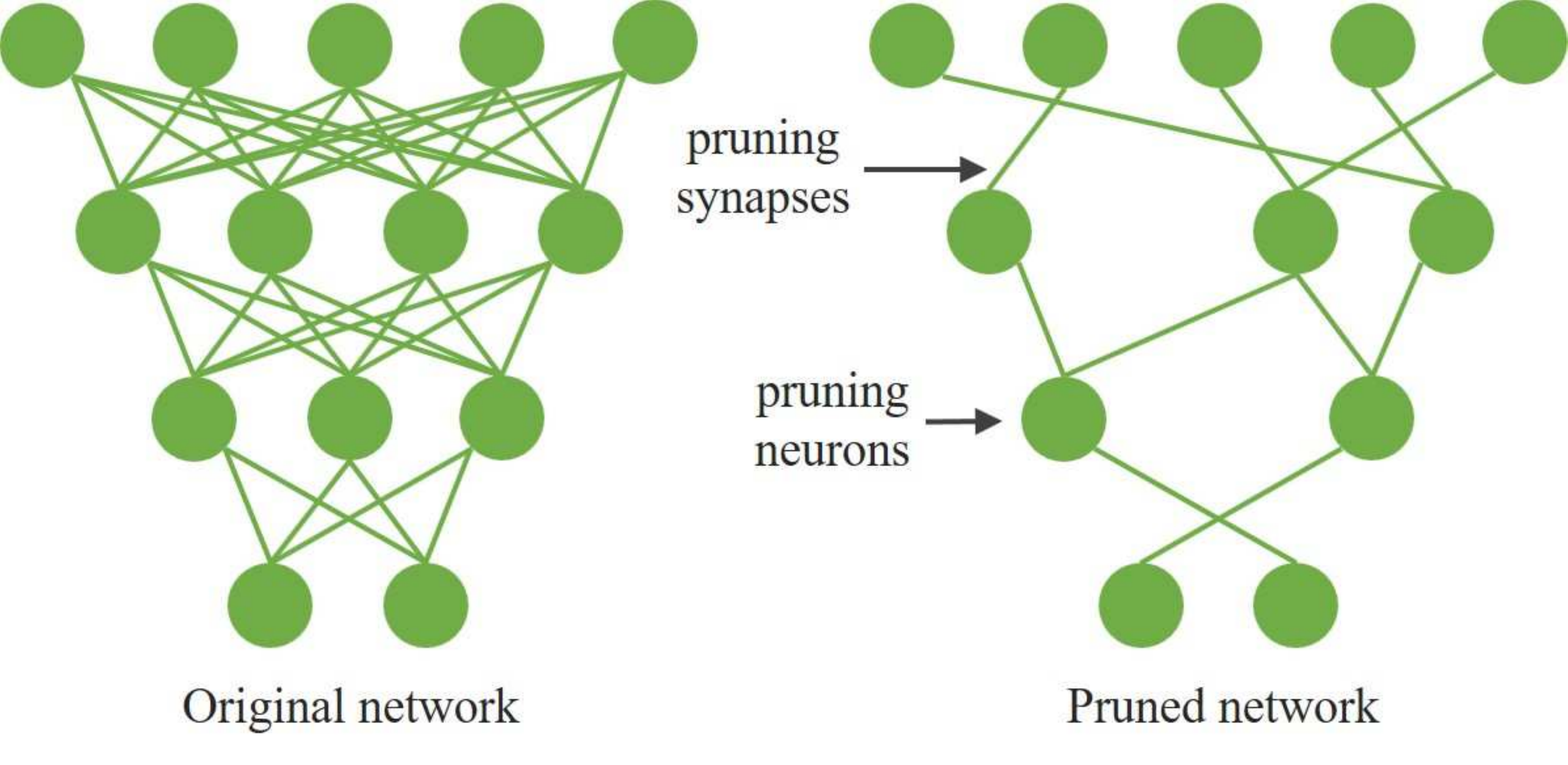}
\end{center}
\caption{Illustration of network pruning. Unimportant synapses and neurons would be deleted to generate a sparse network.}
\label{prune}
\end{figure}

The main idea of network pruning is to delete unimportant parameters, since not all parameters are important in highly precise deep neural networks. Consequently, connections with less weights are removed, which converts a dense network into a sparse one, as shown in Fig.~\ref{prune} There are some works which attempt to compress neural networks by network pruning.

The work \cite{lecun1990optimal} and \cite{hassibi1993second} have taken the earliest steps towards network pruning. They prune neural networks to eliminate unimportant connections by using Hessian loss function. Experiment results prove the efficiency of prunning methods. Subsequent research focuses on how to prune the networks. Han \emph{et al.} propose to prune networks based on a weight threshold \cite{han2015learning}. Practically, they first train a model to learn the weights of each connection. The connections with lower weights than the threshold would then be removed. Afterwards, the network is retrained. The pruning approach is straightforward and simple. A similar approach is also used in \cite{gordon2018morphnet}. In \cite{gordon2018morphnet}, the authors select and delete neurons of low performance, and then use a width multiplier to expand all layer sizes, which could allocate more resources to neurons of high performance. However, the assumption that connections with lower weights contribute less to the results may destroy the structure of the networks.

Identifying an appropriate threshold to prune neural networks usually takes iteratively trained networks, which consumes a lot of resources and time. Moreover, the threshold is shared by all the layers. Consequently, the pruned configuration maybe not the optimal, comparing with the case of identify thresholds for each layer. To break through these limitations, Manessi \emph{et al.} propose a differentiability-based pruning method to jointly optimise the weights and thresholds for each layer \cite{manessi2018automated}. Specifically, the authors propose a set of differentiable pruning functions and a new regulariser. Pruning could be performed during the back propagation phase, which could effectively reduce the training time.

\begin{figure*}[tp!]
\begin{center}
\includegraphics[width=\linewidth]{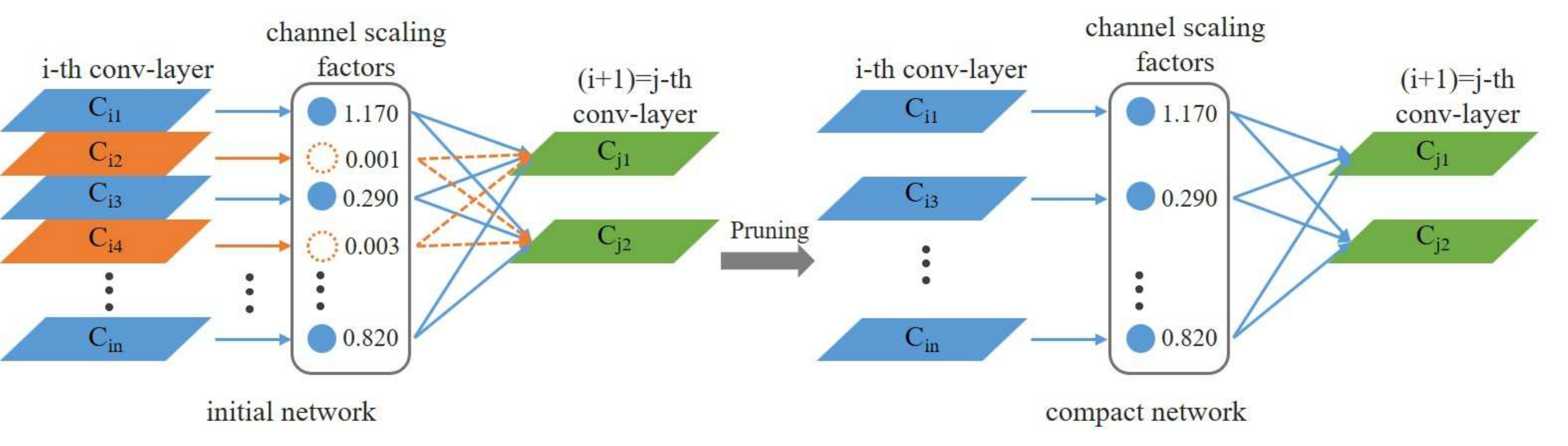}
\end{center}
\caption{Each channel is associated with a scaling factor $\gamma$ in convolutional layers. Then the network is trained to jointly learn weights and scaling factors. After that, the channels with small scaling factors (in orange colour) are pruned, which results in a compact model.}
\label{slim}
\end{figure*}

Molchanov \emph{et al.} propose a new criterion based on the Taylor expansion to identify unimportant neutrons in convolutional layers \cite{molchanov2016pruning}. Specifically, they use the change of cost function to evaluate the result of pruning. They formulate pruning as an optimisation problem, trying to find a weight matrix that minimises the change in cost function. The formulation is approximately converted to its first-degree Taylor polynomial. The gradient and feature map's activation could be easily computed during back-propagation. Therefore, the approach could train the network and prune parameters simultaneously. You \emph{et al.} propose a global filter pruning algorithm, named Gate Decorator, which transforms a CNN module through multiplying its output by the channel-wise scaling factors \cite{you2019gate}. If the scaling factor is set to be 0, the corresponding filter would be removed. They also adopt the Taylor expansion to estimate the change of the loss function caused by the changing of the scaling factor. They rank all global filters based on the estimation and prune according to the rank. Compared with \cite{molchanov2016pruning}, \cite{you2019gate} does not require special operations or structures.

In addition to minimum weight and cost functions, there are efforts trying to prune with the metric of energy consumption. Yang \emph{et al.} propose an energy-aware pruning algorithm to prune CNNs with the goal of minimising the energy consumption \cite{yang2017designing}. The authors model the relationship between data sparsity and bit-width reduction through extrapolating the detailed value of consumed energy from hardware measurements. The pruning algorithm identifies the parts of a CNN that consumes the most energy and prunes the weights to maximise energy reduction.

Yao \emph{et al.} propose to minimise the number of non-redundant hidden elements in each layer whilst retaining the accuracy in sensing applications and propose DeepIoT \cite{yao2017deepiot}. In DeepIoT, the authors compress neural networks through removing hidden elements. This regularisation approach is called dropout. Each hidden element is dropouted with a probability. The dropout probability is initialised with 0.5 for all hidden elements. DeepIoT develops a compressor neural network to learn the optimal dropout probabilities of all elements.

Liu \emph{et al.} propose to identify important channels in CNN and remove unimportant channels to compress networks \cite{liu2017learning}. Specifically, they introduce a scaling factor $\gamma$ for each channel. The output $\hat{z}$ (also the input of the next layer) could be formulated as $\hat{z}=\gamma z+\beta$, where $z$ is the input of the current layer and $\beta$ is min-batch. Afterwards, they jointly train the network weight and scaling factors, with L1 regulation imposed on the latter. Following that, they prune the channels with the small scaling factor $\gamma$. Finally, the model is fine-tuned, which achieves a comparable performance with the full network. Fig.~\ref{slim} presents this slimming process. However, the threshold of the scaling factor is not computed, which requires iterative evaluations to obtain a proper one.

Based on network pruning, the work in \cite{chen2016deep} investigates the data flow inside computing blocks and develops a data reuse scheme to alleviate the bandwidth burden in convolution layers. The data flow of a convolution layer is regular. If the common data could be reused, it is not necessary to load all data to a new computing block. The data reuse is used to parallelise computing threads and accelerate the inference of a CNN model.

\subsubsection{Parameter quantisation}

A very deep neural network usually involves many layers with millions of parameters, which consumes a large amount of storage and slows down the training procedure. However, highly precise parameters in neural networks are not always necessary in achieving high performance, especially when these highly precise parameters are redundant. It has been proved that only a small number of parameters are enough to reconstruct a complete network \cite{denil2013predicting}. In \cite{denil2013predicting}, the authors find that the parameters within one layer could be predicted by 5\% of parameters, which means we could compress the model by eliminating redundant parameters. There are some works exploiting parameter quantisation for model compression.

\begin{figure*}[tp!]
\begin{center}
\includegraphics[width=0.8\linewidth]{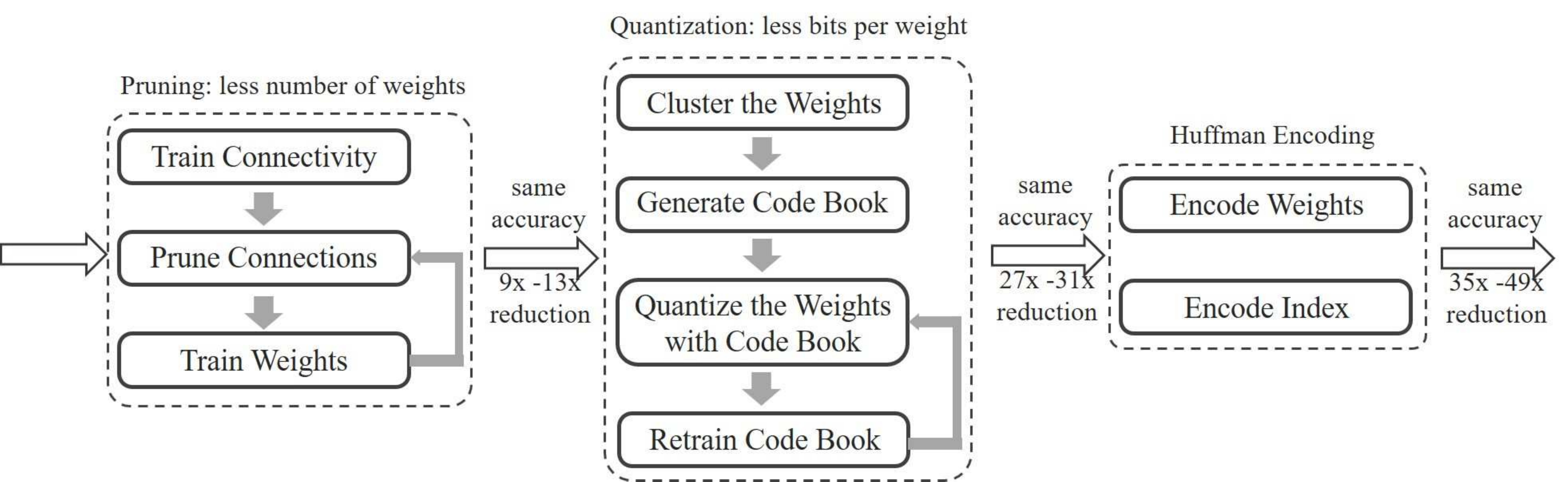}
\end{center}
\caption{Illustration of three-stage compression pipeline. First use pruning to reduce the number of weights by $10\times$, then use quantisation to further compress by $27\times$ and $31\times$. Finally use Huffman coding to get more compression.}
\label{compression}
\end{figure*}

Gong \emph{et al.} propose to use vector quantisation methods to reduce  parameters in CNN \cite{gong2014compressing}. Vector quantisation is often used in lossy data compression, which is based on block coding \cite{cortes2017adanet}. The main idea of vector quantisation is to divide a set of points into groups, which are represented by their central points. Hence, these points could be denoted with fewer coding bits, which is the basis of compression. In \cite{gong2014compressing}, the authors use k-means to cluster parameters and quantise these clusters. They find that this method could achieve $16-24\times$ compression rate of the parameters with the scarification of no more than 1\% of the top-5 accuracy. In addition to k-means, hash method has been utilised in parameter quantisation. In \cite{chen2015compressing}, Chen \emph{et al.} propose to use hash functions to cluster connections into different hash buckets uniformly. Connections in the same hash bucket share the same weight. Han \emph{et al.} combine parameter quantisation and pruning to further compress the neural network without compromising the accuracy \cite{han2015deep}. Specifically, they first prune the neural network through recognising the important connections through all connections. Unimportant connections are ignored to minimise computation. Then, they quantise the parameters, to save the storage of parameters. After these two steps, the model will be retrained. These remaining connections and parameters could be properly adjusted. Finally, they use Huffman coding to further compress the model. Huffman coding is a prefix coding, which effectively reduces the required storage of data \cite{van1976construction}. Fig.~\ref{compression} presents the three-step compression.

For most CNNs, the fully connected layers consume most storage in neural network. Compressing parameters of fully connected layers could effectively reduce the model size. The convolutional layers consume most of the times during training and inference. Wu \emph{et al.} design Q-CNN to quantise both fully connected layers and convolutional layers to jointly compress and accelerate the neural network \cite{wu2016quantized}. Similar to \cite{gong2014compressing}, the authors utilise k-means to optimally cluster parameters in fully connected and convolutional layers. Then, they quantise parameters by minimising the estimated error of response for each layer. They also propose a training scheme to suppress the accumulative error for the quantisation of multiple layers.

Enormous amount of floating point multiplications consumes significant times and computing resources in inference. There are two potential solutions to address this challenge. The first one is to replace floating point with fixed point, and the second one is to reduce the amount of floating point multiplications.

According to the evaluation of Xilinx, fixed point could achieve the same accuracy results as float \cite{malki2003cnn}. Vanhoucke \emph{et al.} evaluate the implementation of fixed point of an 8-bit integer on x86 platform \cite{vanhoucke2011improving}. Specifically, activation and the weights of intermediate layer are quantised into an 8-bit fixed point with the exception of biases that are encoded as 32-bit. The input layer remains floating point to accommodate possible large inputs. Through the quantisation, the total required memory shrinks $3-4\times$. Results show that the quantised model could achieve a 10x speedup over an optimised baseline and a $4\times$ speedup over an aggressively optimised float point baseline without affecting the accuracy. Similarly, Nasution \emph{et al.} convert floating point to 8 and 16 bits to represent weights and outputs of layers, which lowers the storage to $4.16\times$ \cite{nasution2017faster}. Peng \emph{et al.} quantise an image classification CNN model into an 8-bit fixed-point at the cost of 1\% accuracy drop \cite{peng2017running}. Anwar \emph{et al.} propose to use L2 error minimisation to quantise parameters \cite{anwar2015fixed}. They quantise each layer one by one to induce sparsity and retrain the network with the quantised parameters. This approach is evaluated with MNIST and CIRAR-10 dataset. The results shows that the approach could reduce the required memory by 1/10.

In addition to fixed point, posit number could also be utilised to replace floating point numbers to compress neural networks. Posit number is a unique non-linear numerical system, which could represent all numbers in a dynamic range \cite{gustafson2017beating}. The posit number system represents numbers with fewer bits. Float point numbers could be converted into the posit number format to save storage. To learn more about the conversion, readers may refer to \cite{morris1971tapered}. Langroudi \emph{et al.} propose to use the posit number system to compress CNNs with non-uniform data \cite{langroudi2018deep}. The weights are converted into posit number format during the reading and writing operations in memory. During the training or inference, when computing operations are required, the number would be converted back to float point. Because this approach only converts the weight between two number systems, no quantisation occurs. The network does not require to be re-trained.

Network Binarisation is an extreme case of weight quantisation. Weight quantisation indicates that all weights are represented by two possible values (e.g., -1 or 1), which could overwhelmingly compress neural networks \cite{courbariaux2015}. For example, the original network requires 32 bits to store one parameter, while in binary connect based network, only 1 bit is enough, which significantly reduces the model size. Another advantage of binary connect is that replacing multiply-accumulate operations by simple accumulations, which could drastically reduce computation in training. Courbariaux \emph{et al.} extend the work \cite{courbariaux2015} further and proposes Binary Neural Network (BNN), which completely changes the computing style of traditional neural networks \cite{courbariaux2016binarized}. Not only the weights, but also the input of each layer is binarised. Hence, during the training, all multiplication operations are replaced by accumulation operations, which drastically improves the power-efficiency. However, substantial experiments indicate that BNN could only achieve good performance on small scale datasets.

Rastegari \emph{et al.} propose a XNOR-net to reduce storage and improve training efficiency, which is different with \cite{courbariaux2016binarized} in the binarisation method and network structure \cite{rastegari2016xnor}. In Binary-Weight network, all weight values are approximately binarized, e.g., -1 or 1, which reduces the size of network by $32\times$. Convolutions could be finished with only addition and subtraction, which is different with \cite{courbariaux2016binarized}. Hence, the training is speed up $2\times$. With XNOR-net, in addition to weights, the input to convolutional layers are approximately binarised. Moreover, they further simplify the convolution with XNOR operations, which achieves a speed up of $58\times$. The comparison amongst standard convolution, Binary-Weight and XNOR-net is presented as Table.~\ref{xnor}.

\begin{table*}[htp]
    \footnotesize
    \centering
    \caption{The comparison amongst standard convolution, Binary-Weight and XNOR-net.}
    \label{xnor}
\begin{tabular}{c |c| c| c| c| c| c}
    \toprule

     & \bf{Input}  & \bf{Weight} & \bf{\makecell*[c]{Convolution \\ operation}} &\bf{\makecell*[c]{Memory \\saving}}& \bf{\makecell*[c]{Computation \\ saving}} & \bf{\makecell*[c]{Accuracy \\ (imageNet)}} \\
     \hline
    Standard Convolution& Real value& Real value & $\times$, $+$, $-$ &1$\times$ & 1$\times$& 56.7\% \\
     \hline
     Binary-Weight& Real value & Binary value & $+$, $-$ &$\sim$32$\times$ & $\sim$2$\times$& 56.8\% \\
     \hline
     XNOR-Net& Binary value & Binary value & XNOR, bitcount &$\sim$32$\times$ & $\sim$58$\times$& 44.2\% \\
     \hline
\end{tabular}
\end{table*}


Lin \emph{et al.} propose to use binary connect to reduce multiplications in DNN \cite{lin2015neural}. In the forward pass, the authors stochastically binarise weights by binary connect. Afterwards, they quantise the representations at each layer to replace the remaining multiply operations into bit-shifts. Their results show that there is no loss in accuracy in training and sometimes this approach surprisingly achieves even better performance than standard stochastic gradient descent training.

Soudry \emph{et al.} prove that binary weights and activations could be used in Expectation Backprogagation (EBP) and achieves high performance \cite{soudry2014expectation}. This is based on a variational Bayesian approach. The authors test eight binary text classification tasks with EBP-trained multilayer neural networks (MNN). The results show that binary weights always achieve better performance than continuous weights. Esser \emph{et al.} further develop a fully binary network with the same approach to EBP to improve the energy efficiency on neuromorphic chips \cite{esser2015backpropagation}. They perform the experimentation on the MNIST dataset, and the results show that the method achieves 99.42\% accuracy at 108 $\mu J$ per image.

\subsubsection{Applications}

Some efforts try to use these compression techniques on practical applications and prototypes at the edge, including image analysis \cite{mathur2017deepeye,zeng2017mobiledeeppill,wang2018deepsearch}, compression service \cite{liu2018demand}, and automotive \cite{kim2017design,xu2017fast}.

Mathur \emph{et al.} develop a wearable camera, called DeepEye, that runs multiple cloud-scale deep learning models at edge provide real-time analysis on the captured images \cite{mathur2017deepeye}. DeepEye enables the creation of five state-of-the-art image recognition models.
After camera captures an image, the image pre-processing component deals with the image according to the adopted deep model. There is a model compression component inside the inference engine, which applies available compression techniques to reduce energy consumption and the running time. Finally, DeepEye use the optimised BLAS library to optimise the numeric operations on hardware.

\begin{figure}[tp!]
\begin{center}
\includegraphics[width=\linewidth]{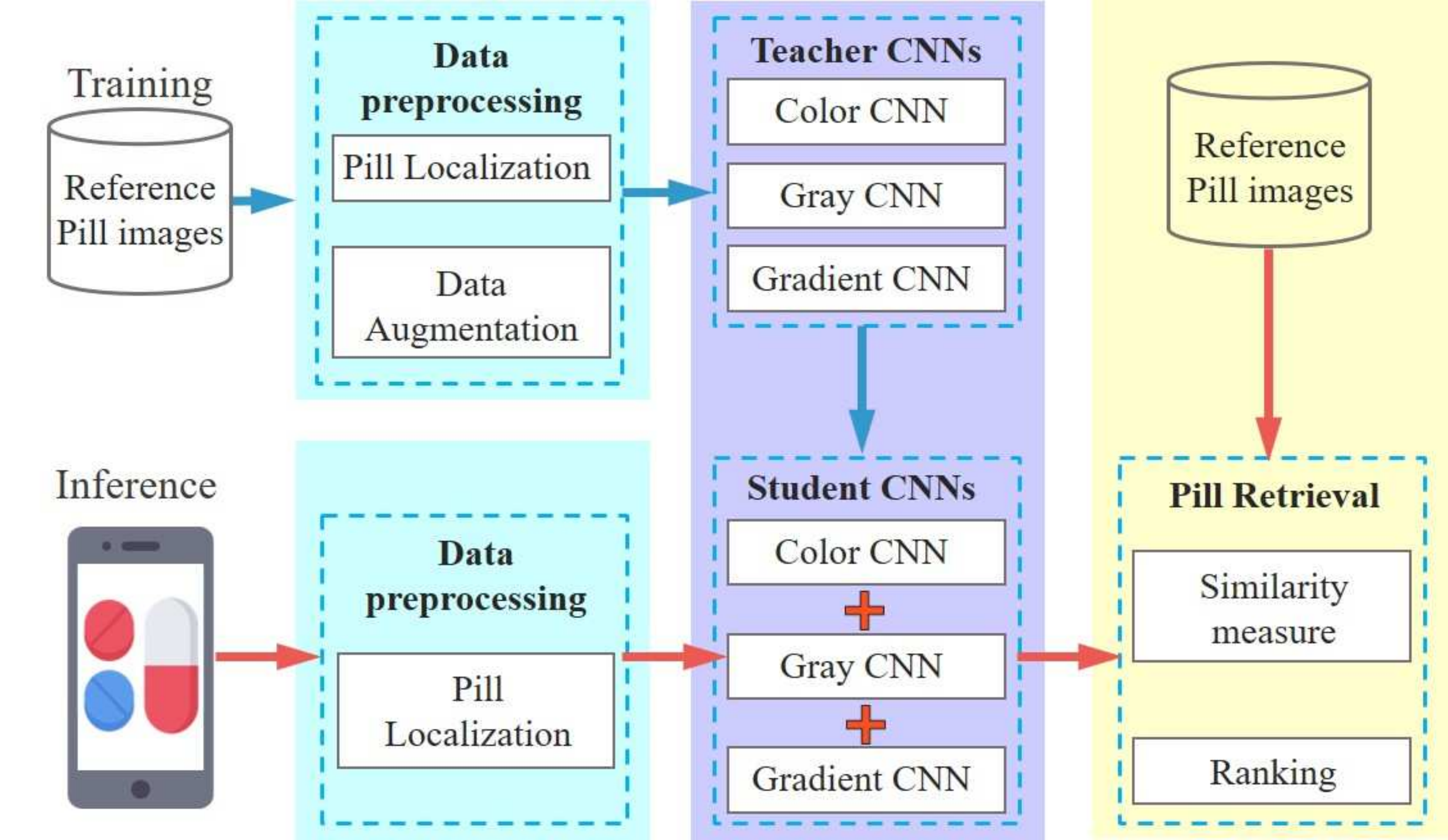}
\end{center}
\caption{The architecture of MobileDeepPill. The blue arrows indicates the flow of the training phase, whilst the red arrows indicate the inference phase.}
\label{deeppill}
\end{figure}

To correctly identify prescription pills for patients based on their visual appearance, Zeng \emph{et al.} develop MobileDeepPill, a pill image recognition system \cite{zeng2017mobiledeeppill}. The pill image recognition model is based on ImageNet \cite{krizhevsky2012imagenet}. Fig.~\ref{deeppill} presents the architecture of MobileDeepPill. In the training phase, the system first localises and splits the pill image in consumer and pill references. The system then enrich samples through running data augmentation module. Finally, the system imports CNNs as the teacher model to supervise the student model. In the inference phase, the system first processes the pill photo and extracts features to perform the student CNNs. As a last step, the system ranks the results according to their possibilities.

Wang \emph{et al.} propose a fast image search framework to implement the content-based image retrieval (CBIR) service from cloud servers to edge devices \cite{wang2018deepsearch}. Traditional CBIR services are based on the cloud, which suffers from high latency and privacy concerns. The authors propose to reduce the resource requirements of the model and to deploy it on edge devices. For the two components consuming most resources, i.e., object detection and feature extraction, the authors use low-rank approximation to compress these two parts. The compressed model achieves $6.1\times$ speedup for inference.

Liu \emph{et al.} develop an on-demand customised compression system, named AdaDeep \cite{liu2018demand}. Various kinds of compression approaches could be jointly used in AdaDeep to balance the performance and resource constraints. Specifically, the authors propose a reinforcement learning based optimiser to automatically select the combination of compression approaches to achieve appropriate trade-offs among multiple metrics such as accuracy, storage, and energy consumption.

With growing interests from the automotive industry, various large deep learning models with high accuracy have been implemented in smart vehicles with the assistance of compression techniques. Kim \emph{et al.} develop a DL based object recognition system to recognise vehicles \cite{kim2017design}. The vehicle recognition system is based on faster-RCNN. To deploy the system on vehicles, the authors apply network pruning and parameter quantisation to compress the network. Evaluations show that these two compression techniques reduce the network size to 16\% and reduce runtime to 64\%. Xu \emph{et al.} propose an RNN based driving behaviour analysis system on vehicles\cite{xu2017fast}. The system uses the raw data collected by a variety of sensors on vehicles to predict the driving patterns. To deploy the system on automobiles, the authors apply parameter quantisation to reduce the energy consumption and model size. After compression, the system size is reduced to 44 KB and the power overhead is 7.7 mW.

\subsection{Inference Acceleration}\label{modelacceleration}
The computing capacities of edge devices have been increased and some embedded devices, such as NVIDIA Jetson TX2 \cite{blanco2019deep} could directly perform CNN. However, it is still difficult for most edge devices to directly run large models. Model compression techniques reduce the required resources to create neural network models and facilitate the performance of these models on edge devices. Model acceleration techniques further speed up the performance of the compressed model on edge devices. The main idea of model acceleration in inference is to reduce the run-time of inference on edge devices and realise real-time responses for specific neural network based applications without changing the structure of the trained model. According to acceleration approaches, research works on inference acceleration could be divided into two categories: hardware acceleration and software acceleration. Hardware acceleration methods focuses on parallelising inference tasks to available hardware, such as CPU, GPU, and DSP. Software acceleration method focuses on optimising resource management, pipeline design, and compiler.

\begin{figure*}[tp!]
\begin{center}
\includegraphics[width=0.7\linewidth]{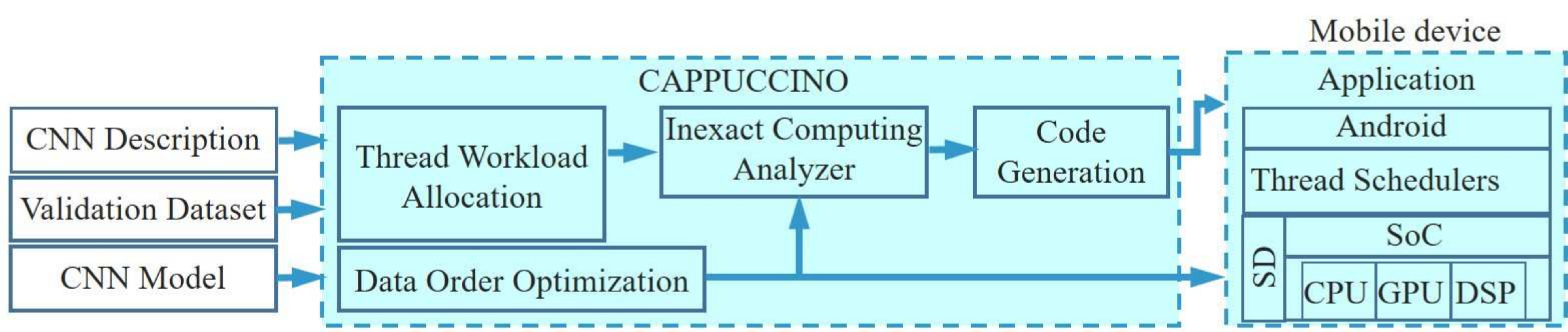}
\end{center}
\caption{The architecture of Cappuccino. Thread workload allocation component optimises the workload of each thread. Data order optimisation component converts data format. Inexact computing analyser determines the tradeoff amongst multiple metrics.}
\label{cappuccino}
\end{figure*}

\subsubsection{Hardware Acceleration}
Recently, mobile devices are becoming increasingly powerful. More and more mobile platforms are equipped with GPUs. Since mobile CPUs are not suitable for the computing of deep neural networks, the embedded GPU could be used to share the computing tasks and accelerate the inference. Table \ref{acceleration-hardware} summaries existing literature on hardware acceleration.

Alzantot \emph{et al.} evaluate the performance of CNNs and RNNs only on CPU, and compares against the execution in parallel on all available computing resources, e.g., CPU, GPU, DSP, etc. \cite{alzantot2017rstensorflow}. Results show that the parallel computing paradigm is much faster. Loukadakis \emph{et al.} propose two parallel implementations of VGG-16 network on ODROID-XU4 board: OpenMP version and OpenCL version \cite{loukadakis2018accelerating}. The former parallelises the inference within the CPU, whilst the latter one parallelises within the Mali GPU. These two approaches achieve $2.8\times$ and $11.6\times$ speedup, respectively. Oskouei \emph{et al.} design a mobile GPU-based accelerator for using deep CNN on mobile platforms, which executes inference in parallel on both CPU and GPU \cite{oskouei2015gpu}. The accelerator achieves $60\times$ speedup. The authors further develop a GPU-based accelerated library for Android devices, called CNNdroid, which could achieve up to $60\times$ speedup and $130\times$ energy reduction Android platforms \cite{latifi2016cnndroid}.

With the consideration that the memory on edge devices are usually not sufficient for neural networks, Tsung \emph{et al.} propose to optimise the flow to accelerate inference \cite{tsung2016high}. They use a matrix multiplication function to improve the cache hit rate in memory, which indirectly speeds up the execution of the model.

Nvidia has developed a parallelisation framework, named Compute Unified Device Architecture (CUDA) for desktop GPUs to reduce the complexity of neural networks and improve inference speed. For example, in \cite{appleyard2016optimizing}, CUDA significantly improves the execution efficiency of RNN on desktop GPUs. Some efforts implement the CUDA framework onto mobile platforms. Rizvi \emph{et al.} propose an approach for image classification on embedded devices based on the CUDA framework \cite{rizvi2016gpgpu}. The approach features the most common layers in CNN models, e.g., convolutions, max-pooling, batch-normalisation, and activation functions. General Purpose Computing GPU (GPGPU) is used to speed up the most computation-intensive operations in each layer. The approach is also used to implement an Italian license plate detection and recognition system on tablets \cite{rizvi2017deep}. They subsequently introduce matrix multiplication to reduce the computational complexity of convolution in a similar system to achieve real-time object classification on mobile devices \cite{rizvi2017optimized}. They also apply the approach in a robotic controller system \cite{rizvi2017general}.

However, the experiments in \cite{cao2017mobirnn} show that directly applying CUDA on mobile GPUs may be ineffective, or even deteriorates the performance. Cao \emph{et al.} propose to accelerate RNN on mobile devices based on a parallelisation framework, called RenderScript \cite{cao2017mobirnn}. RenderScript \cite{guihot2012renderscript} is a component of the Android platform, which provides an API for hardware acceleration. RenderScript could automatically parallelise the data structure across available GPU cores. The proposed framework could reduce latency by $4\times$.

Motamedi \emph{et al.} implement SqueezeNet on mobile and evaluates the performance on three different Android devices based on RenderScript \cite{motamedi2016fast}. Results show that it achieves $310.74\times$ speedup on a Nexus 5. They further develop a general framework, called Cappuccino, for automatic synthesis of efficient inference on edge devices \cite{motamedi2018cappuccino}. The structure of Cappuccino is shown as in Fig.~\ref{cappuccino}. There are three inputs for the framework: basic information of the model, model file, and dataset. There are three kinds of parallelisation: kernel-level, filter bank-level, and output-level parallelisation. The thread workload allocation component allocates tasks by using these three kinds of parallelisation. They specially investigate the optimal degree of concurrency for each task, i.e., the number of threads in \cite{motamedi2017machine}. The data order optimisation component is used to convert the data format. Cappuccino enables imprecise computing in exchange for high speed and energy efficiency. The inexact computing analyser component is used to analyse the effect of imprecise computing and determine the tradeoff amongst accuracy, speed and energy efficiency.

\begin{figure}[tp!]
\begin{center}
\includegraphics[width=0.75\linewidth]{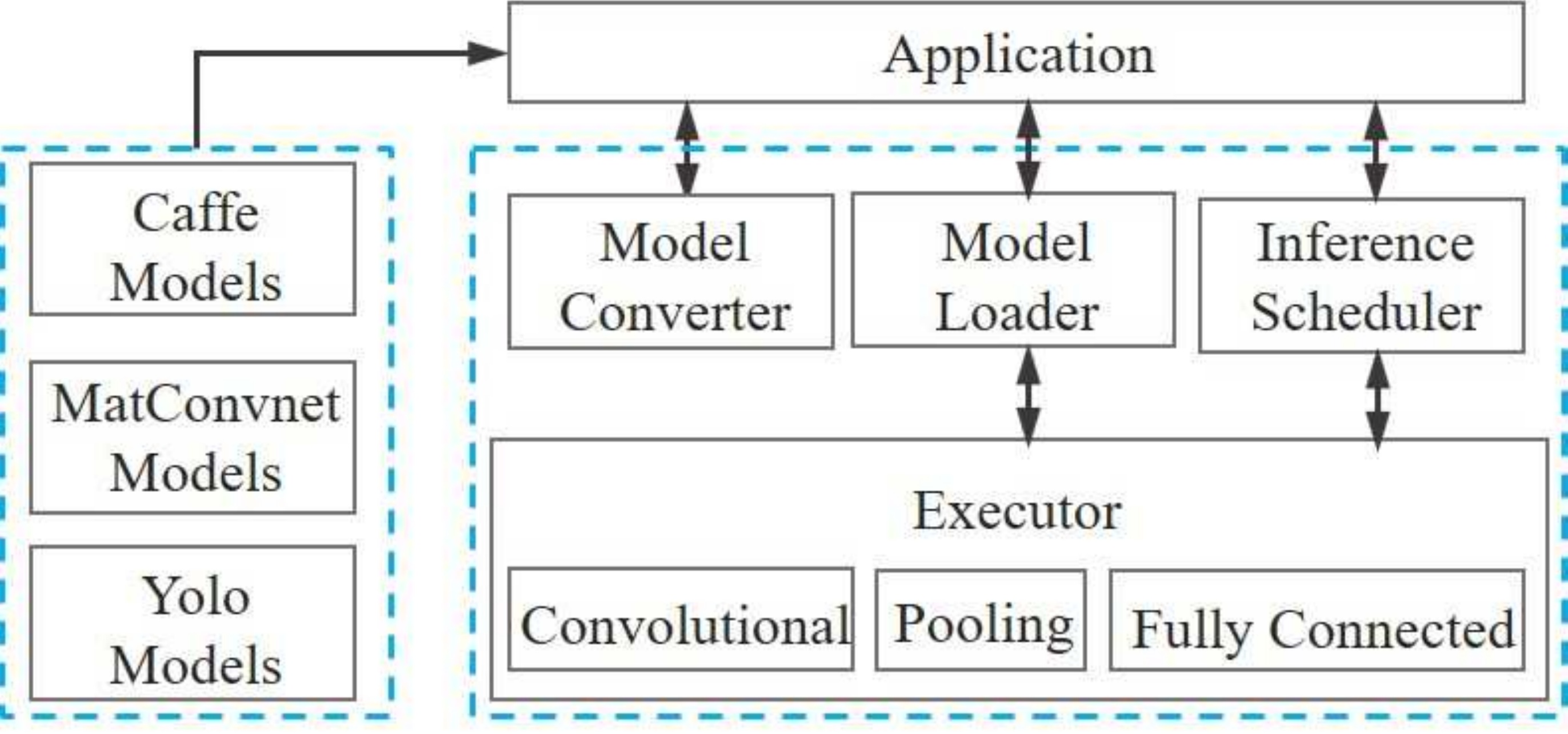}
\end{center}
\caption{The structure of Deepsense. The model converter converts the format of the input model, then the model loader loads the model into memory. Inference scheduler is responsible for task scheduling for GPU. The executor runs the allocated tasks on a GPU.}
\label{deepsense}
\end{figure}

Huynh \emph{et al.} propose Deepsense, a GPU-based CNN framework to run various CNN models in soft real-time on mobile devices with GPUs \cite{huynh2016deepsense}. To minimise the latency, Deepsense applies various optimisation strategies, including branch divergence elimination and memory vectorisation. The structure of Deepsense is shown as in Fig.~\ref{deepsense}. The model converter first converts pre-trained models with different representations into a pre-defined format. Then, the model loader component loads the converted model into memory. When inference starts, the inference scheduler allocates tasks to the GPU sequentially. The executor takes inputted data and the model for executing. During the execution pipeline, CPU is only responsible for padding and intermediate memory allocation, whilst most computing tasks are done by the GPU. The authors further present a demo of the framework in \cite{huynh2017deepmon} for continuous vision sensing applications on mobile devices.

The heterogeneous multi-core architecture, including CPU and GPU on mobile enables the application of neural networks. By reasonably mapping tasks to cores could improve energy efficiency and inference speed. Taylor \emph{et al.} propose a machine learning based approach to map OpenCL kernels onto proper heterogeneous multi-cores to achieve given objectives, e.g., speedup, energy-efficiency or a tradeoff \cite{taylor2017adaptive}. The framework first trains the mapping model with the optimisation setting for each objective, then it uses the learned model to schedule OpenCL kernels based on the information of the application.

Rallapalli \emph{et al.} find that the memory of GPUs severely limits the operation of deep CNNs on mobile devices, and proposes to properly allocate part of computation of the fully-connected layers to the CPU \cite{rallapalli2016very}. The fully-connected layers are split into several parts, which are executed sequentially. Meanwhile, part of these tasks are loaded into the memory of the CPU for processing. They evaluate the method with an object detection model, YOLO \cite{redmon2016you} on Jetson TK1 board and achieve $60\times$ speedup.

In addition to commonly used hardware, i.e., CPUs, mobile GPUs, GPGPU, and DSP, field-programmable gate arrays (FPGAs) could also be used for acceleration. Different from CPUs and GPUs, which run software code, FPGA uses hardware level programming, which means that FPGA is much faster than CPU and GPU. Bettoni \emph{et al.} implement an object recognition CNN model on FPGA via Tiling and Pipelining parallelisation \cite{bettoni2017convolutional}. Ma \emph{et al.} exploit the data reuse and data movement in a convolution loop and proposes to use loop optimisation (including loop unrolling, tiling, and interchange) to accelerate the inference of CNN models in FPGA \cite{ma2017optimizing}. A similar approach is also adopted in \cite{park2018implementation}.

\begin{figure}[tp!]
\begin{center}
\includegraphics[width=\linewidth]{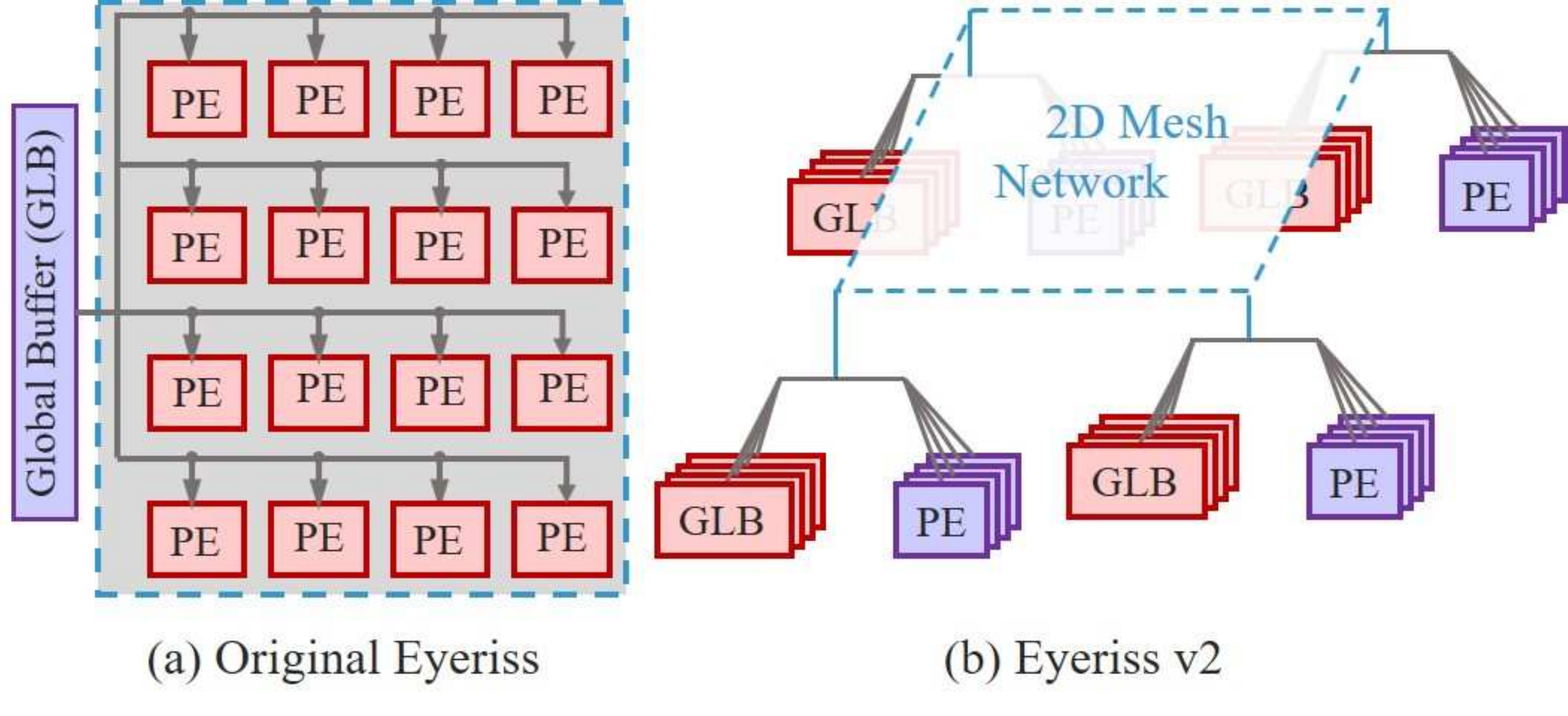}
\end{center}
\caption{The comparison between Eyeriss and Eyeriss v2. Both of them are composed of GLB and PE. Eyeriss v2 adopts a hierarchical structure to reduce communication cost.}
\label{eyerisscomp}
\end{figure}

Lots of literature focus on developing energy-efficiency DNNs. However, the diversity of DNNs makes them inflexible for hardware \cite{chen2018understanding}. Hence, some researchers attempt to design special accelerating chips to flexibly use DNNs. Chen \emph{et al.} develop an energy-efficient hardware accelerator, called Eyeriss \cite{chen2016eyeriss}. Eyeriss uses two methods to accelerate the performance of DNNs. The first method is to exploit data reuse to minimise data movement, whilst the second method is to exploit data statistics to avoid unnecessary reads and computations, which improves energy efficiency. Subsequently, they change the structure of the accelerator and propose a new version, Eyeriss v2, to run compact and sparse DNNs \cite{chen2019eyeriss}. Fig.~\ref{eyerisscomp} shows the comparison between Eyeriss and Eyeriss v2. Both of them consist of an array of processing elements (PE) and global buffers (GLB). The main difference is the structure. Eyeriss v2 is hierarchical, in which PEs and GLBs are grouped to reduce communication cost.

\begin{table*}[htp]
    \footnotesize
    \centering
    \caption{Literature summary of hardware acceleration.}
    \label{acceleration-hardware}
    \begin{tabular}{c c c c c  c}
    \toprule
    \bf{Ref.} & \bf{Model}   & \bf{Executor} &\bf{Strategy}& \bf{Object} & \bf{Performance} \\
    \hline
    \cite{alzantot2017rstensorflow}& CNN, RNN  & CPU, GPU &RenderScript & Feasibility check& $3\times$ faster \\
     \hline
     \cite{loukadakis2018accelerating}& VGG-16  & CPU, GPU &SIMD & Speed up inference& $11.6\times$ faster\\
     \hline
    \cite{oskouei2015gpu} & CNN  & GPU &SIMD & Speed up inference & $60\times$ faster \\
     \hline
    \cite{latifi2016cnndroid}& CNN  & GPU &SIMD & Speed up inference & \makecell{$60\times$ faster\\$130\times$ energy-saving }\\
     \hline
    \cite{tsung2016high}& DNN  & GPU &Flow optimisation & Enable DNN on mobile device & \makecell{$58\times$ faster\\ $104\times$ energy-saving} \\
     \hline
    \cite{rizvi2016gpgpu}& CNN  & GPGPU &CUDA & Maximise throughput & $50\times$ faster \\
     \hline
    \cite{rizvi2017deep}& DNN  & GPU &CUDA & Real-time character detection & 250ms per time \\
     \hline
    \cite{rizvi2017optimized}& DNN & GPU &Matrix multiplication & Real-time character detection & $3\times$ faster \\
     \hline
    \cite{cao2017mobirnn}& LSTM  & GPU &RenderScript & Rnn RNN on mobile platform & $4\times$ reduction on latency \\
     \hline
    \cite{motamedi2016fast}& SqueezeNet  & GPU &RenderScript & Acceleration, energy-efficiency & \makecell{$310.74\times$ faster \\$249.47\times$ energy-saving }\\
     \hline
    \cite{motamedi2017machine}& CNN  & CPU, GPU, DSP &RenderScript & Optimise thread number& $2.37\times$ faster \\
     \hline
    \cite{motamedi2018cappuccino}& CNN  & CPU, GPU, DSP &RenderScript & Automatic speedup & $272.03\times$ faster at most \\
     \hline
    \cite{huynh2016deepsense}& CNN & GPU & Memory vectorisation  & Real-time response & VGG-F 361ms \\
     \hline
    \cite{huynh2017deepmon}& YOLO & GPU &Tucker decomposition & Real-time response & 36\% faster \\
     \hline
    \cite{taylor2017adaptive}& OpenCL  & CPU, GPU &Kernel mapping & Adaptive optimisation & \makecell{$1.2\times$ faster\\ $1.6\times$ energy saving }\\
     \hline
    \cite{rallapalli2016very}& YOLO  & CPU, GPU &Memory optimisation & Enable CNN on mobile device & 0.42s for YOLO \\
     \hline
    \cite{bettoni2017convolutional}& CNN  & FPGA &Tiling, Pipelining & Enable CNN in FPGA & $15\times$ faster \\
     \hline
    \cite{ma2017optimizing}& CNN  & FPGA &Loop optimisation& Memory and data movement & $3.2\times$ faster \\
     \hline
    \cite{park2018implementation}& CNN  & FPGA &Loop optimisation& Improve energy efficiency & \makecell{23\% faster\\ $9.05\times$ energy-saving } \\
     \hline
    \cite{chen2016eyeriss}& DNN  &Eyeriss &Data reuse & Improve energy efficiency & 45\% power saving \\
     \hline
    \cite{chen2019eyeriss}& DNN  &Eyeriss v2&Hierarchical mesh & Hardware processing efficiency & \makecell{$12.6\times$ faster\\ $2.5\times$ energy-saving} \\
    \hline
    \cite{9034111}& CNN  &TPU & Systolic tensor array & Improve systolic array  & $3.14\times$ faster \\

     \bottomrule
    \end{tabular}
    \end{table*}

\subsubsection{Software Acceleration}

Different from hardware acceleration, which depends on the parallelisation of tasks on available hardware, software acceleration mainly focuses on optimising resource management, pipeline design, and compilers. Hardware acceleration methods speed up inference through increasing available computing powers, which usually does not affect the accuracy, whilst software acceleration methods maximise the performance of limited resources for speedup, which may lead to a drop in accuracy with some cases. For example, in \cite{han2016mcdnn}, the authors sacrifice accuracy for real-time response. Table \ref{acceleration-software} summarises existing literature on software acceleration.

Georgiev \emph{et al.} investigate the tradeoff between performance and energy consumption of an audio sensing model on edge devices \cite{georgiev2017accelerating}. Work items need to access different kinds of memory, i.e., global, shared, and private memory. Global memory has the maximum size but minimum speed, whilst private memory is fastest and smallest but exclusive to each work item. Shared memory is between global and private memory. Typical audio sensing models have the maximum parameters, which surpasses the capacity of memory. They use memory access optimisation techniques to speed up the inference, including vectorisation, shared memory sliding window, and tiling.

\begin{figure}[tp!]
\begin{center}
\includegraphics[width=\linewidth]{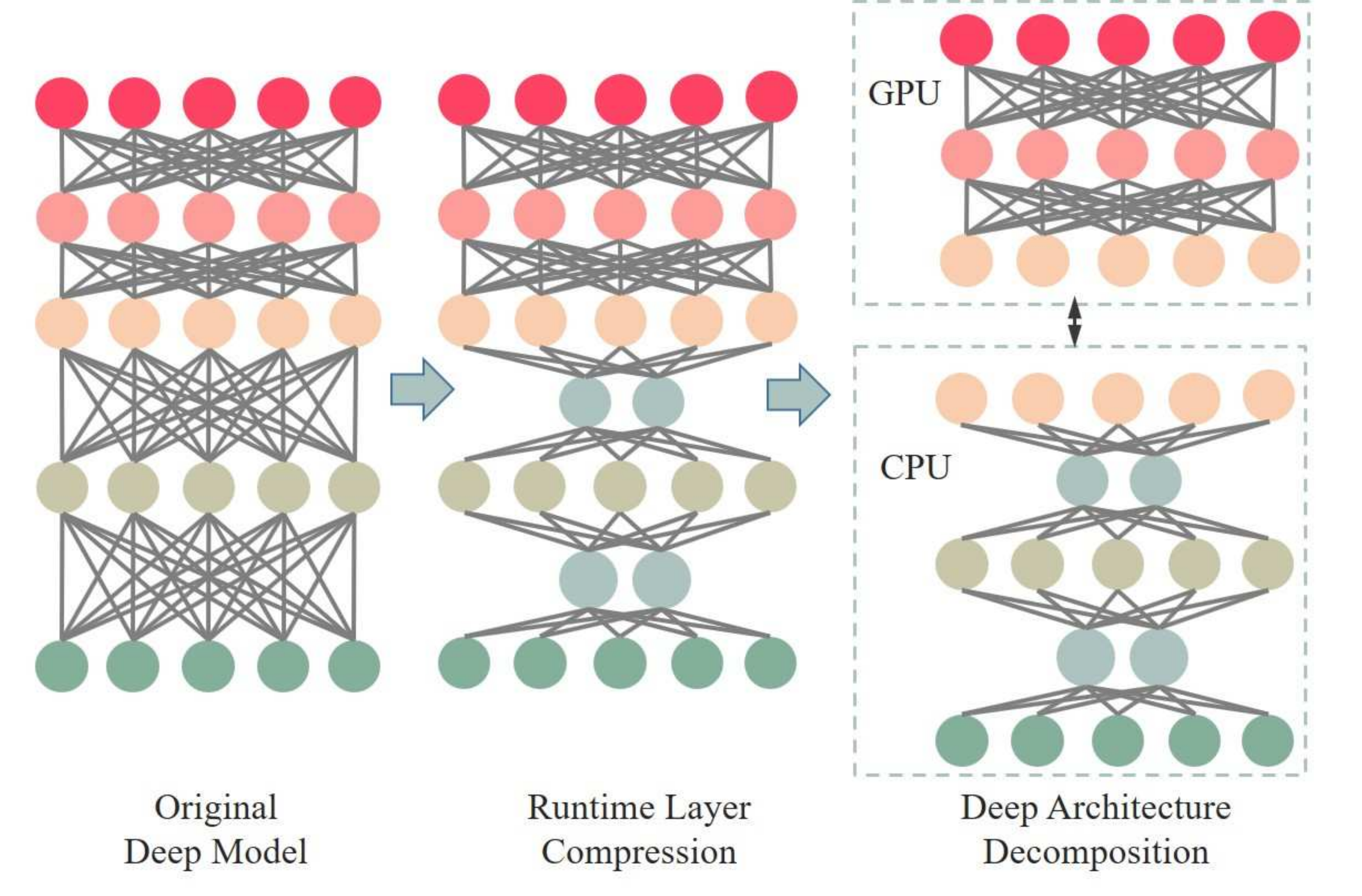}
\end{center}
\caption{The workflow of DeepX. Layer compression could reduce the requirement on resource, whilst the architecture decomposition divides the model into multiple blocks that could be performed in parallel.}
\label{deepx}
\end{figure}

Lane \emph{et al.} propose DeepX to reduce the resource usage on mobile devices based on two resource management algorithms \cite{lane2016deepx}. The first resource management algorithm is for runtime layer compression. The model compression method discussed in Section~\ref{modelcompression} could also be used to remove redundancy from original layers. Specifically, they use a SVD-based layer compression technique to simplify the model. The second algorithm is for architecture decomposition, which decomposes the model into blocks that could be performed in parallel. The workflow of DeepX is shown in Fig.~\ref{deepx}. They further develop a prototype of DeepX on wearable devices \cite{lane2016accelerated}. Subsequently, they develop the DeepX toolkit (DXTK) \cite{lane2016dxtk}. A number of pre-trained and compressed deep neural network models are packaged in DXTK. Users could directly use DXTK for specific applications.

\begin{figure}[tp!]
\begin{center}
\includegraphics[width=\linewidth]{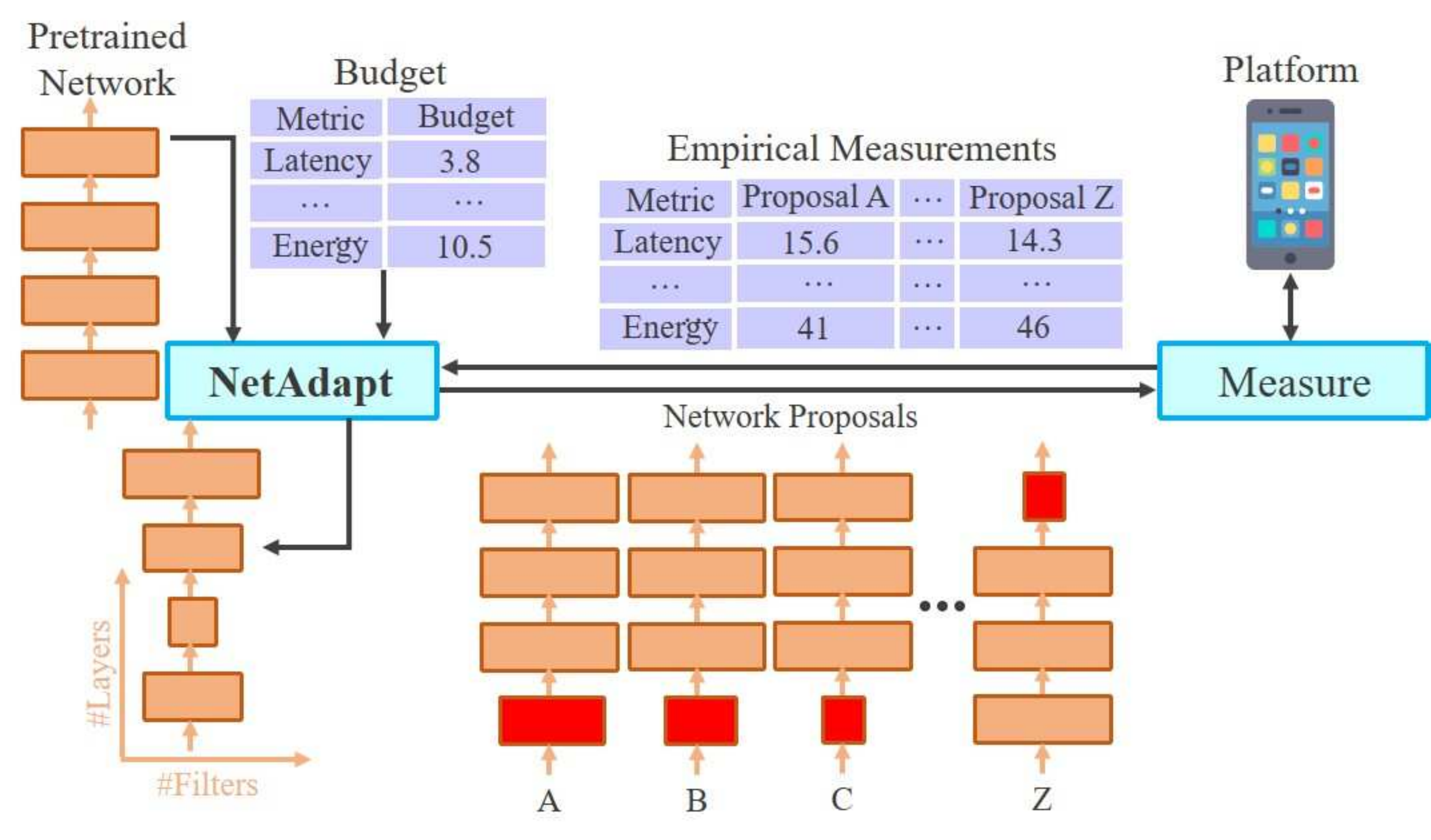}
\end{center}
\caption{The structure of Netadapt. Netadapt caches multiple pre-trained models. When requests arrive, Netadapt selects a specific model and adjusts its structure according to the given budget. Then it chooses the best proposal as the accelerating scheme according to empirical measurement.}
\label{netadapt}
\end{figure}

Yang \emph{et al.} propose an adaptive software accelerator, Netadpt, which could dynamically speed up the model according to specific metrics \cite{yang2018netadapt}. They use empirical measurements on practical devices to evaluate the performance of the accelerator. Fig.~\ref{netadapt} shows the structure of Netadapt. Netadpat adjusts the network according to the given budget, i.e., latency, energy, etc. During iteration, the framework generates many network proposals. Then, Netadapt evaluates these proposals according to direct empirical measurements, and selects one with maximum accuracy. The framework is similar to \cite{liu2018demand}, which caches multiple model compression systems, and compresses the input model according to users' demand.

Ma \emph{et al.} introduce the concept of quality of service (QoS) in model acceleration and develop an accelerator, DeepRT \cite{ma2017deeprt}. The QoS of an accelerated model is defined as a tuple $Q=(d,C)$, where $d$ is a desired response time and $C$ denotes model compression bound. There is a QoS manager component in DeepRT, which controls the system resources to support the QoS during the acceleration.

Liu \emph{et al.} find that fast Fourier transform (FFT) could effectively speed up convolution operation \cite{liu2000versatile}. Abtahi \emph{et al.} applie FFT-based convolution ResNet-20 on NVIDIA Jetson TX1 and evaluates the performance \cite{abtahi2018accelerating}. Results show the inference speed is improved several times. However, FFT-based convolution only works when the convolution kernel is big, e.g., bigger than $9 \times 9 \times 9$. Most models adopt smaller kernels in practice. Hence, there are few applications of FFT-based convolution in practice.

In continuous mobile vision applications, mobile devices are required to deal with continuous videos or images for classification, object recognition, text translation, etc. These continuous videos or images contain large amounts of repeated frames, which are computed through the model again and again during the inference. In such applications, caching mechanisms are promising for acceleration. Xu \emph{et al.} propose CNNCache, a cache-based software accelerator for mobile continuous vision applications, which reuses the computation of similar image regions to avoid unnecessary computation and saves resources on mobile devices \cite{xu2017accelerating}.
Cavigelli \emph{et al.} present a similar framework, named CBinfer \cite{cavigelli2019cbinfer}. The difference is that CBinfer considers the threshold of the pixel size when matching frames. However, CBinfer only matches frames of the same position, which may be ineffective in mobile scenarios. \cite{huynh2017deepmon} also considers reusing the result of the similar input in inference. Different from \cite{xu2017accelerating} and \cite{cavigelli2019cbinfer}, the authors extract histogram-based features to match frames, instead of comparing pixels.

\begin{table*}[htp]
    \footnotesize
    \centering
    \caption{Literature summary of software acceleration.}
    \label{acceleration-software}
    \begin{tabular}{c c c c c c}
    \toprule
    \bf{Ref.} & \bf{Model}  &\bf{Strategy}& \bf{Object} & \bf{Performance} &\bf{Accuracy}\\
    \hline
    \cite{han2016mcdnn}& DNN  &memory access optimisation & Performance-energy tradeoff& 42ms, 83\% accuracy &Lossy\\
    \hline
    \cite{georgiev2017accelerating}& DNN &Resource management & Accelerate inference& $6.5\times$ faster, $3-4\times$ less energy &Lossless\\
    \hline
    \cite{lane2016deepx}& DNN  &Compression, decomposition & Reduce resource use& $5.8\times$ faster &4.9\% loss\\
     \hline
    \cite{lane2016accelerated}& DNN  &Compression, decomposition & Reduce resource use& $5.8\times$ faster &4.9\% loss\\
     \hline
      \cite{lane2016dxtk}& DNN  &Caching & Reduce resource use& $5.8\times$ faster &4.9\% Lossless\\
     \hline
    \cite{yang2018netadapt}& NN  &Caching & Adaptive speedup& $1.7\times$ speedup &4.9\% Lossless\\
     \hline

     \cite{3371154}& DNN  &Caching, model selection & Optimizing DL inference& $1.8\times$ speedup &7.52\% improvement\\
     \hline
    \cite{ma2017deeprt}& DNN  &QoS control & Improve QoS&N/A&N/A\\
     \hline
     \cite{abtahi2018accelerating}& CNN  &FFT-based convolution & Accelerate convolution& $10916\times$ faster at most & N/A \\
     \hline
    \cite{xu2017accelerating}& CNN  &Cache mechanism & Accelerate inference &20.2\% faster &3.51\% drop\\
     \hline
    \cite{cavigelli2019cbinfer}& CNN  &Caching, pixel matching & Accelerate inference &$9.1\times$ faster & 0.1\% drop\\
     \hline
    \cite{huynh2017deepmon}& YOLO  &Caching, feature extraction & Real-time response &36\% faster &3.8\%-6.2\% drop\\
     \hline

     \cite{garofalo2020pulp}& NN  &Optimized computing library & Ultra-low-power computing &Up to $63\times$ faster & Negligible loss\\
     \bottomrule
    \end{tabular}
\end{table*}

\section{Edge Offloading}
Computation is of utmost importance for supporting edge intelligence, which powers the other three components. Most edge devices and edge servers are not as powerful as central servers or computing clusters. Hence, there are two approaches to enable computation-intensive intelligent applications at the edge: reducing the computational complexity of applications and improving the computing power of edge devices and edge servers. The former approach has been discussed in previous sections. In this section, we focus on the latter approach.

Considering the hardware limitation of edge devices, computation offloading \cite{hu2015mobile,abbas2017mobile,braud2020multipath,shatilov2019using,kosta2012thinkair} offers promising approaches to increase computation capability. Literature of this area mainly focuses on designing an optimal offloading strategy to achieve a particular objective, such as latency minimisation, energy-efficiency, and privacy preservation. According to their realisation approaches, these strategies could be divided into five categories: device-to-cloud (D2C), device-to-edge (D2E), device-to-device (D2D), hybrid architecture, and caching.


\subsection{D2C offloading strategy}
It consumes considerable computing resources and energy to deal with streamed AI tasks, such as video analysis and continuous speech translation. Most applications, such as Apple Siri and Google Assistant, adopt pure cloud based offloading strategy, in which devices upload input data, e.g., speech or image to cloud server through cellular or WiFi networks. The inference through a giant neural model with high accuracy is done by powerful computers and the results are transmitted back through the same way. There are three main disadvantages in this procedure: (1) mobile devices are required to upload enormous volumes of data to the cloud, which has proved to be the bottleneck of the whole procedure \cite{kang2017neurosurgeon}. Such a bottleneck increases users' waiting time; (2) the execution depends on the Internet connectivity; once the device is offline, relative applications could not be used; (3) the uploaded data from mobile devices may contain private information of users, e.g., personal photos, which might be attacked by malicious hackers during the inference on cloud server \cite{raval2016you}. There are some efforts trying to solve these problems, which will be discussed next. Table \ref{d2coffloadingstrategy} summarises existing literature on D2C offloading strategy.

\begin{figure}[tp!]
\begin{center}
\includegraphics[width=\linewidth]{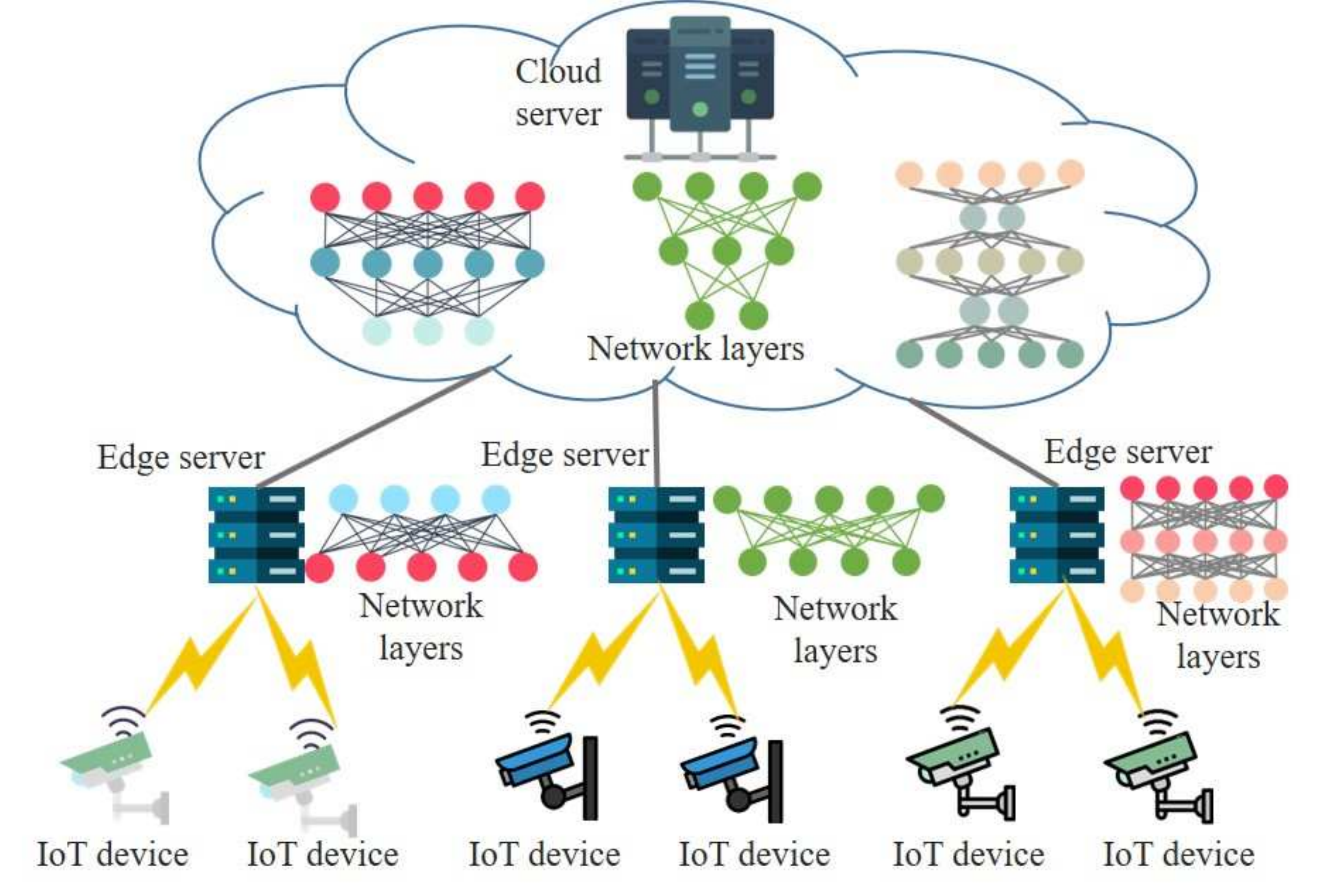}
\end{center}
\caption{The structure of neural network layer scheduling-based offloading. IoT devices first upload collected data to edge server, where few neural network layers are deployed. The raw data is first pre-processed on edge servers. Then  the intermediate results are uploaded to cloud server for further processing.}
\label{learningiot}
\end{figure}

There are usually many layers in a typical deep neural network, which processes the input data layer by layer. The size of intermediate data could be scaled down through layers. Li \emph{et al.} propose a deep neural network layer schedule scheme for the edge environment, leveraging this characteristic of deep neural networks \cite{li2018learning}. Fig.~\ref{learningiot} shows the structure of neural network layer scheduling-based offloading scheme. Edge devices lacking computing resources, such as IoT devices, first upload the collected data to nearby edge server, which would process the original input data through few low network layers. The generated intermediate data would be uploaded to the cloud server for further processing and eventually output the classification results. The framework is also adopted in \cite{huang2017deep}. The authors use edge server to pre-process raw data and extract key features.

The model partitioning and layer scheduling could be designed from multiple perspectives, e.g., energy-efficiency, latency, and privacy. Eshratifar \emph{et al.} propose a layer scheduling algorithm from the perspective of energy-efficiency in a similar offloading framework \cite{eshratifar2018energy}. Kang \emph{et al.} investigate this problem between edge and cloud side \cite{kang2017neurosurgeon}. They propose to partition the computing tasks of DNN between local mobile devices and cloud server and design a system, called Neurosurgeon, to intelligently partition DNN based on the prediction of system dynamics. Osia \emph{et al.} consider the layer scheduling from the perspective of privacy preservation \cite{osia2017hybrid}. They add a feature extractor module to identify private features from raw data, which will be sent to the cloud for further processing. Analogous approaches are also adopted in  \cite{liu2017deeprotect,xu2019edgesanitizer}.

In continuous computer vision analysis, video streams need to be uploaded to the cloud server, which requires a large amount of network resources and consumes battery energy. Ananthanarayanan \emph{et al.} propose a geographically distributed architecture of clouds and edge servers for real-time video analysis \cite{ananthanarayanan2017real}. Fixed (e.g., traffic light) and mobile cameras (e.g., event data recorder) upload video streams to available edge servers for pre-processing. The pre-processed data would be further transmitted to a central cloud server in a geographic location for inference. Similarly, Ali \emph{et al.} leverage all available edge resources to pre-process data for large-scale video stream analytics \cite{ali2018edge}. Deep learning based video analytic applications contain four stages, including motion detection, frame enhancement, object detection based on shallow networks, and object detection based on deep networks. With the traditional cloud-based approach, these four stages are executed on a cloud server. The authors propose to execute the first two stages locally, which does not require much computation capacity. The output is then transmitted to edge servers for further processing (the third stage). The output is then uploaded to the cloud for final recognition.

Some efforts \cite{hoisko2000context,hodges2006sensecam} propose to upload only `interesting' frames to the cloud, which significantly reduces the amount of uploaded data. However, detecting these `interesting' frames also requires intensive computation. Naderiparizi \emph{et al.} develop a framework, Glimpse, to select valid frames by performing coarse visual processing with low energy consumption \cite{naderiparizi2017glimpse}. Glimpse adopts gating sensors and redesigns the processing pipeline to save energy.

The easiest offloading strategy is to offload the inference task to the cloud when the network condition is good, otherwise perform model compression locally. For example, \cite{sanabria2018code} only considers the network condition when offloading healthcare inference tasks. Cui \emph{et al.} characterise the resource requirements of data processing applications on an edge gateway and cloud server \cite{cui2017cross}. Hanhirova \emph{et al.} explore and characterises the performance of some CNNs, e.g., object detection and recognition, both on smartphones and cloud server \cite{hanhirova2018latency}. They find that the latency and throughput are conflicting metrics, in turn, are difficult to be jointly optimised on both mobile devices and cloud server. Some efforts focus on designing an offloading scheme to decide when to use a neural network locally and when to offload the task to the cloud server, instead of always executing on cloud. Considering the capacities of local devices and the network conditions, Qi \emph{et al.} design an adaptive decision scheme to dynamically perform tasks \cite{qi2017dnn}. To enable the local execution, mobile devices adopt compressed models, which achieve lower accuracy than the complete model on the cloud server. If the network condition could not guarantee a real-time response, the inference task would be executed locally. \cite{ran2017delivering} also proposes an offloading decision scheme for the same problem with extra consideration to energy consumption. Ran \emph{et al.} subsequently extend the work with a measurement-driven mathematical framework for achieving a tradeoff between data compression, network condition, energy consumption, latency, and accuracy \cite{ran2018deepdecision}.

Georgiev \emph{et al.} consider a collective offloading scheme for heterogeneous mobile processors and cloud for sensor based applications, which makes best possible use out of different kinds of computing resources on mobile devices, e.g., CPU, GPU, and DSP \cite{georgiev2016leo}. They designed a task scheduler running on low-power co-processor unit (LPU) to dynamically restructure and allocate tasks from applications across heterogeneous computing resources based on fluctuations in device and network.

\begin{table*}[htp]
    \footnotesize
    \centering
    \caption{Literature summary of D2C offloading strategy.}
    \label{d2coffloadingstrategy}
    \begin{tabular}{c c c c c c}
    \toprule
    \bf{Ref.} & \bf{Model}  &\bf{Execution platform}&\bf{Focus and problem} & \bf{Latency} &\bf{Energy efficiency}\\
    \hline
    \cite{li2018learning}& DNN  &Edge and cloud & Layer partitioning to reduce uploaded data &0.2s & N/A\\
    \hline
    \cite{huang2017deep}& DNN  &Edge and cloud & Framework design &$3.23\times$ faster & N/A\\
    \hline
    \cite{eshratifar2018energy}& DNN  &Edge and cloud & Layer partitioning for energy-efficiency &$3.07\times$ faster & $4.26\times$ higher\\
    \hline
    \cite{kang2017neurosurgeon}& DNN  &Edge and cloud & Layer partitioning for latency, energy &$3.1\times$ faster & 140.5\% higher\\
     \hline
     \cite{osia2017hybrid}& DNN  &Local and cloud & Layer partitioning for privacy&N/A& N/A\\
     \hline
     \cite{liu2017deeprotect} & DNN  &Local and cloud & Feature obfuscation for sensitive data &N/A& N/A\\
     \hline
      \cite{xu2019edgesanitizer} & DNN  &Local and cloud & Feature obfuscation for privacy protection &N/A& N/A\\
     \hline
    \cite{ananthanarayanan2017real}& DNN  &Edge and cloud & Resource-accuracy tradeoff  for real-time performance& N/A & N/A\\
     \hline
    \cite{ali2018edge}& CNN  &Edge and cloud & Task allocation for QoS &$3.1\times$ faster & 140.5\% higher\\
     \hline
      \cite{naderiparizi2017glimpse}& CV  &Edge and cloud & Hardware-based energy and computation efficiency &$10-20\times$ faster & $7-25\times$ higher\\
     \hline
     \cite{sanabria2018code}& DNN  &Edge or cloud & Offloading decision for acceleration &N/A &N/A\\
     \hline
     \cite{hanhirova2018latency}& CNN  &Edge or cloud & Performance characterisation and measurement &N/A& N/A\\
     \hline
     \cite{qi2017dnn}& CNN  &Edge or cloud & Latency-accuracy tradeoff for computation-efficiency &N/A& N/A\\
     \hline
     \cite{ran2017delivering}& NN  &Edge or cloud & Multi-objective tradeoff for real-time performance &N/A& N/A\\
     \hline
    \cite{ran2018deepdecision}& NN  &Edge or cloud & Multi-objective tradeoff for real-time performance &N/A& N/A\\
     \hline
     \cite{georgiev2016leo}& N/A  &Local and cloud &Optimal schedule for energy efficiency &Real-time& $1.6-3\times$ higher\\
     \hline
    \end{tabular}
\end{table*}

\subsection{D2E offloading strategy}

Three main disadvantages with the D2C offloading strategy have been discussed, i.e., latency, wireless network dependency, and privacy concern. Although various solutions have been proposed to alleviate these problems, they do not address these fundamental challenges. Users still need to wait for a long time. Congested wireless networks lead to failed inference. Moreover, the potential risk of private information leakage still exists. Hence, some works try to explore the potential of D2E offloading, which may effectively address these three problems. Edge server refers to the powerful servers (more powerful than ordinary edge devices) that is physically near mobile devices. For example, wearable devices could offload the inference tasks to their connected smartphones. Smartphones could offload computing tasks to cloudlets deployed at roadside. Table \ref{d2eoffloadingstrategy} summarises the existing literature on D2E offloading strategy.

First, we review the works that offload inference tasks to specialised edge servers, e.g., cloudlets and surrogates \cite{xu2018survey}, which refer to infrastructure deployed at edge of the network. There are two general problems that need to be considered in this scenario, including (1) which component of the model could be offloaded to the edge; and (2) which edge server should be selected to offload to. Ra \emph{et al.} develop a framework, named Odessa for interactive perception applications, which enables parallel execution of the inference on local devices and edge server \cite{ra2011odessa}. They propose a greedy algorithm to partition the model based on the interactive deadlines. The edge servers and edge devices in Odessa are assumed to be fixed, meaning they do not consider problem (2). Streiffer \emph{et al.} appoint an edge server for mobile devices that requests video frame analytics \cite{streiffer2017eprivateeye}. They evaluate the impact of distance between edge server and mobile devices on latency and packet loss and find that offloading inference tasks to an edge server at a city-scale distance could achieve the similar performance with execution locally on each mobile devices.

Similar to D2C offloading, where the partitioned model layers could be simultaneously deployed on both cloud server and local edge device, the partitioned model layers could also be deployed on edge servers and edge devices. This strategy reduces the transmitted data, which further reduces latency and preserve privacy. Li \emph{et al.} propose Edgent, a device-edge co-inference framework to realise this strategy \cite{li2018edge}. The core idea of Edgen is to run computation-intensive layers on powerful edge servers and run the rest layers on device. They also adopt model compression techniques to reduce the model size and further reduce the latency. Similarly, Ko \emph{et al.} propose a model partitioning approach with the consideration of energy efficiency \cite{ko2018edge}. Due to the difference of available resources between edge devices and edge servers, partitioning the network at a deeper layer would reduce the energy efficiency. Hence, they propose to partition the network at the end of the convolution layers. The output features through the layers on edge device would be compressed before transmitted to edge server to minimise the bandwidth usage.

Some efforts \cite{gilad2016cryptonets,chabanne2017privacy,hesamifard2017cryptodl} attempt to encrypt the sensitive data locally before uploading. On cloud side, non-linear layers of a model are converted into linear layers, and then they use homomorphic encryption to execute inference over encrypted input data. This offloading paradigm could also be adopted on edge servers. However, the encryption operation is also computation-intensive. Tian \emph{et al.} propose a private CNN inference framework, LEP-CNN, to offload most inference tasks to edge servers and to avoid privacy breaches \cite{tian2019lep}. The authors propose an online/offline encryption method to speed up the encryption, which trades offline computation and storage for online computation speedup. The execution of the inference over encrypted input data on edge server addresses privacy issues.

Mobility of devices introduces a challenge during the offloading, e.g., in autonomous driving scenarios. Mobile devices may lose the connection with edge server before the inference is done. Hence, selecting an appropriate edge server according to users' mobility pattern is crucial. Zhang \emph{et al.} use reinforcement learning to decide when and which edge server to offload to \cite{zhang2019task}. A deep Q-network (DQN) based approach is used to automatically learn the optimal offloading scheme from previous experiences. If one mobile device moves away before the edge server finishes the execution of the task, the edge server must drop the task, which wastes the computing resources. Jeong \emph{et al.} propose to move the execution state of the task back to the mobile device from the edge server before the mobile device moves away in the context of web apps \cite{jeong2018computation}. The mobile device continues the execution of the task in this way.

Since the number of edge servers and computation capacity of edge servers are limited, edge devices may compete for resources on edge servers. Hence, proper task scheduling and resource management schemes should be proposed to provide better services at edge. Yi \emph{et al.} propose a latency-aware video edge analytic (LAVEA) system to schedule the tasks from edge devices \cite{yi2017lavea}. For a single edge server, they adopt Johnson's rule \cite{johnson1954optimal} to partition the inference task into a two-stage job and prioritise all received offloading requests from edge devices. LAVEA also enables cooperation among different edge servers. They propose three inter-server task scheduling algorithms based on transmission time, scheduling time, and queue length, respectively.

\begin{table*}[htp]
    \footnotesize
    \centering
    \caption{Literature summary of D2E offloading strategy.}
    \label{d2eoffloadingstrategy}
    \begin{tabular}{c c c c c c c}
    \toprule
    \bf{Ref.} & \bf{Model} &\bf{Problem}& \bf{Object} & \bf{Latency} &\bf{Energy consumption}\\
    \hline

    \cite{ra2011odessa}& Object recognition  & Model partitioning & Responsiveness and accuracy &$3\times$ & N/A\\
    \hline
    \cite{streiffer2017eprivateeye}& Object recognition & Model partitioning & Responsiveness and accuracy &$3\times$ & N/A\\
    \hline
    \cite{li2018edge}& DNN  & Model partitioning & Reduce latency &100-1000ms & N/A\\
    \hline
    \cite{ko2018edge}& DNN  & Model partitioning & Energy-efficiency &N/A & $4.5\times$ enhanced\\
    \hline
    \cite{tian2019lep}& CNN & online/offline encryption & Privacy and latency & $35\times$ & 95.56\% saved\\
    \hline
    \cite{zhang2019task}& N/A  & Edge server selection & Optimal task migration &N/A & N/A\\
    \hline
    \cite{jeong2018computation}& DNN  & Execution state migration & Computation resource &N/A & N/A\\
    \hline
    \end{tabular}
    \end{table*}

\subsection{D2D offloading strategy}

Most neural networks could be executed on mobile devices after compression and achieve a compatible performance. For example, the width-halved GoogLeNet on unmanned aerial vehicles achieves 99\% accuracy \cite{lee2017deep}. Some works consider a more static scenario, where edge devices, such as smart watches are linked to smartphones or home gateways. Wearable devices could offload their model inference tasks to connected powerful devices. There are two kinds of offloading paradigms in this scenario, including binary decision-based offloading and partial offloading. Binary decision offloading refers to executing the task either on a local device or through offloading. This paradigm is similar to D2C offloading. Partial offloading means dividing the inference task into multiple sub-tasks and offloading some of them to associated devices. In fact, although the associated devices are more powerful, the performance of the complete offloading is not necessarily better than partial offloading. Because complete offloading is required to transmit complete input data to the associated device, which increases the latency. Table \ref{d2doffloadingstrategy} summarises the existing literature for D2D offloading strategy.

Xu \emph{et al.} present CoINF, an offloading framework for wearable devices, which offloads partial inference tasks to associated smartphones \cite{xu2017enabling}. CoINF partitions the model into two sub-models, in which the first sub-model could be executed on the wearable devices, while the second sub-model could be performed on smartphones. They find that the performance of partial offloading outperforms the complete offloading in some scenarios. They further develop a library and provide API for developers. Liu \emph{et al.} also propose EdgeEye, an open source edge computing framework to provide real-time service of video analysis, which provides a task-specific API for developers \cite{liu2018edgeeye}. Such APIs help developers focus on application logic. Similar methods are also adopted in \cite{thomas2018pushing}.

\begin{figure}[tp!]
\begin{center}
\includegraphics[width=\linewidth]{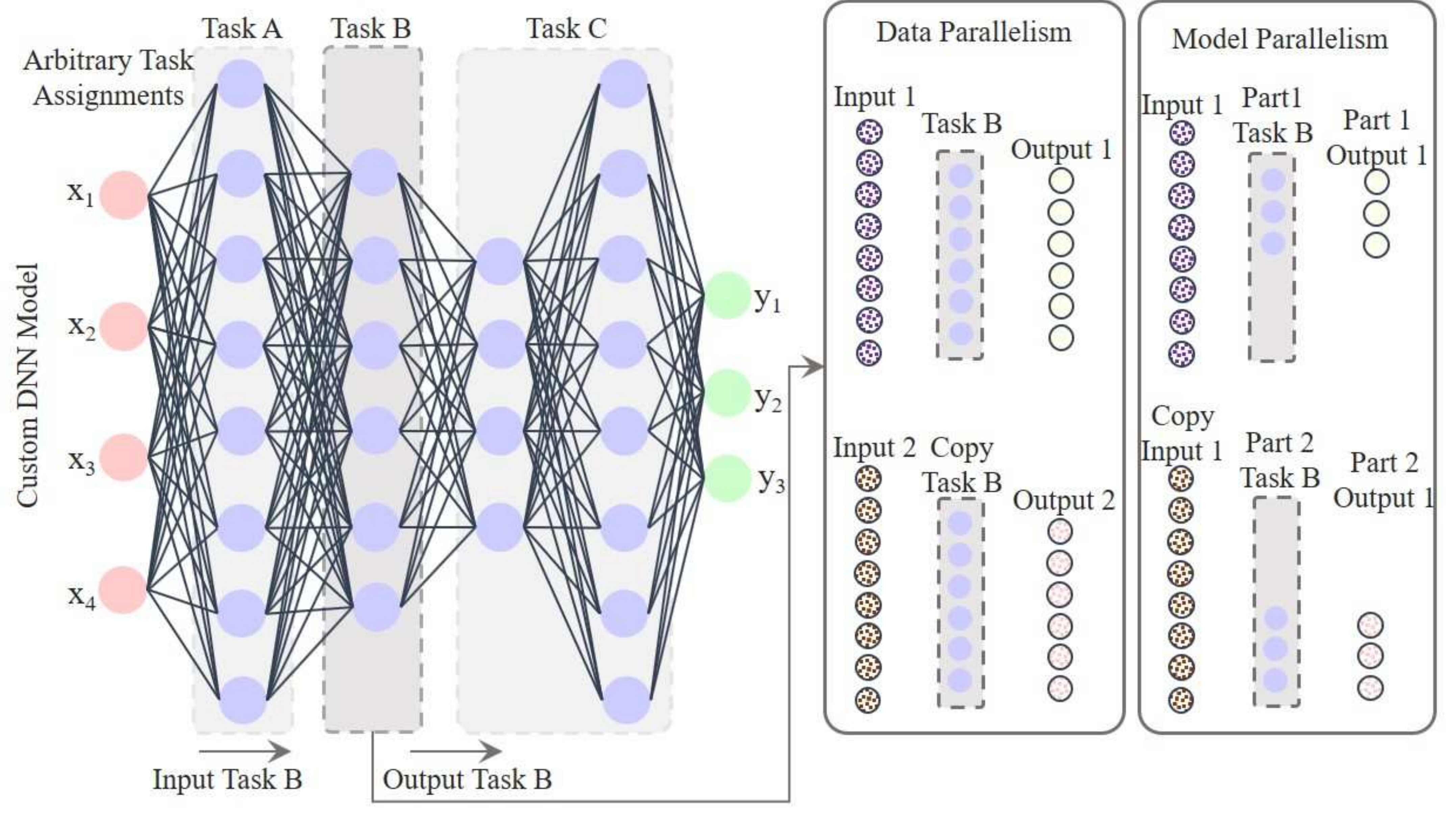}
\end{center}
\caption{The parallelism structure of Musical Chair. Task B is a layer-level task, which are further partitioned into two sub-tasks on two devices. These two devices adopt different input to double the system throughput.}
\label{muscial}
\end{figure}

If one edge device is not powerful enough to provide real-time response for model inference, a cluster of edge devices could cooperate and help each other to provide enough computation resources. For example, if a camera needs to perform image recognition task, it could partition the CNN model by layers, and transmit the partitioned tasks to other devices nearby.
In this scenario, a cluster of edge devices could be organised as a virtual edge server, which could execute inference tasks from both inside and outside of the cluster.
Hadidi \emph{et al.} propose Musical Chair, an offloading framework that harvests available computing resources in an IoT network for cooperation \cite{hadidi2018musical}. In Musical Chair, the authors develop data parallelism and model parallelism scheme to speed up the inference. Data parallelism refers to duplicating devices that performs the same task whilst model parallelism is about performing different sub-tasks of a task on different devices. Fig.~\ref{muscial} shows the parallel structure for a layer-level task. Talagala \emph{et al.} use a graph-based overlay network to specify the pipeline dependencies in neural networks and propose a server/agent architecture to schedule computing tasks amongst edge devices in similar scenarios \cite{talagala2018eco}.

\begin{figure}[tp!]
\begin{center}
\includegraphics[width=\linewidth]{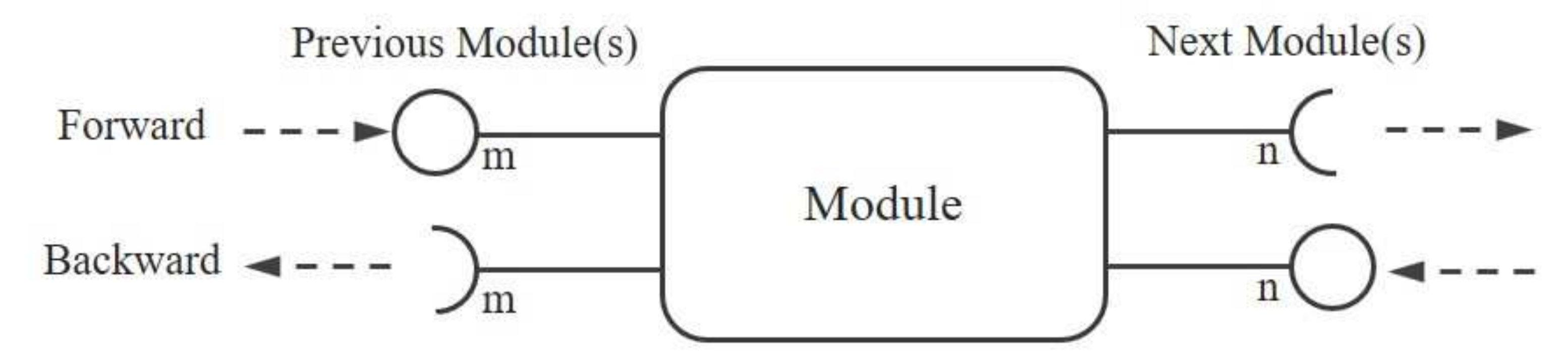}
\end{center}
\caption{The illustration of a DIANNE module. Each module has references to its predecessors and successors for feedforward and back-propagation during training.}
\label{dianne}
\end{figure}

Coninck \emph{et al.} develop DIANNE, a modular distributed framework, which treats neural network layers as directed graphs \cite{de2018dianne}. As shown in Fig.~\ref{dianne}, each module provides forward and backward methods, corresponding to feedforward and back-propagation, respectively. Each module is a unit deployed on an edge device. There is a job scheduler component, which assigns learning jobs to devices with spare resources. Fukushima \emph{et al.} propose the MicroDeep framework, which assigns neurons of CNN to wireless sensors for image recognition model training \cite{fukushima2018microdeep}. The structure of MicroDeep is similar to DIANNE. Each CNN unit is allocated to a wireless sensor, which executes the training of the unit parameters. In feedforward phase, sensors exchange their output data. Once a sensor receives the necessary input, it executes its unit and broadcasts its output for subsequent layer. If a sensor with an output layer unit obtains its input and ground-truth, it starts the back-propagation phase. They adopt a 2D coordinate based approach to approximately allocate a CNN unit to sensors.

Distributed solo learning enables edge devices or edge servers to train models with local data. Consequently, each model may become local experts that are good at predicting local phenomena. For example, RSUs use local trained models to predict local traffic condition. However, users are interested in the traffic condition of places they plan to visit, in addition to the local traffic condition. Bach \emph{et al.} propose a routing strategy to forward the queries to devices that have the specific knowledge \cite{bach2017knowledge}. The strategy is similar to the routing strategy in TCP/IP networks. Each device maintains a routing table to guide the forwarding. The strategy achieve 95\% accuracy in their experiments. However, latency is a big problem in such frameworks.

\begin{table*}[htp]
    \footnotesize
\centering
\caption{Literature summary of D2D offloading strategy.}
\label{d2doffloadingstrategy}
\begin{tabular}{c c c c c c c}
\toprule
\bf{Ref.} & \bf{Model} &\bf{Problem}& \bf{Object} & \bf{Latency} &\bf{Energy consumption}\\
\hline
\cite{xu2017enabling}& DNN  & Model partition & Acceleration, save energy &$23\times$  & 85.5\% reduction\\
\hline
\cite{liu2018edgeeye}& DNN  & Open source framework & Enable edge inference &N/A & N/A\\
\hline
\cite{song2018situ}& AlexNet,VGGNet  & Incremental training & Improve accuracy&$1.4-3.3\times$ & 30\%-70\% saving\\
\hline
 \cite{yi2017lavea}& DNN  & Task scheduling & Minimise latency &$1.2-1.7\times$  & N/A\\
\hline
\cite{hadidi2018musical}& DNN  & Data and task parallelism & Computing power &$90\times$  & $200\times$ reduction\\
\hline
\cite{talagala2018eco}& N/A & Execution management & ML deployments at edge &N/A & N/A\\
\hline
\cite{de2018dianne} &AlexNet & Data, model parallelism & Modular architecture & N/A& N/A\\
\hline
\cite{fukushima2018microdeep} & CNN & Neuron assignment & Enable training/inference &N/A & N/A\\
\hline
\cite{bach2017knowledge}& Bayesian & Knowledge retrieval& Routing strategy &N/A & N/A\\
\hline
\end{tabular}
\end{table*}

\subsection{Hybrid offloading}
The hybrid offloading architecture, also named osmotic computing \cite{villari2016osmotic}, refers to the computing paradigm that is supported by the seamless collaboration between edge and cloud computing resources, along with the assistance of data transfer protocols. The hybrid computing architecture takes advantage of cloud, edge, and mobile devices in a holistic manner. There are some efforts focusing on distributing deep learning models in such environments. Table \ref{hybrid-computing} presents a summary of these efforts.

Morshed \emph{et al.} investigate `deep osmosis' and analyses the challenges involved with the holistic distributed deep learning architecture, as well as the data and resource architecture \cite{morshed2017deep}. Teerapittayanon \emph{et al.} propose distributed deep neural networks (DDNNs) based on the holistic computing architecture, which maps sections of a DNN onto a distributed computing hierarchy \cite{teerapittayanon2017distributed}. All sections are jointly trained in the cloud to minimise communication and resource usage for edge devices. During inference, each edge device performs local computation and then all outputs are aggregated to output the final results.

There is always the risk that the physical nodes i.e., edge devices and edge servers may fail, which results in the failure of DNN units deployed on these physical nodes. Yousefpour \emph{et al.} introduce `deepFogGuard' to make the distributed DNN inference failure-resilient \cite{yousefpour2019guardians}. Similar to residual connections \cite{he2016deep}, which skips DNN layers to reduce the runtime, `deepFogGuard' skips physical nodes to minimise the impact of failed DNN units. The authors also verify the resilience of `deepFogGuard' on sensing and vision applications.

\begin{table*}[htp]
    \footnotesize
\centering
\caption{Literature summary of hybrid offloading strategy.}
\label{hybrid-computing}
\begin{tabular}{c c c c c c c}
\toprule
\bf{Ref.} &\bf{Contribution}& \bf{Solution} & \bf{Performance} \\
\hline
\cite{morshed2017deep}&   Challenge analysis in deep osmosis& N/A &N/A \\
\hline
\cite{teerapittayanon2017distributed}&   DDNN frame & Joint training of DNN sections &$20\times$ cost reduction \\
\hline
\cite{yousefpour2019guardians}&   deepFogGuard & Skip hyperconnections &16\% improvement on accuracy \\
\hline
\end{tabular}
\end{table*}

\subsection{Applications}

There exists some works applying the above mentioned offloading strategy to practical applications, such as intelligent transportation \cite{ferdowsi2019deep}, smart industry \cite{li2018deep}, smart city \cite{tang2017incorporating}, and healthcare \cite{liu2017new} \cite{muhammed2018ubehealth}. Specifically, in \cite{ferdowsi2019deep}, the authors design an edge-centric architecture for intelligent transportation, where roadside smart sensors and vehicles could work as edge servers to provide low latency deep learning based services. \cite{li2018deep} proposes a deep learning based classification model to detect defective products in assembly lines, which leverages an edge server to provide computing resources. Tang \emph{et al.} develop a hierarchical distributed framework to support data intensive analytics in smart cities \cite{tang2017incorporating}. They develop a pipeline monitoring system for anomaly detection. Edge devices and servers provide computing resources for the execution of these detection models. Liu \emph{et al.} design an edge based food recognition system for dietary assessment, which splits the recognition tasks between nearby edge devices and cloud server to solve the latency and energy consumption problem \cite{liu2017new}. Muhammed \emph{et al.} develop a ubiquitous healthcare framework, called UbeHealth, which makes full use of deep learning, big data, and computing resources \cite{muhammed2018ubehealth}. They use big data to predict the network traffic, which in turn is used to assist the edge server to make the optimal offloading decision.

\section{Future Directions and Open Challenges}\label{S5}
We present a thorough and comprehensive survey on the literature surrounding edge intelligence. The benefits of edge intelligence are obvious - it paves the way for the last mile of AI and to provide high-efficient intelligent services for people, which significantly lessens the dependency on central cloud servers, and can effectively protect data privacy. It is worth recapping that there are still some unsolved open challenges in realising edge intelligence. It is crucial to identify and analyze these challenges and seek for novel theoretical and technical solutions. In this view, we discuss some prominent challenges in edge intelligence along some possible solutions. These challenges include data scarcity at edge, data consistency on edge devices, bad adaptability of statically trained model, privacy and security issues, and Incentive mechanism.

\subsection{Data scarcity at edge}\label{S5.1}
For most machine learning algorithms, especially supervised machine learning, the high performance depends on sufficiently high-quality training instances. However, it often does not work in edge intelligence scenarios, where the collected data is sparse and unlabelled, e.g., in HAR and speech recognition applications. Different from traditional cloud based intelligent services, where the training instances are all gathered in a central database, edge devices use the self-generated data or the data captured from surrounding environments to generate models. High-quality training instances, e.g., good image features are lacking in such datasets. Most existing works ignore this challenge, assuming that the training instances are of good quality. Moreover, the training dataset is often unlabelled. Some works \cite{xing2018enabling,shahmohammadi2017smartwatch} propose to use active learning to solve the problem of unlabelled training instances, which requires manual intervention for annotation. Such an approach could only be used in scenarios with few instances and classifications. Federated learning approaches leverage the decentralised characteristic of data to effectively solve the problem. However, federated learning is only suitable for collaboration training, instead of the solo training needed for personalised models. 

We discuss several possible solutions for this problem as follows. 
\begin{itemize}
\item Adopt shallow models, which could be trained with only a small dataset. Generally, the simpler the machine learning algorithm is, the better the algorithm will learn from the small datasets. A simple model, e.g., Naive Bayes, linear model, and decision tree, are enough to deal with the problem in some scenarios, compared with complicated models, e.g., neural network, since they are essentially trying to learn less. Hence, adopting an appropriate model should be taken into consideration when dealing with practical problems. 
\item Incremental learning based methods. Edge devices could re-train a commonly-used pre-trained model in an incremental fashion to accommodate their new data. In such a manner, only few training instances are required to generate a customised model. 
\item Transfer learning based methods, e.g., few-shot learning. Transfer learning uses the learned knowledge from other models to enhance the performance of a related model, typically avoiding the cold-start problem and reducing the amount of required training data. Hence, transfer learning could be a possible solution, when there is not enough target training data, and the source and target domains have some similarities.
\item Data augmentation based methods. Data augmentation enables a model to be more robust by enriching data during the training phase \cite{mathur2018using}. For example, increasing the number of images without changing the semantic meaning of the labels through flipping, rotation, scaling, translation, cropping, etc. Through the training on augmented data, the network would be invariant to these deformations and have better performance to unseen data. 
\end{itemize}

\subsection{Data consistency on edge devices}\label{S5.3}
Edge intelligence based applications, e.g., speech recognition, activity recognition, emotion recognition, etc., usually collect data from large amounts of sensors distributed at the edge network. Nevertheless, the collected data may not be consistent. Two factors contribute to this problem: different sensing environments, and sensor heterogeneity. The environment (e.g., street and library) and its conditions (e.g., raining, windy) add background noise to the collected sensor data, which may have an impact on the model accuracy. The heterogeneity of sensors (e.g., hardware and software) may also cause the unexpected variation in their collected data. For example, different sensors have different sensitivities, sampling rates, and sensing efficiencies. Even the sensor data collected from the same source may vary on different sensors. Consequently, the variation of the data would result in the variation on the model training, e.g., the parameters of features \cite{das2014fingerprinting,stisen2015smart,mathur2018robustness}. Such variation is still a challenge for existing sensing applications. 

This problem could be solved easily if the model is trained in a centralised manner. The centralised large training set guarantees that the invariant features to the variations could be learned. However, this is not the scope of edge intelligence. Future efforts of this problem should focus on how to block the negative effect of the variation on model accuracy. To this end, two possible research directions maybe considered: data augmentation, and representation learning. Data augmentation could enrich the data during the model training process to enable the model to be more robust to noise. For example, adding various kinds of background noises to block the variation caused by the environments in speech recognition applications on mobile devices. Similarly, the noise caused by the hardware of sensors could also be added to deal with the inconsistency problem. Through the training of the augmented data, the models are more robust with these variations. 

Data representation heavily affects the performance of models. Representation leaning focuses on learning the representation of data to extract more effective features when building models \cite{bengio2013representation}, which could also be used to hide the differences between different hardware. For this problem, if we could make a `translation' on the representations between two sensors which are working on the same data source, the performance of the model would be improved significantly. Hence, representation learning is a promising solution to diminish the impact of data inconsistency. Future efforts could be made on this direction, e.g., design more effective processing pipelines and data transformations.



\subsection{Bad adaptability of statically trained model}\label{S5.2}
In most edge intelligence based AI applications, the model is first trained on a central server, then deployed on edge devices. The trained model will not be retrained, once the training procedure is finished. These statically trained models cannot effectively deal with the unknown new data and tasks in unfamiliar environments, which results in low performance and bad user experience. On the other hands, for models trained with a decentralised learning manner, only the local experience is used. Consequently, such models may become experts only in their small local areas. When the serving area broadens, the quality of service decreases.

To cope with this problem, two possible solutions may be considered: lifelong machine learning and knowledge sharing. Lifelong machine learning (LML)\cite{chen2016lifelong} is an advanced learning paradigm, which enables continuous knowledge accumulation and self-learning on new tasks. Machines are taught to learn new knowledge by themselves based on previously learned knowledge, instead of being trained by humans. LML is slightly different from meta learning \cite{vilalta2002perspective}, which enables machines to automatically learn new models. Edge devices with a series of learned tasks could use LML to adapt to changing environments and to deal with unknown data. It is worth recapping that the LML is not primarily designed for edge devices, which means that the machines are expected to be computationally powerful. Accordingly, model design, model compression, and offloading strategies should be also considered if LML is applied.

Knowledge sharing \cite{soller2002machine} enables the knowledge communication between different edge servers. When there is a task submitted to an edge server that does not have enough knowledge to provide a good service, the server could send knowledge queries to other edge servers. Since the knowledge is allocated on different edge servers, the server with the required knowledge responds to the query and performs the task for users. A knowledge assessment method and knowledge query system are required in such a knowledge sharing paradigm.

\subsection{Privacy and security issues}\label{S5.6}
To realise edge intelligence, heterogeneous edge devices and edge servers are required to work collaboratively to provide computing powers. In this procedure, the locally cached data and computing tasks (either training or inference tasks) might be sent to unfamiliar devices for further processing. The data may contain users' private information, e.g. photos and tokens, which leads to the risk of privacy leakage and attacks from malicious users. If the data is not encrypted, malicious users could easily obtain private information from the data. Some efforts \cite{kang2017neurosurgeon,raval2016you,li2018learning,eshratifar2018energy} propose to do some preliminary processing locally, which could hide private information and reduce the amount of transmitted data. However, it is still possible to extract private information from the processed data \cite{yang2019securing}. Moreover, malicious users could also attack and control a device that provides computing power through inserting a virus in the computing tasks. The key challenge is the lack of relevant privacy preserving and security protocols or mechanisms to protect users' privacy and security from being attacked. 

Credit system maybe a possible solution. This is similar with the credit system used in banks, which authenticates each user participated in the system and checks their credit information. Users with bad credit records would be deleted from the system. Consequently, all devices that provide computing powers are credible and all users are safe. 

Encryption could be used to protect privacy, which is already applied in some works \cite{konevcny2015federated,konevcny2016federated}. However, the encrypted data need to be decrypted before the training or inference tasks are executed, which requires an increase in the amount of computation needed. To cope with the problem, future efforts could pay more attention to homomorphic encryption \cite{gentry2009fully}. Homomorphic encryption refers to an encryption method that allows direct computation on ciphertexts and generate encrypted results. After decryption, the result is the same as the result achieved by computation on the unencrypted data. Hence, by applying homomorphic encryption, the training or inference task could be direct executed on encrypted data. 

\subsection{Incentive mechanism}\label{S5.5}

Data collection and model training/inference are two utmost important steps for edge intelligence. For data collection, it is challenging to ensure the quality and usability of information of the collected data. Data collectors consume their own resources, e.g., battery, bandwidth, and the time to sense and collect data. It is not realistic to assume that all data collectors are willing to contribute, let alone for preprocessing data cleaning, feature extraction and encryption, which further consumes more resources. For model training/inference in a collaborative manner, all participants are required to unselfishly work together for a given task. For example, the architecture proposed in \cite{huang2018task} consists of one master and multi workers. Workers recognise objects in a particular mobile visual domain and provides training instances for masters through pipelines. Such architecture works in private scenarios, e.g., at home, where all devices are inherently motivated to collaboratively create a better intelligent model for their master, i.e., their owner. However, it would not work well in public scenarios, where the master initialises a task and allocates sub-tasks to unfamiliar participants. In this context, additional incentive issue arises, which not typically considered in smart environments where all devices are not under the ownership of a single master. Participants need to be incentivised to perform data collection and task execution. 

Reasonable incentive mechanisms should be considered for future efforts. On one hand, participants have different missions, e.g., data collection, data processing, and data analysis, which have different resource consumptions. All participants hope to get as much reward as possible. On the other hand, the operator hopes to achieve the best model accuracy with as a low cost as possible. The challenges of designing the optimal incentive mechanism are how to quantify the workloads of different missions to match corresponding rewards and how to jointly optimise these two conflicting objectives. Future efforts could focus on addressing these challenges.
\section{Conclusions}\label{S6}
In this paper, we present a thorough and comprehensive survey on the literature surrounding edge intelligence. Specifically, we identify critical components of edge intelligence: edge caching, edge training, edge inference, and edge offloading. Based on this, we provide a systematic classification of literature by reviewing research achievements for each components and present a systematical taxonomy according to practical problems, adopted techniques, application goals, etc. We compare, discuss and analyse literature in the taxonomy from multi-dimensions, i.e., adopted techniques, objectives, performance, advantages and drawbacks, etc. Moreover, we also discuss important open issues and present possible theoretical research directions. Concerning the era of edge intelligence, We believe that this is only the tip of iceberg. Along with the explosive development trend of IoT and AI, we expect that more and more research efforts would be carried out to completely realize edge intelligence in the following decades.

\ifCLASSOPTIONcaptionsoff
  \newpage
\fi

\begin{IEEEbiography}[{\includegraphics[width=1in,height=1.25in,clip,keepaspectratio]{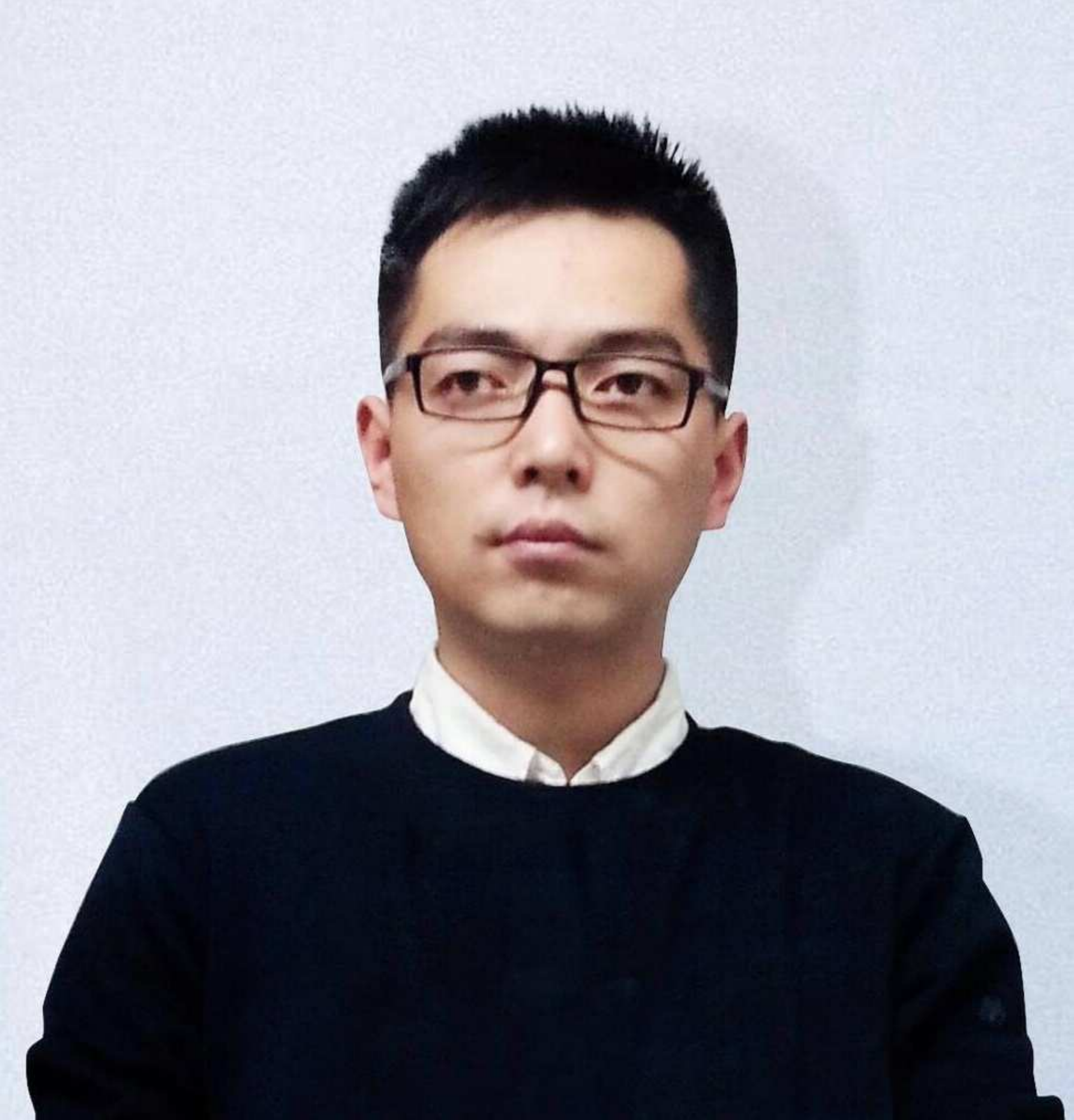}}]{Dianlei Xu}
 received the B.S. degree from Anhui University, Hefei, China, and is currently a joint doctoral student in the Department of Computer Science, Helsinki, Finland and Beijing National Research Center for Information Science and Technology (BNRist), Department of Electronic Engineering, Tsinghua University, Beijing, China.

 His research interests include edge/fog computing and AIoT.
\end{IEEEbiography}

\begin{IEEEbiography}[{\includegraphics[width=1in,height=1.25in,clip,keepaspectratio]{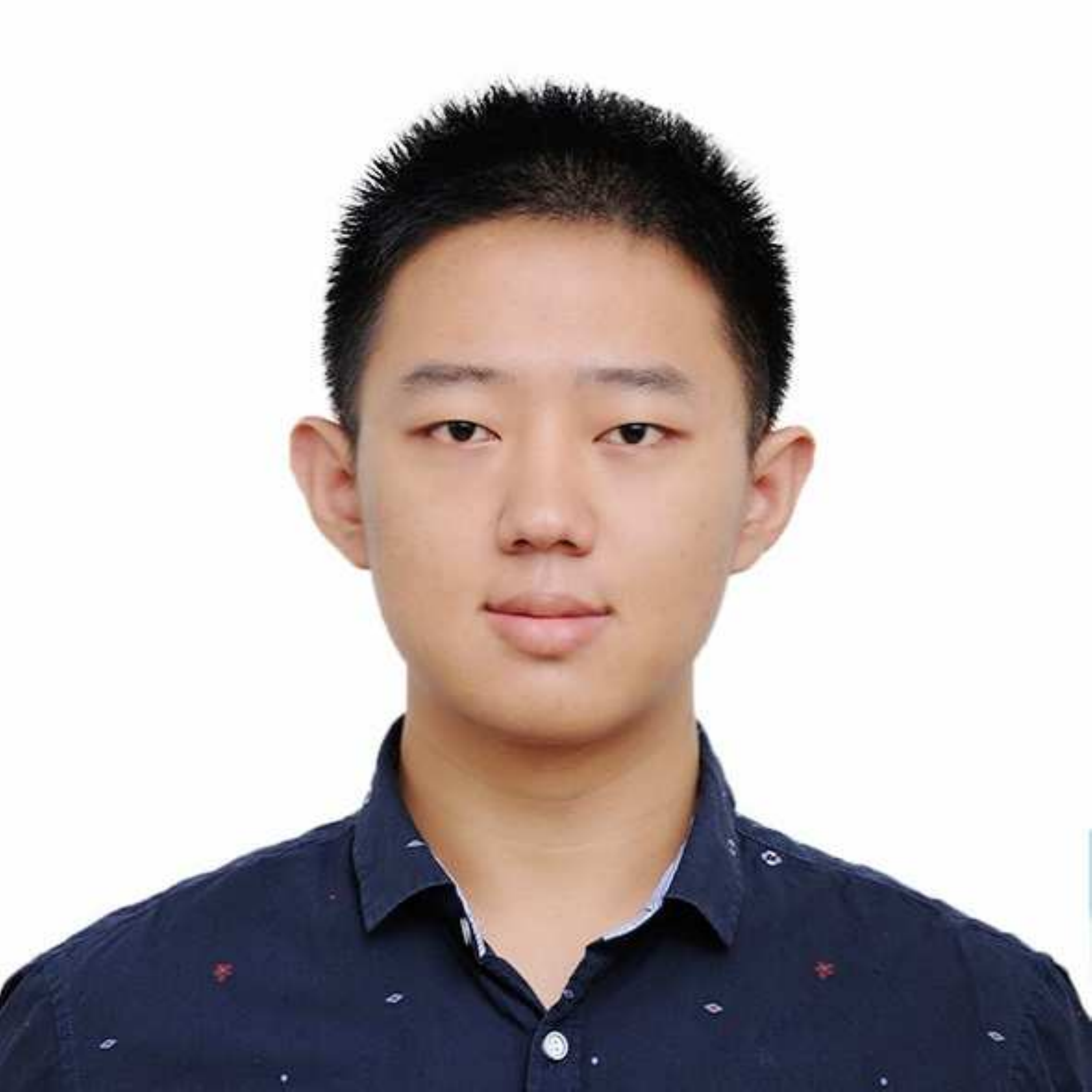}}]{Tong Li}
received the B.S. degree and M.S. degree in communication engineering from Hunan University, China, in 2014 and 2017.
At present, he is a dual Ph.D. student at the Hong Kong University of Science and Technology and the University of Helsinki. His research interests include distributed systems, edge network, and data-driven network. He is an IEEE student member.\
\end{IEEEbiography}

\begin{IEEEbiography}[{\includegraphics[width=1in,height=1.25in,clip,keepaspectratio]{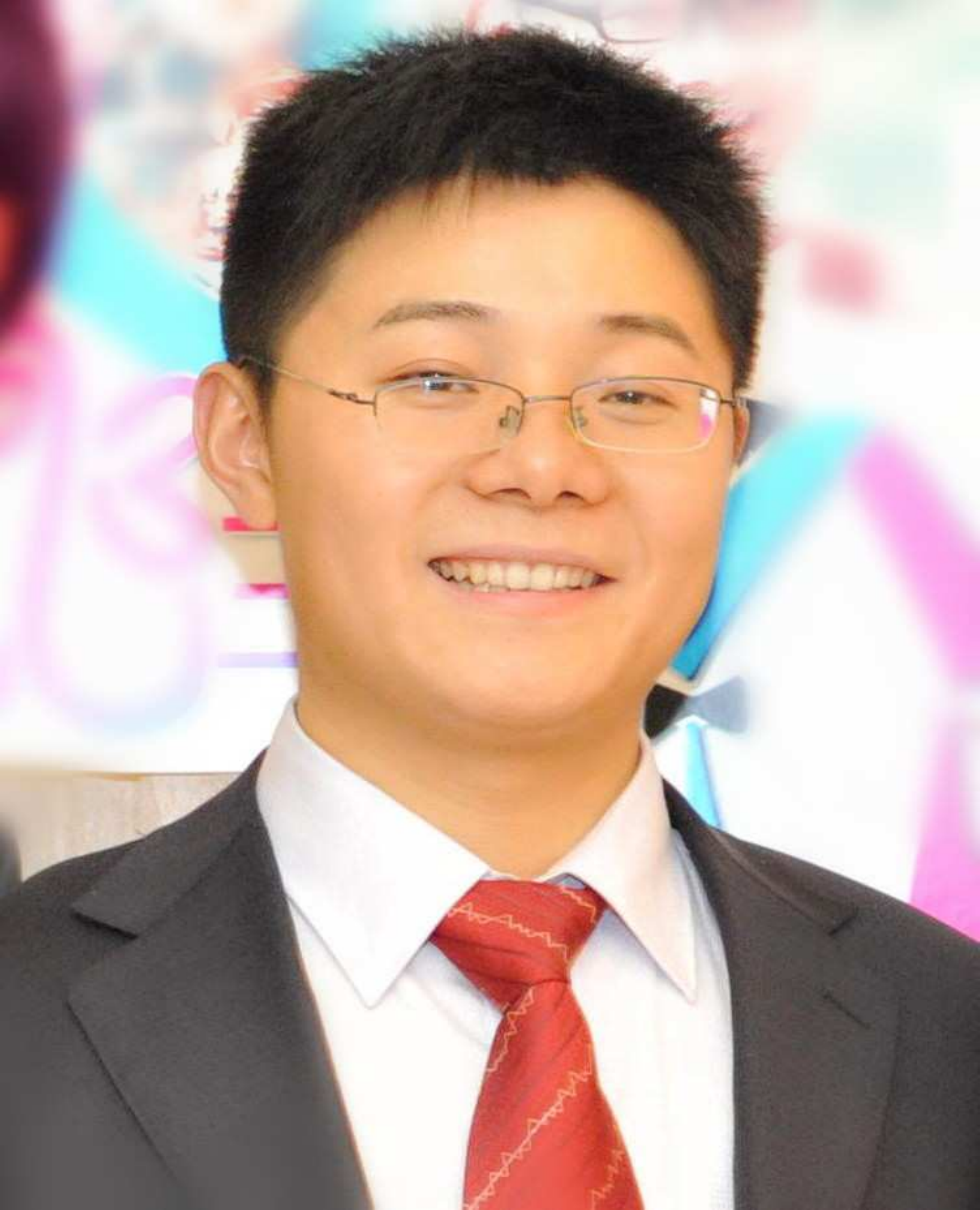}}]{Yong Li}(M'09-SM'16)
received the B.S. degree in electronics and information engineering from Huazhong University of Science and Technology, Wuhan, China, in 2007 and the Ph.D. degree in electronic engineering from Tsinghua University, Beijing, China, in 2012. He is currently a Faculty Member of the Department of Electronic Engineering, Tsinghua University.

Dr. Li has served as General Chair, TPC Chair, SPC/TPC Member for several international workshops and conferences, and he is on the editorial board of two IEEE journals. His papers have total citations more than 6900. Among them, ten are ESI Highly Cited Papers in Computer Science, and four receive conference Best Paper (run-up) Awards. He received IEEE 2016 ComSoc Asia-Pacific Outstanding Young Researchers, Young Talent Program of China Association for Science and Technology, and the National Youth Talent Support Program.

 \end{IEEEbiography}

\begin{IEEEbiography}[{\includegraphics[width=1in,height=1.25in,clip,keepaspectratio]{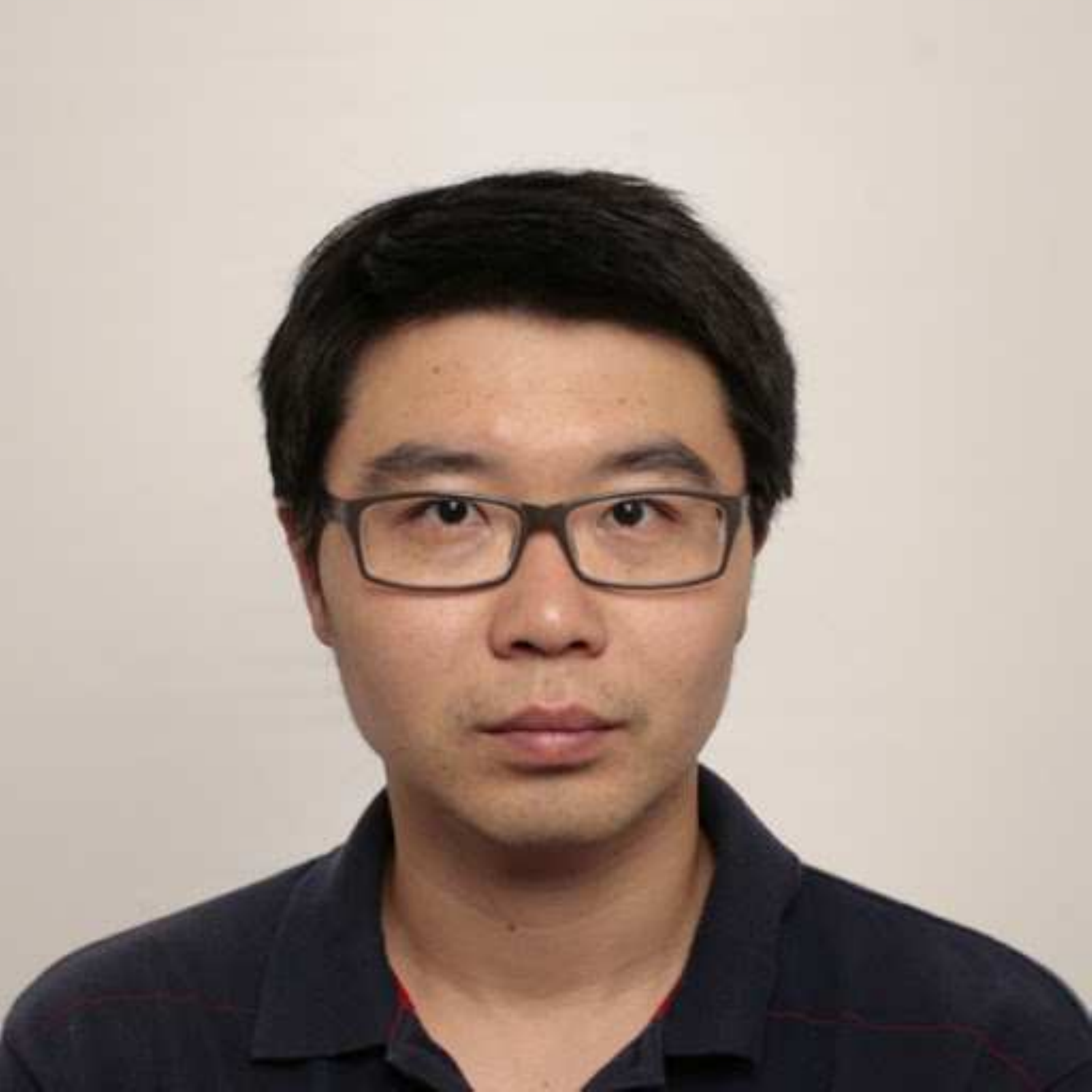}}]{Xiang Su}
received his Ph.D. in technology from the University of Oulu in 2016. He is currently an Academy of Finland postdoc fellow and a senior postdoctoral researcher in computer science in the University of Helsinki. Dr. Su has extensive expertise on Internet of Things, edge computing, mobile augmented reality, knowledge representations, and context modeling and reasoning.

\end{IEEEbiography}

\begin{IEEEbiography}[{\includegraphics[width=1in,height=1.25in,clip,keepaspectratio]{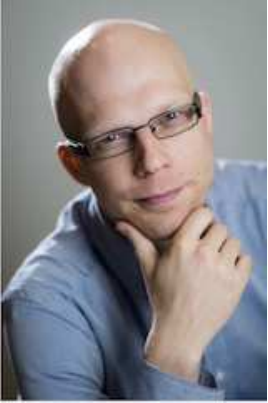}}]{Sasu Tarkoma}
received the MSc and PhD degrees in computer science from the Department of Computer Science, University of Helsinki. He is a full professor at the Department of Computer Science, University of Helsinki, and the deputy head of the department. He has managed and participated in national and international research projects at the University of Helsinki, Aalto University, and the Helsinki Institute for Information Technology. His research interests include mobile computing, Internet technologies, and middleware. He is a senior member of the IEEE.
\end{IEEEbiography}

\begin{IEEEbiography}[{\includegraphics[width=1in,height=1.25in,clip,keepaspectratio]{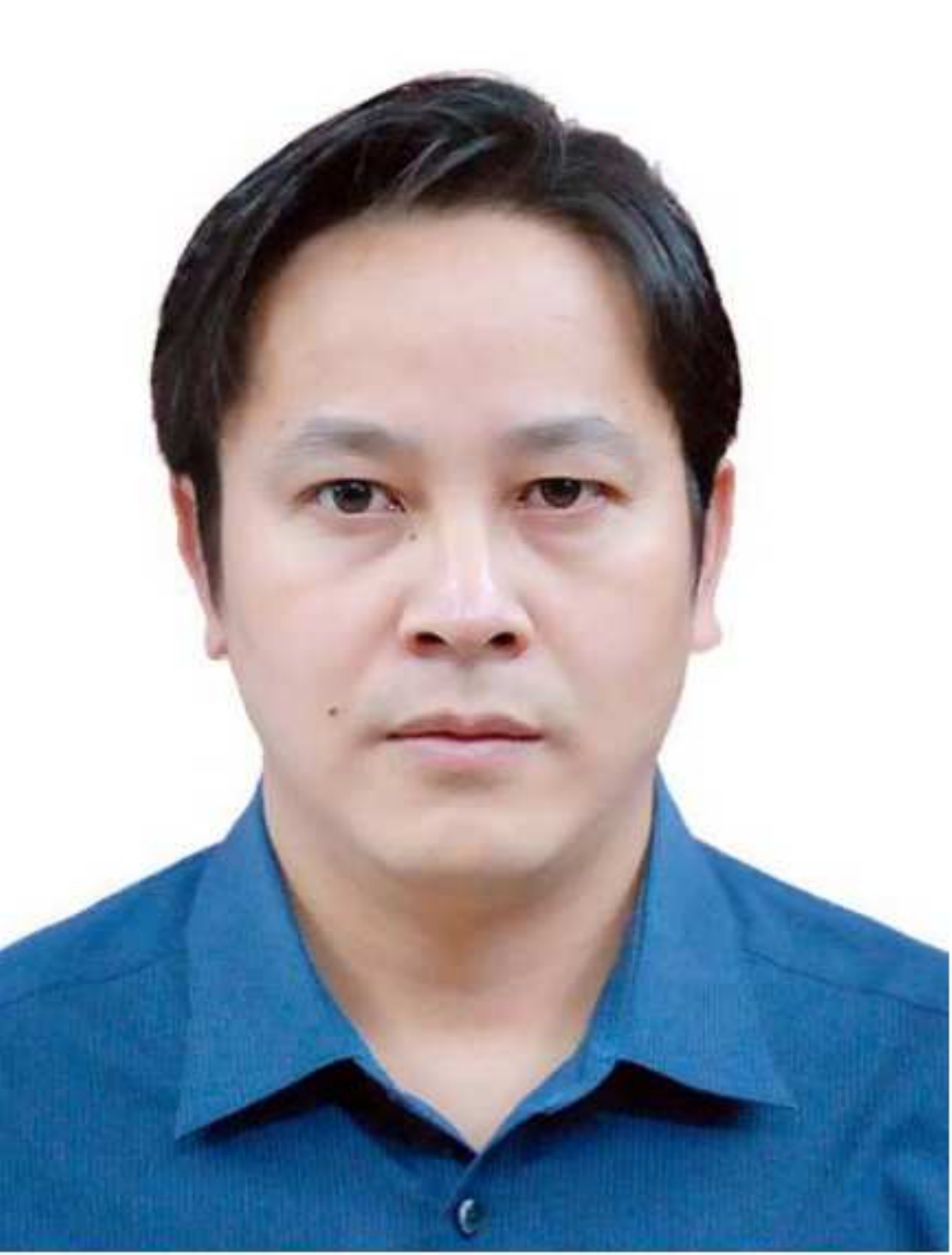}}]{Tao Jiang}
(M'06-SM'10-F'19) received the Ph.D. degree in information and communication engineering from the Huazhong University of Science and Technology, Wuhan, China, in April 2004. From August 2004 to December 2007, he worked in some universities, such as Brunel University and University of Michigan-Dearborn, respectively. He is currently a Distinguished Professor with the Wuhan National Laboratory for Optoelectronics and School of Electronics Information and Communications, Huazhong University of Science and Technology. He has authored or coauthored more than 300 technical articles in major journals and conferences and nine books/chapters in the areas of communications and networks. He served or is serving as symposium technical program committee membership of some major IEEE conferences, including INFOCOM, GLOBECOM, and ICC. He was invited to serve as a TPC Symposium Chair for the IEEE GLOBECOM 2013, the IEEEE WCNC 2013, and ICCC 2013. He is served or serving as an Associate Editor of some technical journals in communications, including in the IEEE NETWORK, the IEEE TRANSACTIONS ON SIGNAL PROCESSING, the IEEE COMMUNICATIONS SURVEYS AND TUTORIALS, the IEEE TRANSACTIONS ON VEHICULAR TECHNOLOGY, and the IEEE INTERNET OF THINGS JOURNAL. He is the Associate Editor-in-Chief of China Communications.
\end{IEEEbiography}

\begin{IEEEbiography}[{\includegraphics[width=1in,height=1.25in,clip,keepaspectratio]{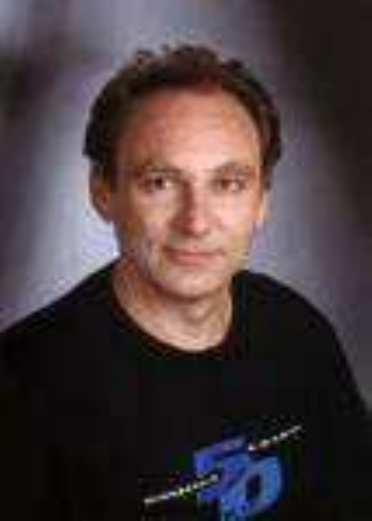}}]{Jon Crowcroft}
(SM'95-F'04) graduated in physics from Trinity College, Cambridge University, United Kingdom, in 1979, and received the MSc degree in computing in 1981 and the PhD degree in 1993 from University College London (UCL), United Kingdom. He is currently the Marconi Professor of Communications Systems in the Computer Lab at the University of Cambridge, United Kingdom. Professor Crowcroft is a fellow of the United Kingdom Royal Academy of Engineering, a fellow of the ACM, and a fellow of IET. He was a recipient of the ACM Sigcomm Award in 2009.
\end{IEEEbiography}

\begin{IEEEbiography}[{\includegraphics[width=1in,height=1.25in,clip,keepaspectratio]{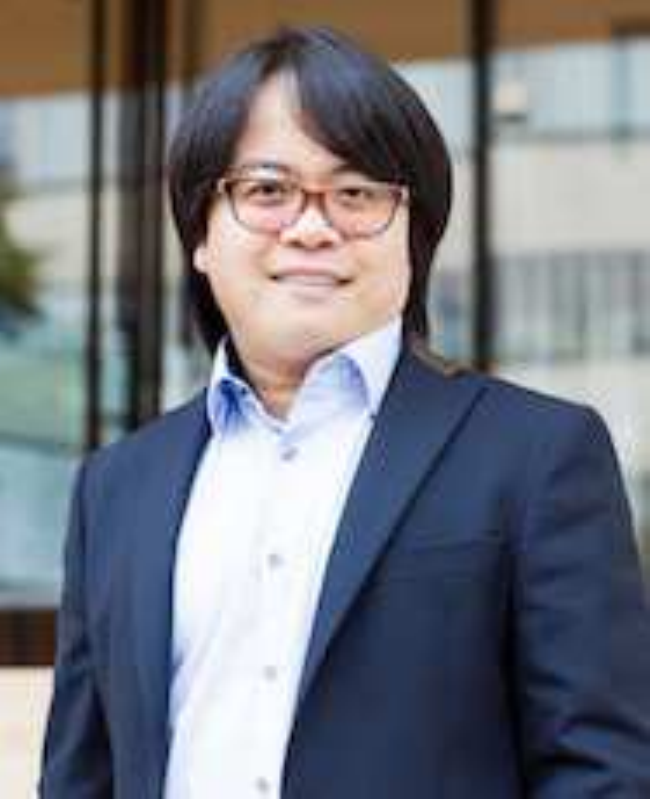}}]{Pan Hui}
 (SM'14-F'18) received his PhD from the Computer Laboratory at University of Cambridge, and both his Bachelor and MPhil degrees from the University of Hong Kong.

 He is the Nokia Chair Professor in Data Science and Professor of Computer Science at the University of Helsinki. He is also the director of the HKUST-DT System and Media Lab at the Hong Kong University of Science and Technology. He was an adjunct Professor of social computing and networking at Aalto University from 2012 to 2017. He was a senior research scientist and then a Distinguished Scientist for Telekom Innovation Laboratories (T-labs) Germany from 2008 to 2015.  His industrial profile also includes his research at Intel Research Cambridge and Thomson Research Paris from 2004 to 2006. His research has been generously sponsored by Nokia, Deutsche Telekom, Microsoft Research, and China Mobile. He has published more than 300 research papers and with over 17,500 citations. He has 30 granted and filed European and US patents in the areas of augmented reality, data science, and mobile computing.

 He has founded and chaired several IEEE/ACM conferences/workshops, and has served as track chair, senior program committee member, organising committee member, and program committee member of numerous top conferences including ACM WWW, ACM SIGCOMM, ACM Mobisys, ACM MobiCom, ACM CoNext, IEEE Infocom, IEEE ICNP, IEEE ICDCS, IJCAI, AAAI, and ICWSM. He is an associate editor for the leading journals IEEE Transactions on Mobile Computing and IEEE Transactions on Cloud Computing. He is an IEEE Fellow, an ACM Distinguished Scientist, and a member of Academia Europaea.
\end{IEEEbiography}

\end{document}